\documentclass[%
twocolumn,
superscriptaddress,
amsmath,amssymb,
prb,
]{revtex4-1}

\usepackage{graphicx}
\usepackage{dcolumn}
\usepackage{bm}
\usepackage{hyperref}
\usepackage{float}
\usepackage{color}
\usepackage{braket}
\usepackage{subcaption}
\usepackage{mathtools}
\usepackage{bbold}
\usepackage{multirow}
\usepackage{calligra} 
\usepackage{yfonts} 
\usepackage{physics}
\usepackage{tikz-feynman}
\usetikzlibrary{
  decorations.pathmorphing,
  positioning,
  arrows.meta,
  fit,
  backgrounds,
  decorations.pathreplacing,
  calc
}
\usepackage[top=1cm, bottom=1cm, left=1cm, right=1cm]{geometry}
\newcommand{\editor}[2]{%
  \expandafter\newcommand\csname #1note\endcsname[1]{%
    \textcolor{#2}{(\textbf{#1:} \textit{##1})}}%
  \expandafter\newcommand\csname #1\endcsname[1]{%
    \textcolor{#2}{##1}}%
  \expandafter\newcommand\csname #1cancel\endcsname[1]{%
    \textcolor{#2}{\sout{##1}}}%
  \expandafter\newcommand\csname #1change\endcsname[2]{%
    \textcolor{#2}{\sout{##1} ##2}}%
  \newenvironment{#1text}{\color{#2}}{\color{black}}
}
\input{F_diagrams.tex}

\usepackage[color = teal]{changes}
\definecolor{BluBondi}{rgb}{0.00,0.58,0.71}
\definecolor{Orange}{rgb}{1.00,0.45,0.00}
\editor{AG}{BluBondi}
\editor{FM}{blue}

\begin{document}

\preprint{}

\title{The principle of detailed balance between electrons and phonons in presence of excitonic effects}

\author{Alberto Guandalini}
\thanks{These authors contributed equally to this work.}
\affiliation{Dipartimento di Fisica, Università di Roma La Sapienza, Roma, Italy}
\author{Giovanni Caldarelli}
\thanks{These authors contributed equally to this work.}
\affiliation{Theory and Simulation of Materials (THEOS), and National Centre for Computational Design and Discovery of Novel Materials (MARVEL), \'Ecole Polytechnique F\'ed\'erale de Lausanne, Lausanne, Switzerland}
\author{Francesco Mauri}
\affiliation{Dipartimento di Fisica, Università di Roma La Sapienza, Roma, Italy}
\author{Francesco Macheda}
\email{francesco.macheda@unimore.it}
\affiliation{Dipartimento di Scienze e Metodi dell’Ingegneria, Università di Modena e Reggio Emilia, Reggio Emilia, Italy}


\begin{abstract}
Based on a many-body formulation, we derive an electron-phonon coupling including excitonic effects that preserves thermodynamic detailed balance between electronic and phononic scattering processes. We start from the microscopic electron-nucleus Hamiltonian, expand around the Born-Oppenheimer equilibrium geometry, and construct an effective action for the electronic and phononic propagators. From the same effective action, we derive both electronic and phononic self-energies in terms of a nonlocal vertex \(\mathcal G^{\textrm{s}}\) including excitonic effects, which generalizes the usual local interaction vertex \(g^{\textrm{s}}\). When electrons and phonons are well-defined quasiparticles in the screened-exchange approximation and \(\mathcal G^{\textrm{s}}\) is taken in its static limit, both self-energies reduce to Fermi-golden-rule expressions containing the same \(\mathcal G^{\textrm{s}}\), thereby ensuring detailed balance. As an application, we compute electronic and phononic linewidths in graphene and illustrate this common-vertex construction. We analyze the competition between the reduced scattering phase space induced by the screened-exchange band structure and the enhancement of the electron-phonon vertex due to excitonic effects, finding that the vertex enhancement can compensate for and overcome the phase-space reduction in both electronic and phononic linewidths.
\end{abstract}

\maketitle 

\section{Introduction}

The electron-phonon interaction is one of the most studied couplings in condensed-matter physics. It controls phonon linewidths, electronic lifetimes, charge and heat transport, Raman spectra, hot-carrier relaxation, temperature-dependent electronic structures, and phonon-mediated superconductivity \cite{Grimvall1981TheEI,ziman2001electrons,Giustino2017}. In first-principles calculations, the electron-phonon coupling is usually defined as the derivative of a self-consistent static electronic potential with respect to the atomic displacements \cite{RevModPhys.73.515}. This definition is extremely successful because it naturally leads to Fermi golden rule expressions for electron-phonon scattering rates and can be efficiently combined with density-functional perturbation theory (DFPT) and Wannier interpolation \cite{PhysRevB.76.165108,PONCE2016116,Pizzi2020,Lee2020,MARINI2024108950,Lee2023,ZHOU2021107970}. 

Electron-hole interaction and the associated vertex corrections in the electronic response are not explicitly included in standard first-principles electron-phonon calculations. This approximation is problematic when screening is weak, which is typical, for example, of two-dimensional materials, where charges interact through electric fields extending far outside the material thickness \cite{PhysRevB.84.085406,PhysRevB.88.045318,PhysRevLett.113.076802,PhysRevB.92.245123,PhysRevLett.111.216805,PhysRevB.93.235435}. Indeed, several works have shown that many-body effects can strongly modify electron-phonon couplings. In the prototypical examples of graphene and graphite, nonlocal exchange-correlation effects enhance the electron-phonon coupling of zone-boundary optical phonons beyond the local and semilocal density-functional approximations, improving the description of Kohn anomalies, Raman spectra, and phonon lifetimes \cite{PhysRevB.78.081406,Venanzi2023,PhysRevB.84.035433,Graziotto2024,y1dn-m6pc}. More generally, many-body perturbation theory (MBPT) has been used to compute electron-phonon matrix elements by differentiating effective quasiparticle Hamiltonians or self-energies with respect to atomic displacements \cite{PhysRevLett.122.186402,Li2024GWPT,Laricchia2025QSGW}. These approaches demonstrate the importance of many-body corrections, but they generally rely on a one-particle construction of the electron-phonon vertex. 

While MBPT has already been extended to include electron and lattice excitations on an equal footing \cite{kadanoff1976quantum,PhysRevX.13.031026,Giustino2017}, its practical applications are historically restricted to the renormalization of the real parts of electronic and phononic self-energies, rather than imaginary parts \cite{Antonius2014ZPR,Monserrat2016Correlation,PhysRevLett.98.067005,PhysRevLett.105.265501,6m5p-mmg3}. Extending the same framework to the imaginary parts of the self-energies is more delicate: electronic and phononic linewidths describe inverse scattering processes and therefore cannot, in general, be approximated independently. A consistent treatment must relate the decay of a phonon into an electron-hole pair to the corresponding electronic process involving phonon emission or absorption.  

From a many-body perspective, electron-hole interaction is explicitly included at the level of the two-particle response, through Bethe-Salpeter ladder diagrams and the associated vertex corrections \cite{PhysRev.139.A796,RevModPhys.74.601,PhysRevB.62.4927,PhysRevB.29.5718,Reining_16,stefanucci2013nonequilibrium,PhysRevB.106.125403}. Their consistent inclusion in electron-phonon physics is timely, as recent developments have addressed dynamically screened electron-phonon couplings and positive phonon-linewidth approximations \cite{PhysRevB.111.024307}, excitonic Bloch equations \cite{SciPostPhys.18.1.009}, nonperturbative electron-phonon spectral functions \cite{PhysRevLett.134.186401}, vertex-corrected transport \cite{mtjf-ylf7,83vn-7zyw}, and superconductivity \cite{Mishra_2025}. Electron-hole correlations may be represented in different but complementary ways. Recent approaches to excited-state dynamics introduce excitons as explicit quasiparticles obtained from a Bethe-Salpeter problem and subsequently describe their interaction with electronic or lattice degrees of freedom \cite{PerfettoWuStefanucci2024,StefanucciPerfetto2026,PhysRevB.105.085111,PhysRevB.102.045136,PhysRevB.106.125403}. This representation is convenient when the physical observables are governed by the formation, propagation, or relaxation of bounded excitons. In this work, we adopt a different perspective, appropriate when electron-hole interaction influence mainly the continuous part of the electron excitations, e.g. graphene\cite{guandalini_nanoletter}. In our framework the electron-hole correlations are transferred from the interacting two-particle response to the interaction vertices. The intermediate electron and hole then propagate as independent electronic quasiparticles, while excitonic corrections are retained in a nonlocal electron-phonon vertex. This organization is particularly convenient for electronic and phononic scattering rates because it preserves the conventional quasiparticle interpretation of the initial and final states while incorporating electron-hole correlations in the scattering amplitude.

A fundamental question then arises: if the electron-phonon vertex is dressed by excitonic corrections, which vertex should be used in the phonon self-energy, and which one should be used in the electronic self-energy? These two self-energies are often discussed separately. The imaginary part of the phonon self-energy gives the decay rate of a phonon into an electron-hole pair, whereas the imaginary part of the electronic self-energy gives the rate for electron scattering through phonon absorption or emission. At thermal equilibrium, however, these are inverse microscopic processes. Therefore, when electrons and phonons are well-defined quasiparticles, the corresponding Fermi golden rule expressions must be mutually consistent and satisfy detailed balance. Within MBPT, the electronic and phononic self-energies must be built from Feynman diagrams that preserve detailed balance. To the best of our knowledge, this consistency requirement has never been addressed.

In this work, we devise a compact theoretical scheme for interacting electrons and phonons that enforces the required detailed-balance relation. We start from the microscopic electron-nucleus Hamiltonian and expand it around the Born-Oppenheimer equilibrium geometry \cite{doi:10.1119/1.1934059,PhysRevB.1.910,PhysRevB.13.694,Marini_1}. This procedure identifies the reference electronic and phononic Hamiltonians, the electron-electron interaction, the linear electron-phonon interaction, and the static correction associated with the Born-Oppenheimer force constants. Starting from the microscopic Hamiltonian makes the content of the approximation explicit and helps avoid ambiguities associated with empirical vertex modifications or double counting. We then reformulate the theory in terms of an effective action functional for the electronic and phononic propagators, \(G\) and \(D\). The electron-phonon part of the functional is constructed by selecting, in a controlled way, the sector that contains a single phonon propagator \(D\). This functional construction is essential because the electronic and phononic self-energies are obtained from the same approximate functional, following the conserving approach of Luttinger, Ward, Baym, and Kadanoff \cite{PhysRev.118.1417,PhysRev.127.1391,PhysRevX.13.031026,BaymKadanoff1961}. In other words, the vertex corrections entering electron and phonon scattering are not introduced separately, but arise from the same diagrammatic approximation. As a result, the two subsystems are treated coherently. 

Within this framework, electronic screening first transforms the bare electron-phonon coupling \(g\) into the screened vertex \(g^{\textrm{s}}\). Electron-hole ladder corrections further dress this quantity into the nonlocal vertex \(\mathcal G^{\textrm{s}}\). The same \(\mathcal G^{\textrm{s}}\) enters both the phonon self-energy and the electron-phonon contribution to the electronic self-energy. In the quasiparticle limit, using a static screened-exchange approximation for the electronic self-energy and the static limit of the electron-phonon vertex, the imaginary part of both self-energies reduce to Fermi golden rule expressions containing the same interaction matrix element. Electron and phonon scattering processes therefore satisfy detailed balance by construction.

As an application, we consider graphene. Graphene is an ideal test case because electron-phonon coupling is responsible for Kohn anomalies, phonon linewidths, Raman features, and intrinsic resistivity \cite{PhysRevLett.97.266407,PhysRevLett.99.086804,PhysRevB.90.125414}, while electron-electron and electron-hole interactions strongly reshape its electronic response \cite{RevModPhys.84.1067,Gonzalez1994,Bostwick2007,Elias2011,PhysRevB.111.075118,PhysRevB.111.075137}. Building on our previous results \cite{PhysRevB.111.075118,PhysRevB.111.075137,y1dn-m6pc}, we determine the screened-exchange electronic bands and phonon frequencies and obtain the excitonically enhanced electron-phonon coupling by solving the Bethe-Salpeter equation. The comparison among different theory levels allows us to separate the effects of the quasiparticle spectrum from those of the vertex correction. These contributions act in opposite directions. The screened-exchange band structure increases the electronic velocity and reduces the available scattering phase space, thereby suppressing the linewidth. The Bethe-Salpeter vertex correction instead enhances the electron-phonon matrix element. The resulting comparison shows, for both electronic and phononic linewidths, how the same vertex enhancement competes with and can overcome the phase-space reduction induced by the screened-exchange band structure.

Finally, our results provide a controlled route for extending many-body perturbation theory from the renormalization of quasiparticle energies and phonon frequencies to the imaginary parts of electronic and phononic self-energies. The resulting static quasiparticle vertices can be incorporated directly into calculations of electronic transport \cite{PhysRevB.98.201201,PhysRevB.102.094308,Macheda2020,PhysRevB.100.085204,PhysRevResearch.3.043022,Ponce2020,PhysRevB.94.201201,PhysRevB.97.045201,PhysRevResearch.2.033055,macheda2026coupleddynamicalboltzmanntransport,dinar2026electronictransportbnencasulatedgraphene}, Raman and infrared spectra  \cite{PhysRevB.103.134304,Gruneis2009,Ferrari2013}, excited-carrier and exciton relaxation \cite{PhysRevB.84.075449,Harb2006,Bernardi2014,Bernardi2016,PhysRevLett.125.107401,Betz2013,Tielrooij2018,Marini2021,Mocatti2026}, and superconductivity \cite{Bardeen1957,Migdal1958,Eliashberg1960,PhysRevB.6.2577,PhysRevB.12.905,Schrieffer1999,Marsiglio2020}. 
%
%

\begin{figure}[t]
\centering

\begin{tikzpicture}[
    font=\sffamily\small,
    node distance=0.62cm,
    arrow/.style={
        -{Latex[length=2.2mm,width=1.5mm]},
        line width=0.8pt,
        draw=black
    },
    box/.style={
        draw=black,
        fill=white,
        rounded corners=2pt,
        align=center,
        inner sep=7pt,
        minimum height=1.55cm,
        text width=0.84\columnwidth
    },
    widebox/.style={
        draw=black,
        fill=white,
        rounded corners=2pt,
        align=center,
        inner sep=7pt,
        minimum height=1.88cm,
        text width=0.84\columnwidth
    },
    eqlabel/.style={
        font=\sffamily\scriptsize,
        text=black
    },
    seclabel/.style={
        font=\sffamily\small,
        rotate=90,
        anchor=center,
        inner sep=0pt
    }
]

\node[box] (ham) {
    \textbf{Low-energy reduction}\\[0.48em]
    {\large
    \(\displaystyle
    H
    \longrightarrow
    H^{\textrm{low-en}}
    \)}
    \\[0.42em]
    { Eq.~\eqref{eq:lowenH}}
};

\node[box, below=of ham] (phi) {
    \textbf{Single-phonon functional}\\[0.48em]
    {\large
    \(\displaystyle
    \Phi[G,D,v,g]
    \longrightarrow
    \widetilde{\Phi}[G,D,v,g]
    \)}
    \\[0.42em]
    {
    Eqs.~\eqref{eq:tildePhidecomp} and \eqref{eq:phiephdef}}
};

\node[widebox, below=of phi] (self) {
    \textbf{Self-energies from the single-phonon functional}\\[0.50em]
    {\large
    \(\displaystyle
    \frac{\delta\widetilde{\Phi}}{\delta G}
    \longrightarrow
    \Sigma
    \qquad
    -2\frac{\delta\widetilde{\Phi}}{\delta D}
    \longrightarrow
    \Pi
    \)}
    \\[0.42em]
    {
    Eqs.~\eqref{eq:Sigmaderphitilde} and
    \eqref{eq:Piderphitilde}}
};

\node[widebox, below=of self] (vertex) {
    \textbf{Screened and excitonic vertex}\\[0.50em]
    {\large
    \(\displaystyle
    g
    \xrightarrow{\ \varepsilon^{-1}\ }
    g^{\textrm{s}}
    \qquad
    g^{\textrm{s}}
    \xrightarrow{\ \Gamma'\ }
    \mathcal G^{\textrm{s}}
    \)}
    \\[0.42em]
    {
    Eqs.~\eqref{eq:gsdef} and
    \eqref{eq:def-Gcal-right}}
};

\node[widebox, below=1.00cm of vertex] (fgr) {
    \textbf{Quasiparticle linewidths}\\[0.50em]
    {\large
    \(\displaystyle
    \begin{aligned}
    \Gamma^{\textrm{el,SX}}
    &=
    \Gamma^{\textrm{el}}
    [\mathcal G^{\textrm{s,SX}},
    \varepsilon^{\textrm{SX}},
    \omega^{\textrm{SX}}],
    \\[0.55em]
    \Gamma^{\textrm{ph,SX}}
    &=
    \Gamma^{\textrm{ph}}
    [\mathcal G^{\textrm{s,SX}},
    \varepsilon^{\textrm{SX}},
    \omega^{\textrm{SX}}].
    \end{aligned}
    \)}
    \\[0.40em]
    {
    Eqs.~\eqref{eq:electron_linewidth_FGR_SX} and
    \eqref{eq:linewidth_FGR_explicit_static}}
};

\draw[arrow] (ham) -- (phi);
\draw[arrow] (phi) -- (self);
\draw[arrow] (self) -- (vertex);
\draw[arrow]
    (vertex) --
    node[midway, right=0.14cm, font=\sffamily\small, align=left]
    {SX approximation}
    (fgr);

\node[seclabel] at ([xshift=-0.48cm]ham.west)
    {Sec.~\ref*{sec:Ham}};

\node[seclabel] at ([xshift=-0.48cm]phi.west)
    {Sec.~\ref*{sec:action}};

\node[seclabel] at ([xshift=-0.48cm]fgr.west)
    {Sec.~\ref*{sec:fgr}};

\draw[
    decorate,
    decoration={brace,amplitude=5pt,mirror},
    line width=0.75pt
]
([xshift=-0.16cm,yshift=0.03cm]self.north west)
--
([xshift=-0.16cm,yshift=-0.03cm]vertex.south west);

\node[seclabel] at
    ($(self.west)!0.5!(vertex.west)+(-0.68cm,0)$)
    {Sec.~\ref*{sec:selfenergies}};

\end{tikzpicture}

\caption{
Workflow of the theoretical construction. The microscopic Hamiltonian
is reduced to the low-energy model and used to define the
single-phonon functional. Its functional derivatives generate the electronic and
phononic self-energies. Screening and electron-hole ladder corrections
produce, in turn, the dressed vertices \( g^{\textrm{s}}\) and \(\mathcal G^{\textrm{s}}\). In the
screened-exchange quasiparticle approximation and static-vertex limit,
the electronic and phononic linewidths depend on the same vertex,
electronic energies, and phonon frequencies.}
\label{fig:theory-workflow}
\end{figure}
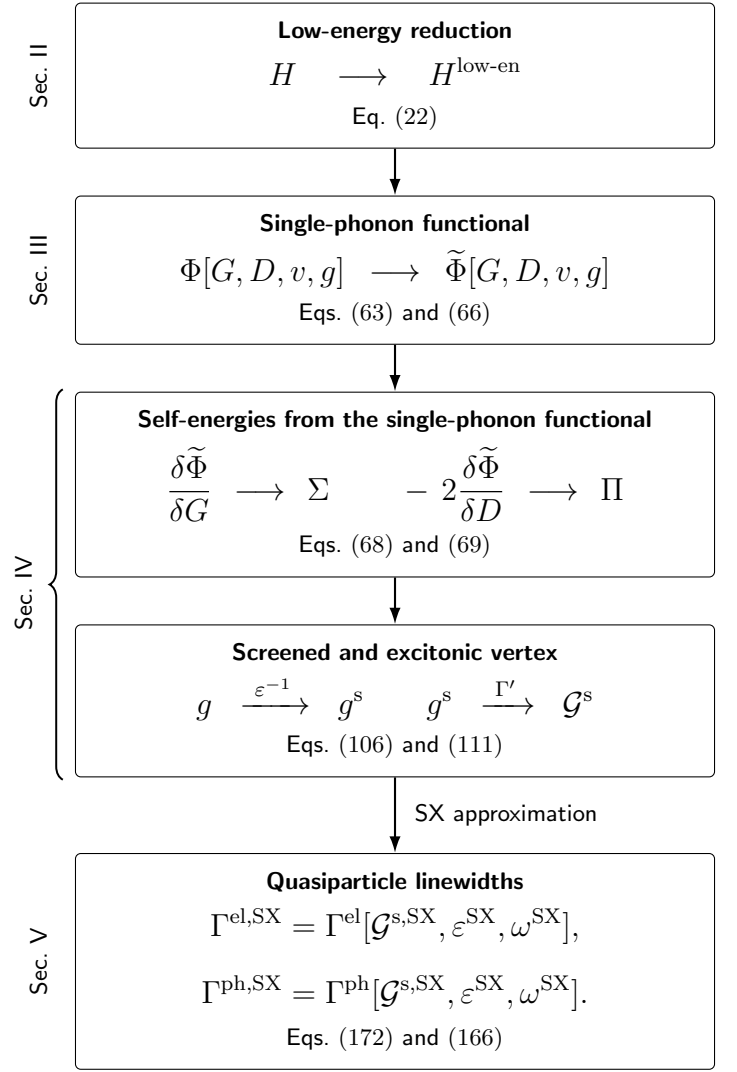

The paper is organized as follows. In Sec.~\ref{sec:Ham}, we derive a low-energy electron-phonon Hamiltonian by expanding the microscopic electron-nucleus problem around the Born-Oppenheimer equilibrium geometry. In Sec.~\ref{sec:action}, we formulate the theory in terms of an effective action and introduce the controlled approximation retaining the sector that contains a single phonon propagator. In Sec.~\ref{sec:selfenergies}, we derive the electronic and phononic self-energies and identify the screened vertex \(g^{\textrm{s}}\) and the excitonically dressed nonlocal vertex \(\mathcal G^{\textrm{s}}\). In Sec.~\ref{sec:fgr}, we specialize the theory to the screened-exchange approximation and derive the corresponding quasiparticle linewidths in the static-vertex limit. The logical structure of the derivation, from the microscopic Hamiltonian to the quasiparticle linewidths, is summarized in Fig.~\ref{fig:theory-workflow}. In Sec.~\ref{sec:graphene}, we apply the formalism to graphene and disentangle the effects of band and vertex renormalization. Finally, Sec.~\ref{sec:conclusions} summarizes our results and discusses their implications.

\section{Reduction to a low-energy Hamiltonian}
\label{sec:Ham}
We begin with the microscopic reduction represented by the first step of
Fig.~\ref{fig:theory-workflow}. Starting from the exact Hamiltonian of
interacting electrons and nuclei, we construct a low-energy Hamiltonian by
expanding around a reference nuclear configuration and its corresponding
electronic ground state. We choose as reference the Born--Oppenheimer (BO)
equilibrium geometry and the electronic BO ground state evaluated at that
geometry. The expansion is truncated at second order in deviations from this
reference. Other reference configurations could in principle be used, including
the exact equilibrium geometry of the complete electron-nuclear problem. The
BO reference, however, provides a well-defined and computationally accessible
starting point that can be connected directly to first-principles calculations.

We consider a system of \(N_{\textrm e}\) electrons and \(N_{\textrm n}\)
nuclei. The electronic coordinates are denoted by \(\mathbf r_i\). The nuclear
coordinate and momentum operators are denoted by
\(\hat{\mathbf R}_I\) and \(\hat{\mathbf P}_I\). In a crystal, \(I\) is a
composite index specifying both the Bravais-lattice vector and the atom within
the unit cell. We denote by \(\hat{\mathbf R}\) the
\(3N_{\textrm n}\)-dimensional vector containing all nuclear coordinate
operators. Global center-of-mass degrees of freedom are understood to have
been separated when necessary \cite{PhysRevB.69.115110}. The following treatment does not explicitly include spin degrees of freedom, but its generalization to include spin is straightforward.

The microscopic Hamiltonian is
\begin{align}
\hat H
&=
\hat H_{\textrm{el}}
+
\hat H_{\textrm n}
+
\hat H_{\textrm{el-n}} .
\label{eq:Hstart}
\end{align}
In terms of the Coulomb interaction
$v(\mathbf r,\mathbf r')
=
\frac{e^2}{|\mathbf r-\mathbf r'|}$,
the electronic Hamiltonian is
\begin{align}
\hat H_{\textrm{el}}
&=
\sum_{i=1}^{N_{\textrm e}}
\frac{\hat{ \mathbf{p}}_i^2}{2m_{\textrm e}}
+
\frac{1}{2}
\sum_{\substack{i,j=1\\i\neq j}}^{N_{\textrm e}}
v(\hat{\mathbf r}_i,\hat{\mathbf r}_j),
\end{align}
where \(m_{\textrm e}\) is the electron mass. Defining the nuclear-nuclear interaction energy
\begin{align}
E_{\textrm{n-n}}(\mathbf R)
&\equiv
\frac{1}{2}
\sum_{\substack{I,J=1\\I\neq J}}^{N_{\textrm n}}
Z_I Z_J
v(\mathbf R_I,\mathbf R_J),
\end{align}
the nuclear Hamiltonian is
\begin{align}
\hat H_{\textrm n}
&=
\sum_{I=1}^{N_{\textrm n}}
\frac{\hat{\mathbf P}_I^2}{2M_I}
+
E_{\textrm{n-n}}(\hat{\mathbf R}),
\end{align}
where \(M_I\) and \(Z_I\) are respectively the mass and atomic number of
nucleus \(I\). We also introduce the one-electron potential generated by a nuclear configuration
\(\mathbf R\),
\begin{align}
V(\mathbf r,\mathbf R)
&\equiv
-
\sum_{J=1}^{N_{\textrm n}}
Z_J
v(\mathbf r,\mathbf R_J).
\end{align}
The electron-nucleus interaction can then be written  as
\begin{align}
\hat H_{\textrm{el-n}}
&=
\sum_{i=1}^{N_{\textrm e}}
V(\hat{\mathbf r}_i,\hat{\mathbf R}).
\end{align}

For fixed nuclear positions, the BO electronic Hamiltonian is
\begin{align}
\hat H^{\textrm{BO}}(\mathbf R)
&=
\sum_{i=1}^{N_{\textrm e}}
\left[
\frac{\hat{ \mathbf{p}}_i^2}{2m_{\textrm e}}
+
V(\hat{\mathbf r}_i,\mathbf R)
\right]
+
\frac{1}{2}
\sum_{\substack{i,j=1\\i\neq j}}^{N_{\textrm e}}
v(\hat{\mathbf r}_i,\hat{\mathbf r}_j).
\end{align}
Its electronic ground state satisfies
\begin{align}
\hat H^{\textrm{BO}}(\mathbf R)
\ket{\Psi^{\textrm{BO}}_0(\mathbf R)}
&=
E^{\textrm{BO}}_{\textrm{el},0}(\mathbf R)
\ket{\Psi^{\textrm{BO}}_0(\mathbf R)}.
\end{align}
The total BO energy is
\begin{align}
E^{\textrm{BO}}_0(\mathbf R)
&=
E^{\textrm{BO}}_{\textrm{el},0}(\mathbf R)
+
E_{\textrm{n-n}}(\mathbf R).
\end{align}

The BO equilibrium geometry \(\mathbf R^{0,\textrm{BO}}\) is determined by
\begin{align}
\left.
\frac{\partial E^{\textrm{BO}}_0(\mathbf R)}
{\partial R_{I,\alpha}}
\right|_{\mathbf R^{0,\textrm{BO}}}
&=
0.
\label{eq:BOequilibrium}
\end{align}
We define the nuclear displacement operators relative to this geometry as
\begin{align}
\hat U_{I,\alpha}
&=
\hat R_{I,\alpha}
-
R^{0,\textrm{BO}}_{I,\alpha}.
\end{align}

For any electronic observable \(\hat O\), we use the notation
\begin{align}
\langle \hat O\rangle_{\textrm{BO}}
&\equiv
\bra{\Psi^{\textrm{BO}}_0(\mathbf R^{0,\textrm{BO}})}
\hat O
\ket{\Psi^{\textrm{BO}}_0(\mathbf R^{0,\textrm{BO}})}
\label{eq:BOexpectation}
\end{align}
for its expectation value in the electronic BO ground state at the equilibrium
geometry. The BO density at an arbitrary nuclear configuration is obtained directly
from the normalized BO wavefunction as
\begin{align}
n^{\textrm{BO}}(\mathbf r,\mathbf R)
&\equiv
N_{\textrm e}
\int d\mathbf r_2\cdots d\mathbf r_{N_{\textrm e}}\,
\left|
\Psi^{\textrm{BO}}_0
(\mathbf r,\mathbf r_2,\ldots,\mathbf r_{N_{\textrm e}};\mathbf R)
\right|^2.
\label{eq:BOdensityWavefunction}
\end{align}
Its value at the BO equilibrium geometry is denoted by
\begin{align}
n^{0,\textrm{BO}}(\mathbf r)
&\equiv
n^{\textrm{BO}}(\mathbf r,\mathbf R^{0,\textrm{BO}}).
\label{eq:BOdensityReference}
\end{align}

Because the BO forces vanish at equilibrium, the expansion of the BO energy
starts at second order:
\begin{align}
E^{\textrm{BO}}_0(\hat{\mathbf R})
&=
E^{\textrm{BO}}_0(\mathbf R^{0,\textrm{BO}})
\nonumber\\
&\quad+
\frac{1}{2}
\sum_{I\alpha,J\beta}
\hat U_{I,\alpha}
C_{I,\alpha;J,\beta}
\hat U_{J,\beta}
+
\mathcal O(\hat U^3),
\end{align}
where the BO force-constant matrix is
\begin{align}
C_{I,\alpha;J,\beta}
&\equiv
\left.
\frac{\partial^2E^{\textrm{BO}}_0(\mathbf R)}
{\partial R_{I,\alpha}\partial R_{J,\beta}}
\right|_{\mathbf R^{0,\textrm{BO}}}.
\label{eq:CdefHessian}
\end{align}

We now expand the electron-nucleus potential around the same geometry:
\begin{align}
V(\hat{\mathbf r}_i,\hat{\mathbf R})
&=
V^{\textrm{BO}}(\hat{\mathbf r}_i)
+
\sum_{I\alpha}
g_{I,\alpha}(\hat{\mathbf r}_i)\hat U_{I,\alpha}
\nonumber\\
&\quad+
\frac{1}{2}
\sum_{I\alpha,J\beta}
g^{\textrm{DW}}_{I,\alpha;J,\beta}(\hat{\mathbf r}_i)
\hat U_{I,\alpha}\hat U_{J,\beta}
+
\mathcal O(\hat U^3).
\label{eq:Vexpansion}
\end{align}
Here
\begin{align}
V^{\textrm{BO}}(\hat{\mathbf r}_i)
&\equiv
V(\hat{\mathbf r}_i,\mathbf R^{0,\textrm{BO}}),
\\
g_{I,\alpha}(\hat{\mathbf r}_i)
&\equiv
\left.
\frac{\partial V(\hat{\mathbf r}_i,\mathbf R)}
{\partial R_{I,\alpha}}
\right|_{\mathbf R^{0,\textrm{BO}}}
=
Z_I
\frac{\partial}{\partial \hat r_{i,\alpha}}
v(\hat{\mathbf r}_i,\mathbf R_I^{0,\textrm{BO}}),
\\
g^{\textrm{DW}}_{I,\alpha;J,\beta}(\hat{\mathbf r}_i)
&\equiv
\left.
\frac{\partial^2V(\hat{\mathbf r}_i,\mathbf R)}
{\partial R_{I,\alpha}\partial R_{J,\beta}}
\right|_{\mathbf R^{0,\textrm{BO}}}
\nonumber\\
&=
-\delta_{IJ}Z_I
\frac{\partial^2}
{\partial \hat r_{i,\alpha}\partial \hat r_{i,\beta}}
v(\hat{\mathbf r}_i,\mathbf R_I^{0,\textrm{BO}}).
\end{align}
The superscript DW denotes the Debye--Waller coupling. The nuclear-nuclear
energy is expanded consistently:
\begin{align}
E_{\textrm{n-n}}(\hat{\mathbf R})
&=
E_{\textrm{n-n}}(\mathbf R^{0,\textrm{BO}})
+
\sum_{I\alpha}
\left.
\frac{\partial E_{\textrm{n-n}}(\mathbf R)}
{\partial R_{I,\alpha}}
\right|_{\mathbf R^{0,\textrm{BO}}}
\hat U_{I,\alpha}
\nonumber\\
&\quad+
\frac{1}{2}
\sum_{I\alpha,J\beta}
\left.
\frac{\partial^2E_{\textrm{n-n}}(\mathbf R)}
{\partial R_{I,\alpha}\partial R_{J,\beta}}
\right|_{\mathbf R^{0,\textrm{BO}}}
\hat U_{I,\alpha}\hat U_{J,\beta}
+
\mathcal O(\hat U^3).
\label{eq:EnnExpansion}
\end{align}
Terms independent of all electronic and nuclear operators are omitted in the
following, since they only shift the zero of energy%
\footnote{In a periodic system, the \(\mathbf G=0\) component of the
electrostatic potential is defined only up to an arbitrary additive constant.
Periodic boundary conditions fix the homogeneous electric field to zero but
leave this constant undetermined. It shifts the one-particle energies while
leaving the total energy of a neutral unit cell unchanged. Here the
zeroth-order electrostatic contribution is therefore omitted by fixing this
arbitrary energy reference.}.


Substituting these expansions into the microscopic Hamiltonian gives
\begin{align}
\hat H^{\textrm{low-en}}
&=
\hat H_{0,\textrm{el}}
+
\hat H_{0,\textrm{ph}}
+
\hat H_{\textrm{el-el}}
+
\hat H_1
+
\hat H_2.
\label{eq:lowenH}
\end{align}
Here \(\hat H_{0,\textrm{el}}\) and \(\hat H_{0,\textrm{ph}}\) define the
reference (free) Hamiltonians for the electronic and phononic propagators, whereas
\(\hat H_{\textrm{el-el}}\), \(\hat H_1\), and \(\hat H_2\) generate the
many-body corrections. The reference electronic Hamiltonian is
\begin{align}
\hat H_{0,\textrm{el}}
&=
\sum_{i=1}^{N_{\textrm e}}
\left[
\frac{\hat{ \mathbf{p}}_i^2}{2m_{\textrm e}}
+
V^{\textrm{BO}}(\hat{\mathbf r}_i)
\right].
\end{align}
For practical calculations, \(\hat H_{0,\textrm{el}}\) may instead contain a
local density-functional Hartree-exchange-correlation potential or a nonlocal
static self-energy, such as screened exchange. Such a reorganization requires
the corresponding subtraction from the interaction sector to avoid double
counting. In the formal derivation below, we retain the bare reference
Hamiltonian so that the origin of every term remains explicit. We define the harmonic
phonon Hamiltonian as follows:
\begin{align}
\hat H_{0,\textrm{ph}}
&=
\sum_{I=1}^{N_{\textrm n}}
\frac{\hat{\mathbf P}_I^2}{2M_I}
+
\frac{1}{2}
\sum_{I\alpha,J\beta}
\hat U_{I,\alpha}
C_{I,\alpha;J,\beta}
\hat U_{J,\beta},
\label{eq:harmph}
\end{align}
where
\begin{align}
C_{I,\alpha;J,\beta}
=
\left.
\frac{\partial^2 E_{\textrm{n-n}}(\mathbf R)}
{\partial R_{I,\alpha}\partial R_{J,\beta}}
\right|_{\mathbf R^{0,\textrm{BO}}}&+
\int d\mathbf r\,
\left.
\frac{\partial n^{\textrm{BO}}(\mathbf r,\mathbf R)}
{\partial R_{J,\beta}}
\right|_{\mathbf R^{0,\textrm{BO}}}
g_{I,\alpha}(\mathbf r)\nonumber\\
&+\left\langle
\sum_{i=1}^{N_{\textrm e}}
g^{\textrm{DW}}_{I,\alpha;J,\beta}(\hat{\mathbf r}_i)
\right\rangle_{\textrm{BO}}
.
\label{eq:Cdef0}
\end{align}
The electron-electron interaction retains its original Coulomb form,
\begin{align}
\hat H_{\textrm{el-el}}
&=
\frac{1}{2}
\sum_{\substack{i,j=1\\i\neq j}}^{N_{\textrm e}}
v(\hat{\mathbf r}_i,\hat{\mathbf r}_j).
\end{align}

Before using the BO equilibrium condition, the term linear in the nuclear
displacements is
\begin{align}
\hat H_1
&=
\sum_{I\alpha}
\left[
\left.
\frac{\partial E_{\textrm{n-n}}(\mathbf R)}
{\partial R_{I,\alpha}}
\right|_{\mathbf R^{0,\textrm{BO}}}
+
\sum_{i=1}^{N_{\textrm e}}
g_{I,\alpha}(\hat{\mathbf r}_i)
\right]
\hat U_{I,\alpha}.
\label{eq:H1FirstQuantized}
\end{align}
The Hellmann--Feynman theorem gives
\begin{align}
\left.
\frac{\partial E^{\textrm{BO}}_0(\mathbf R)}
{\partial R_{I,\alpha}}
\right|_{\mathbf R^{0,\textrm{BO}}}
&=
\left.
\frac{\partial E_{\textrm{n-n}}(\mathbf R)}
{\partial R_{I,\alpha}}
\right|_{\mathbf R^{0,\textrm{BO}}}
+
\left\langle
\sum_{i=1}^{N_{\textrm e}}
g_{I,\alpha}(\hat{\mathbf r}_i)
\right\rangle_{\textrm{BO}}
=
0.
\label{eq:HFforce}
\end{align}
Then, Eq.~\eqref{eq:H1FirstQuantized} becomes
\begin{align}
\hat H_1
&=
\sum_{I\alpha}
\hat U_{I,\alpha}
\left[
\sum_{i=1}^{N_{\textrm e}}
g_{I,\alpha}(\hat{\mathbf r}_i)
-
\left\langle
\sum_{i=1}^{N_{\textrm e}}
g_{I,\alpha}(\hat{\mathbf r}_i)
\right\rangle_{\textrm{BO}}
\right].
\label{eq:H1Centered}
\end{align}

We next consider the quadratic term. Before separating the reference phonon
Hamiltonian, its coefficient is
\begin{align}
\hat K_{I,\alpha;J,\beta}
&\equiv
\left.
\frac{\partial^2E_{\textrm{n-n}}(\mathbf R)}
{\partial R_{I,\alpha}\partial R_{J,\beta}}
\right|_{\mathbf R^{0,\textrm{BO}}}
+
\sum_{i=1}^{N_{\textrm e}}
g^{\textrm{DW}}_{I,\alpha;J,\beta}(\hat{\mathbf r}_i)
\label{eq:KdefFirstQuantized}
\end{align}

Because the BO force constants \(C\) have already been included in
\(\hat H_{0,\textrm{ph}}\) \cite{PhysRevB.69.115110}, the remaining quadratic interaction is
\begin{align}
\hat H_2
&=
\frac{1}{2}
\sum_{I\alpha,J\beta}
\hat U_{I,\alpha}
\left[
\hat K_{I,\alpha;J,\beta}
-
C_{I,\alpha;J,\beta}
\right]
\hat U_{J,\beta}.
\label{eq:H2KC}
\end{align}

Combining Eqs. \eqref{eq:H2KC} and \eqref{eq:Cdef0}, we obtain
\begin{align}
\hat H_2
&=
-\frac{1}{2}
\sum_{I\alpha,J\beta}
\hat U_{I,\alpha}
\int d\mathbf r\,
\left.
\frac{\partial n^{\textrm{BO}}(\mathbf r,\mathbf R)}
{\partial R_{J,\beta}}
\right|_{\mathbf R^{0,\textrm{BO}}}
g_{I,\alpha}(\mathbf r)
\hat U_{J,\beta}+
\nonumber\\
&+\frac{1}{2}\sum_{I\alpha,J\beta}\left[\sum_{i=1}^{N_{\textrm e}}
g^{\textrm{DW}}_{I,\alpha;J,\beta}(\hat{\mathbf r}_i)
-\left\langle
\sum_{i=1}^{N_{\textrm e}}
g^{\textrm{DW}}_{I,\alpha;J,\beta}(\hat{\mathbf r}_i)
\right\rangle_{\textrm{BO}}\right]\hat U_{I,\alpha}\hat U_{J,\beta}.
\label{eq:H2B0}
\end{align}

The above equations can be conveniently rewritten in terms of the electronic density operator
\begin{align}
\hat n(\mathbf r)
&\equiv
\sum_{i=1}^{N_{\textrm e}}
\delta(\mathbf r-\hat{\mathbf r}_i)
\end{align}
and the corresponding density-fluctuation operator relative to the BO
reference:
\begin{align}
\Delta\hat n(\mathbf r)
&\equiv
\hat n(\mathbf r)
-
n^{0,\textrm{BO}}(\mathbf r).
\label{eq:DeltaNDefinition}
\end{align}
The connection between the density defined from the BO wavefunction in
Eq.~\eqref{eq:BOdensityWavefunction} and the density operator is
\begin{align}
\langle\hat n(\mathbf r)\rangle_{\textrm{BO}}
&=
n^{0,\textrm{BO}}(\mathbf r).
\label{eq:DensityExpectationConnection}
\end{align}
For an arbitrary one-electron function \(f(\mathbf r)\), one consequently has
\begin{align}
\sum_{i=1}^{N_{\textrm e}}f(\hat{\mathbf r}_i)
&=
\int d\mathbf r\,
f(\mathbf r)\hat n(\mathbf r),
\\
\left\langle
\sum_{i=1}^{N_{\textrm e}}f(\hat{\mathbf r}_i)
\right\rangle_{\textrm{BO}}
&=
\int d\mathbf r\,
f(\mathbf r)n^{0,\textrm{BO}}(\mathbf r).
\label{eq:braketsumBO}
\end{align}
Hence,
\begin{align}
\sum_{i=1}^{N_{\textrm e}}g_{I,\alpha}(\hat{\mathbf r}_i)
-
\left\langle
\sum_{i=1}^{N_{\textrm e}}
g_{I,\alpha}(\hat{\mathbf r}_i)
\right\rangle_{\textrm{BO}}
&=
\int d\mathbf r\,
g_{I,\alpha}(\mathbf r)\Delta\hat n(\mathbf r).
\end{align}
The linear electron-phonon interaction may therefore be written as
\begin{align}
\hat H_1
&=
\sum_{I\alpha}
\int d\mathbf r\,
g_{I,\alpha}(\mathbf r)
\Delta\hat n(\mathbf r)
\hat U_{I,\alpha}.
\label{eq:H1Density}
\end{align}
Equation~\eqref{eq:H1Density} is simply the density representation of the first-quantized operator in Eq.~\eqref{eq:H1Centered}, and is a useful step to introduce the second quantization needed for the effective action calculation \cite{SI}. Defining
\begin{align}
\mathcal B_{I,\alpha;J,\beta}
&\equiv
-
\int d\mathbf r\,
\left.
\frac{\partial n^{\textrm{BO}}(\mathbf r,\mathbf R)}
{\partial R_{J,\beta}}
\right|_{\mathbf R^{0,\textrm{BO}}}
g_{I,\alpha}(\mathbf r),
\label{eq:Bdef}
\end{align}
$\hat H_2$ can be similarly rewritten as
\begin{align}
\hat H_2
&=\frac{1}{2}
\sum_{I\alpha,J\beta}
\hat U_{I,\alpha}
\left[ \mathcal B_{I,\alpha;J,\beta}+\int d\mathbf{r} g^{\textrm{DW}}_{I,\alpha;J,\beta}(\mathbf{r})\Delta \hat n(\mathbf{r}) \right]
\hat U_{J,\beta},
\label{eq:H2deltan}
\end{align}
This second term of $\hat H_2$ is typically responsible for real electron energies renormalizations, but it can also induce multi-phonon scattering;  in the following we neglect it, and $\hat H_2$ simplifies to
\begin{align}
\hat H_2
&\simeq\frac{1}{2}
\sum_{I\alpha,J\beta}
\hat U_{I,\alpha}
\mathcal B_{I,\alpha;J,\beta}
\hat U_{J,\beta}.
\label{eq:H2B}
\end{align}
This simplified \(\hat H_2\) interaction produces only a static contribution to the phonon self-energy,
as shown below. Its role is to subtract the static electronic response already
contained in the BO force constants and thereby avoid counting the same
screening contribution twice when electronic fluctuations are treated
explicitly. Finally, using Eq. \eqref{eq:braketsumBO}, Eq. \eqref{eq:Cdef0} assumes the standard form \cite{RevModPhys.73.515}
\begin{align}
C_{I,\alpha;J,\beta}
=
\left.
\frac{\partial^2 E_{\textrm{n-n}}(\mathbf R)}
{\partial R_{I,\alpha}\partial R_{J,\beta}}
\right|_{\mathbf R^{0,\textrm{BO}}}&+
\int d\mathbf r\,
\left.
\frac{\partial n^{\textrm{BO}}(\mathbf r,\mathbf R)}
{\partial R_{J,\beta}}
\right|_{\mathbf R^{0,\textrm{BO}}}
g_{I,\alpha}(\mathbf r)\nonumber \\
&+
\int d\mathbf r\,
n^{0,\textrm{BO}}(\mathbf r)\,
g^{\textrm{DW}}_{I,\alpha;J,\beta}(\mathbf r).
\label{eq:Cdef}
\end{align}

We conclude this section by discussing the choice of reference geometry. In
the following, we neglect possible shifts of the equilibrium positions away
from \(\mathbf R^{0,\textrm{BO}}\). Two distinct effects can in principle
produce such shifts. First, after the low-energy Hamiltonian in
Eq.~\eqref{eq:lowenH} is truncated and treated within an approximate many-body
scheme, its interacting equilibrium geometry may differ from the BO reference,
as discussed in the Supplemental Material \cite{SI}. Second, the BO density
\(n^{0,\textrm{BO}}\) is not in general identical to the exact ground-state
density \(n^0\) of the complete electron-nuclear problem. Defining the
density-fluctuation operator relative to \(n^{0,\textrm{BO}}\), rather than
\(n^0\), neglects the residual terms associated with
\(n^0-n^{0,\textrm{BO}}\neq0\), which may shift the nuclei toward the exact
equilibrium geometry of the full Hamiltonian in Eq.~\eqref{eq:Hstart}
\cite{PhysRevX.13.031026}. This distinction is particularly important when electronic correlations
strongly renormalize the equilibrium structure, as in polaron formation
\cite{PhysRevB.106.075119,PhysRevB.111.144309}. Here, we assume that these
shifts vanish or are negligible for the systems of interest, whose equilibrium
atomic positions are often constrained by symmetry. Moreover, effects omitted
from Eq.~\eqref{eq:lowenH}, such as anharmonicity, may play a more important
role in determining the actual equilibrium geometry. Finally, using the BO
force constants as the reference harmonic problem provides, for a stable BO
configuration, a bounded noninteracting phonon Hamiltonian with a well-defined
vacuum, as required for the effective-action formulation developed in the next
section. From this point onward, hats on operators are omitted whenever their operator
character is unambiguous.

\section{Effective action and single-phonon functional} 
\label{sec:action}

In this section we construct the single-phonon functional corresponding to the
second step of Fig.~\ref{fig:theory-workflow}. The Hamiltonian derived in the previous section naturally separates the problem into a free-particle theory, describing independent electrons and harmonic phonons, and interaction terms, describing electron-electron correlations, the linear electron-phonon coupling, and the static quadratic correction
\(\mathcal B\). This structure allows us to formulate the theory in terms of an action: the free-particle sector defines the bare propagators \(G_0\) and \(D_0\), while the interactions generate the many-body corrections embedded in the Dyson's equations. As summarized in the Supplemental Material \cite{SI}, the effective action is then written in terms of $G$ and $D$, and is obtained by introducing sources coupled to the electronic and phononic two-point functions and then performing a Legendre transform. The connection with perturbation theory is encoded in the interaction sector of the effective action, whose diagrammatic expansion reorganizes the conventional perturbative series into closed two-particle-irreducible diagrams constructed from the fully dressed propagators \(G\) and \(D\). Functional differentiation of this diagrammatic functional opens propagator lines and generates the corresponding self-energies, while the stationarity conditions resum their insertions through the Dyson equations, obtaining the same final results as the perturbative theory \cite{PhysRevX.13.031026}. We do not introduce sources that couple linearly with electrons and phonons coherently with the assumption that the associated field operators always assume the same equilibrium value independently from interaction---see Supplemental Material \cite{SI}. The Dyson's equations of the physical propagators then follow from the stationarity of the effective action.

In order to introduce the effective action, we first need to define the non-interacting propagators for electrons and phonons, $G_0$ and $D_0$, and matrix multiplication rules. This is matter of the next paragraphs.

\subsection{Non-interacting propagators}
Let's introduce the variables
\begin{align}
1&\equiv(\mathbf{r}_1,t_1), 
\quad 
2\equiv(\mathbf{r}_2,t_2),
\quad \ldots ,
\\
a&\equiv(I,\alpha,t_a),
\quad
b\equiv(J,\beta,t_b),
\quad \ldots .
\end{align}
We introduce the non-interacting Green's function
\begin{equation}
    G_0^{-1}(1,2)
    =
    \left[
    i\hbar\frac{\partial}{\partial t_1}
    -
    h_0(\mathbf{r}_1)
    \right]\delta(1,2),
    \label{eq:G0def}
\end{equation}
with
\begin{equation}
    h_0(\mathbf{r})
    =
    -\frac{\hbar^2\nabla^2}{2m_{\textrm{e}}}
    +
    V^{\textrm{BO}}(\mathbf{r})
    \label{eq:h0def}.
\end{equation} 

We also introduce the non-interacting phonon propagator as
\begin{equation}
    D_0^{-1}(a,b)
    =
    -
    \left[
    M_I\delta_{IJ}\delta_{\alpha\beta}
    \frac{\partial^2}{\partial t_a^2}
    +
    C_{I,\alpha;J,\beta}
    \right]
    \delta(t_a-t_b).
    \label{eq:D0def}
\end{equation}

\subsection{Matrix multiplication notation}
Let's introduce the notation
\begin{align}
\int d1
&\equiv
\int d\mathbf{r}_1dt_1,
\quad
\int da
\equiv
\sum_{I\alpha}\int dt_a.
\end{align}
Let's also treat electronic and atomic coordinates on the same footing, defining
\begin{align}
g(1,a)\equiv g_{I,\alpha}(\mathbf{r_1})\delta(t_1-t_a).
\end{align}
Let's consider two-point generic matrices $A,B,C$, respectively functions
of electronic coordinates only, electronic and atomic coordinates, and
atomic coordinates only. We introduce the matrix multiplication rules 
\begin{align}
(AA)(1,2)
&\equiv
A(1,\bar 1)A(\bar 1,2)
=
\int d\bar 1\,
A(1,\bar 1)A(\bar 1,2),
\\
(AB)(1,a)
&\equiv
A(1,\bar 1)B(\bar 1,a)
=
\int d\bar 1\,
A(1,\bar 1)B(\bar 1,a),
\\
(BC)(1,a)
&\equiv
B(1,\bar a)C(\bar a,a)
=
\int d\bar a\,
B(1,\bar a)C(\bar a,a),
\\
(CC)(a,b)
&\equiv
C(a,\bar a)C(\bar a,b)
=
\int d\bar a\,
C(a,\bar a)C(\bar a,b).
\end{align}
Coherently with the above definitions, in the following repeated barred indexes are intended as integrated upon. Analogous rules are understood for all other possible matrix products. Further, to simplify notation, we will sometimes use juxtaposition of matrices without explicit indexes; in that case, they are intended to be multiplied following the above described rules. Finally, we define the trace operator as
\begin{align}
\textrm{Tr}\lbrace AA \rbrace= \lim_{\eta \rightarrow 0} \int d1 A(1,\bar 2) A(\bar 2, 1^+)=A(\bar 1,\bar 2) A(\bar 2, \bar 1^+)\\
\textrm{Tr}\lbrace CC \rbrace= \lim_{\eta \rightarrow 0} \int da C(a,\bar b) C(\bar b, a^+)= C(\bar a,\bar b) C(\bar b, \bar a^+),
\end{align}
where $1^+$ means $t_1^+=t_1+0^+$, and similarly for atomic indexes; for other combinations of matrices the trace is defined accordingly. The equal time prescription is performed in order to match the frequency space definition at zero temperature \cite{Reining_16}. Finite temperature extensions are obtained by standard methods.

\subsection{The effective action functional and the self-energies}
The effective action is written as \cite{SI}
\begin{align}
    \Gamma[G,D]&=i\hbar \Phi[G,D,v,g]-i\hbar \textrm{Tr}\left\{
    G_0^{-1}G+\ln[G_0G^{-1}]-I\right\}\nonumber\\
    &\quad+i\hbar \frac12 \textrm{Tr}\left\{D_0^{-1}D-\ln[D_0^{-1}D]-I-\mathcal B D\right\}+\textrm{const.} \,.
\end{align}
The functional \(\Phi[G,D,v,g]\) contains the interaction contribution and is given by the exact sum of closed, two-particle-irreducible skeleton diagrams constructed from the fully dressed propagators \(G\) and \(D\). We display the interaction vertices \(v\) and \(g\) as explicit arguments to keep track of the corresponding couplings, although they are fixed parameters rather than variational variables. The remaining terms originate from the noninteracting electronic and phononic parts of the Hamiltonian, while \(\mathcal B D\) accounts for the static quadratic correction introduced in Sec.~\ref{sec:Ham}. Terms independent of \(G\) and \(D\) are collected in `\(\textrm{const.}\)', since they do not contribute to the stationarity conditions that determine the physical propagators. Finally, \(I\) denotes the identity operator, \(I(1,2)=\delta(1,2)\).

Following standard methods \cite{PhysRevD.10.2428}, we define the electronic and phononic self-energies as
\begin{align}
    \Sigma(1,2)
    &=
    \frac{\delta \Phi[G,D,v,g]}
    {\delta G(2,1^+)},
   \label{eq:Sigmaderphi} \\
    \Pi(a,b)
    &=
    -2
    \left.
    \frac{\delta \Phi[G,D,v,g]}
    {\delta D(a,b^+)}
    \right|_{\textrm{S}} .
    \label{eq:Piderphi}
\end{align}
The symbol $|_{\textrm{S}}$ denotes the symmetrized derivative $\frac{\partial}{\partial f(1,2)}|_{\textrm{S}}=\frac12 \frac{\partial}{\partial f(1,2)}+\frac12 \frac{\partial}{\partial f(2,1)}$ \cite{PhysRevX.13.031026}. Imposing the stationarity of $\Gamma[G,D]$ with respect to $G$ and $D$, one obtains the exact Dyson
equations
\begin{align}
    G^{-1}(1,2)-G_0^{-1}(1,2)+\Sigma(1,2)
    &=0, \label{eq:GDyson}
    \\
    D^{-1}(a,b)-D_0^{-1}(a,b)+\Pi^{\textrm{NA}}(a,b)
    &=0, \label{eq:DDyson}\\
    \Pi^{\textrm{NA}}=\Pi+\mathcal{B}.\label{eq:PiNA}
\end{align}
NA stand for non adiabatic. We will show that $\mathcal{B}$ is a trivial term, and we will therefore concentrate on the expression for $\Pi$ in the following. In the next section, we perform an approximation to the exact functional $\Phi$.

\begin{figure}[htbp]
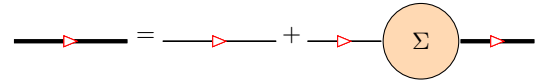

  \centering
  \[
  \G
  =
  \Gzero
  +
  \GzeroSigmaG
  \]
  \caption{Dyson's equation for the electronic Green's function, Eq. \eqref{eq:GDyson}.}
  \label{fig:dyson-electronic}
\end{figure}

\begin{figure}[htbp]
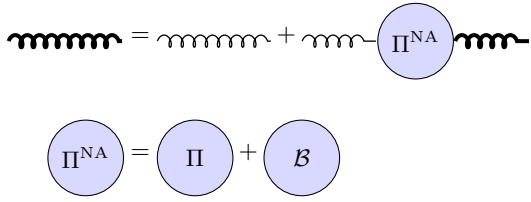

  \centering
  \[
  \begin{aligned}
  \Dphonon
  &=
  \Dzero
  +
  \DzeroPiD
  \\[1.2em]
  \PiNABubble
  &=
  \PiBubble
  +
  \BBubble
  \end{aligned}
  \]
  \caption{Dyson's equation for the phonon propagator and decomposition of the nonadiabatic phonon self-energy, Eqs. \eqref{eq:DDyson} and \eqref{eq:PiNA}.}
  \label{fig:dyson-phonon}
\end{figure}

\subsection{Approximating the $\Phi$ functional to a single-phonon functional}
The exact \(\Phi\) functional is defined by the perturbative construction outlined in the Supplemental Material \cite{SI}. Although formally compact, its complete diagrammatic expansion is not suitable for practical calculations. We therefore introduce an approximation that retains a physically selected subclass of the exact diagrams. Our aim is to describe elementary processes in which an electron scatters by emitting or absorbing a single phonon. We consequently restrict the electron-phonon sector of the functional to diagrams containing a single phonon propagator \(D\). Contributions involving higher powers of \(D\), which describe multiphonon processes, could be incorporated systematically but lie beyond the scope of the present work. To implement this truncation, we separate the functional into a purely electronic contribution and an electron-phonon contribution. The former is kept formally exact at this stage (and still indicated with the letter $\Phi$), whereas the latter is approximated by retaining only terms with one phonon propagator (and indicated with a different letter $\widetilde \Phi$). At zero temperature, we define
\begin{align}
\widetilde{\Phi}[G,D,v,g]
&=
\Phi^{\textrm{el}}[G,v]
+
\widetilde{\Phi}^{\textrm{el-ph}}[G,D,v,g],\label{eq:tildePhidecomp}
\\
\Phi^{\textrm{el}}[G,v]
&=
\Phi^{\textrm{el}}_{\textrm{H}}[G,v]
+
\Phi^{\textrm{el}}_{\textrm{xc}}[G,v],\label{eq:tildePhidecomp1}
\\
\Phi^{\textrm{el}}_{\textrm{H}}[G,v]
&=
-\frac{i\hbar}{2}
G(\bar 1,\bar 1^+)
v(\bar 1,\bar 2)
G(\bar 2,\bar 2^+), \label{eq:tildePhidecomp2}
\\
\widetilde{\Phi}^{\textrm{el-ph}}[G,D,v,g]
&=
\left.
\frac{d}{d\eta}
\Phi^{\textrm{el}}_{\textrm{xc}}
[G,v+\eta U_D]
\right|_{\eta=0}. 
\label{eq:phiephdef}
\end{align}
Here,
\begin{align}
U_D(1,2)
=
g(1,\bar a)
D(\bar a,\bar b)
g(2,\bar b)
\label{eq:UDdef}
\end{align}
is the effective electron-electron interaction mediated by a single phonon propagator.

The functional \(\Phi^{\textrm{el}}[G,v]\), which has the same functional as in the clamped-nuclei case, contains two contributions. The Hartree term is generated by \(\Phi^{\textrm{el}}_{\textrm{H}}[G,v]\), whereas \(\Phi^{\textrm{el}}_{\textrm{xc}}[G,v]\) generates the exact exchange-correlation electronic self-energy. The latter is the sum of closed electronic two-particle-irreducible diagrams constructed from the fully dressed propagator \(G\). Such diagrams cannot be separated into disconnected components by cutting two electronic propagator lines and are therefore skeleton diagrams, with no self-energy insertions \cite{stefanucci2013nonequilibrium,PhysRevX.13.031026}. It is important to stress that \(\Phi_{\rm xc}^{\rm el}[G,v]\) does not play the same role as the exchange-correlation energy functional of density-functional theory. In particular, a truncation of \(\Phi_{\rm xc}^{\rm el}\) does not automatically define a static one-particle potential that generates both a reference Hamiltonian \(H^0\) and its consistent first-order variation \(H^1\). Instead, different functional derivatives of \(\Phi_{\rm xc}^{\rm el}\) generate distinct many-body quantities. For example, an RPA ring approximation to \(\Phi_{\rm xc}^{\rm el}\) generates \(\chi_{\rm el}^{\rm RPA}\) through differentiation with respect to \(v\), whereas differentiation with respect to \(G\) generates a \(GW^{\rm RPA}\)-type electronic self-energy. Complete self-consistence between the one- and two- particle sectors is obtained only considering the exact functional.

The derivative in Eq.~\eqref{eq:phiephdef} replaces, in turn,
one Coulomb interaction line in each diagram of
\(\Phi^{\textrm{el}}_{\textrm{xc}}\) by the phonon-mediated
interaction \(U_D\). The resulting functional
\(\widetilde{\Phi}^{\textrm{el-ph}}\) therefore selects the single-phonon-propagator sector of the exact electron-phonon functional. We show below that this construction generates the corresponding contributions to both the electronic and phononic self-energies with the correct diagrammatic structure and symmetry factors. Before proceeding we notice that, consistently with neglecting the interaction-induced shift of the equilibrium nuclear positions, we discard the phonon-mediated Hartree contribution, which is proportional to the average density difference
\(\langle\Delta\hat n\rangle
=
\langle\hat n\rangle-n^{0,\textrm{BO}}\)
and describes a residual force on the nuclei. In the electronic self-energy, the same contribution would generate the static potential
\(gD(\omega=0)g\langle\Delta\hat n\rangle\),
which is real and can only shift the electronic energies, without contributing to the linewidths considered here. Finally, we also notice that the exchange-correlation contribution is unaffected by the BO density subtraction, because subtracting the fixed reference density changes only disconnected Hartree contributions and leaves all connected density correlations unchanged.
In terms of the approximated functional, the self-energies (indicated with the same letters) are of course given as
\begin{align}
    \Sigma(1,2)
    &=
    \frac{\delta \widetilde \Phi[G,D,v,g]}
    {\delta G(2,1^+)},
   \label{eq:Sigmaderphitilde} \\
    \Pi(a,b)
    &=
    -2
    \left.
    \frac{\delta \widetilde \Phi[G,D,v,g]}
    {\delta D(a,b^+)}
    \right|_{\textrm{S}} .
    \label{eq:Piderphitilde}
\end{align}

We now show that \(\widetilde{\Phi}^{\textrm{el-ph}}\) contains diagrams that are two-particle irreducible with respect to cuts of the combined set of \(G\) and \(D\) lines, and possess the correct symmetry factors \cite{NegeleOrland1988}. Indeed, the following facts hold: (i) the diagrams in \(\widetilde \Phi^{\textrm{el-ph}}[G,D,v,g]\) are two-particle irreducible (2PI) with respect to the cut of two \(G\) lines, since they inherit this property from \(\Phi_{\textrm{xc}}^{\textrm{el}}[G,v]\). This implies that \(\widetilde \Phi^{\textrm{el-ph}}[G,D,v,g]\) contains diagrams that are skeletonic in \(G\); (ii) no interaction line \(v\) in a diagram of \(\Phi_{\textrm{xc}}^{\textrm{el}}[G,v]\) can be a bridge. Indeed, if cutting a single \(v\) line separated the diagram into two connected components, then cutting the incoming and outgoing \(G\) lines at either endpoint of that interaction would disconnect the original diagram, contradicting its 2PI character. The only exception is the Hartree diagram,\footnote{The Hartree diagram is in fact the only possible exception to this argument. If a \(v\) line were a bridge, cutting the two electronic propagators attached to either of its endpoints would disconnect the diagram unless those two propagators coincide and form a one-propagator fermionic loop. This can occur only for the Hartree topology.} for which the incoming and outgoing electronic legs at each endpoint belong to the same one-propagator fermionic loop; this diagram is excluded from \(\Phi_{\textrm{xc}}^{\textrm{el}}\) and hence from the construction of \(\widetilde \Phi^{\textrm{el-ph}}\). Therefore, replacing any \(v\) line by \(U_D=gDg\) cannot produce a diagram that is disconnected by cutting the corresponding \(D\) line; (iii) the diagrams in \(\widetilde \Phi^{\textrm{el-ph}}[G,D,v,g]\) are consequently 2PI with respect to the cut of two \(D\) lines: each diagram contains only one \(D\) line, and its removal does not disconnect the diagram by observation (ii). This also implies that \(\widetilde \Phi^{\textrm{el-ph}}[G,D,v,g]\) is skeletonic in \(D\); (iv) the diagrams are also 2PI with respect to the cut of one \(D\) and one \(G\) line. After cutting the unique \(D\) line, the diagram remains connected by observation (ii). Moreover, every \(G\) propagator belongs to a closed fermionic cycle, which is unaffected by the removal of the \(D\) line. Hence no \(G\) propagator is a bridge of the remaining graph, and cutting one additional \(G\) line cannot disconnect the diagram; (v) each diagram in \(\Phi^{\textrm{el}}_{\textrm{xc}}[G,v]\) appears multiplied by the inverse of the order of its symmetry group. Replacing one interaction line \(v\) by the distinguishable line \(U_D\) generally reduces the symmetry group of the diagram. However, the replacement is performed over all \(v\) lines: replacements of lines related by symmetries of the parent diagram generate equivalent copies of the same mixed diagram. The multiplicity of these equivalent replacements exactly compensates for the reduction of the symmetry group, yielding the correct symmetry factor for the resulting diagram. A rigorous derivation based on the orbit-stabilizer theorem is provided in the Supplemental Material \cite{SI}.

In conclusion, the diagrams contained in $\widetilde \Phi^{\textrm{el-ph}}[G,D,v,g]$ are connected closed vacuum diagrams that are 2PI with respect to the combined set of $G$ and $D$ lines, skeletonic in $G$ and $D$, written in terms of the interactions $v$ and $g$, and have correct symmetry prefactors. Therefore, they are a subclass of the diagrams contained in the exact $\Phi^{\textrm{el-ph}}[G,D,v,g]$.

Finally, we introduce the interacting electronic response generated by the electronic exchange-correlation functional as
\begin{align}
\chi_{\rm el}[G,v](1,2)
=
-2\frac{\delta\Phi_{\rm xc}^{\rm el}[G,v]}
{\delta v(2,1)}
\bigg|_{\rm S}.
\label{eq:chi_el_from_Phi}
\end{align}
At this stage, Eq.~\eqref{eq:chi_el_from_Phi} provides a functional definition of the reducible electronic response. We introduce the corresponding irreducible polarization \(P_{\textrm{el}}\) and screened Coulomb interaction \(W_{\textrm{el}}\) through the Dyson equations
\begin{align}
\chi_{\textrm{el}}(1,2)
&=
P_{\textrm{el}}(1,2)
+
P_{\textrm{el}}(1,\bar 1)
v(\bar 1,\bar 2)
\chi_{\textrm{el}}(\bar 2,2),
\label{eq:WPW1}
\\
W_{\textrm{el}}(1,2)
&=
v(1,2)
+
v(1,\bar 1)
P_{\textrm{el}}(\bar 1,\bar 2)
W_{\textrm{el}}(\bar 2,2).
\label{eq:WPW2}
\end{align}
In Sec.~\ref{sec:selfenergies}, we construct \(P_{\textrm{el}}\) explicitly from the Bethe--Salpeter kernel generated by the same electronic functional and show how the resulting two-particle propagator recovers the reducible response \(\chi_{\textrm{el}}\) introduced in Eq.~\eqref{eq:chi_el_from_Phi}.

\begin{figure}[htbp]
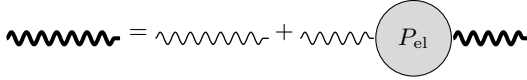

  \centering
  \[
  \Wel
  =
  \vbare
  +
  \vPelWelnoSX
  \]
  \caption{Dyson equation for the screened electronic interaction $W_{\textrm{el}}$, Eq. \eqref{eq:WPW2}.}
  \label{fig:dyson-Pel}
\end{figure}

Applying Eq. \eqref{eq:Sigmaderphitilde}, the electronic self-energy becomes 
\begin{align}
\Sigma(1,2)&=\frac{\delta \Phi^{\textrm{el}}[G,v]}{\delta G(2,1^+)}+\frac{\delta \widetilde \Phi^{\textrm{el-ph}}[G,D,v,g]}{\delta G(2,1^+)}\nonumber \\
&\equiv \Sigma^{\textrm{el}}(1,2)+\Sigma^{\textrm{el-ph}}(1,2). \label{eq:fullSigma}
\end{align}
The clamped nuclei term can be further split in
\begin{align}
\Sigma^{\textrm{el}}(1,2)&=\frac{\delta \Phi^{\textrm{el}}_{\textrm{H}}[G,v]}{\delta G(2,1^+)}+\frac{\delta\Phi^{\textrm{el}}_{\textrm{xc}}[G,v]}{\delta G(2,1^+)} \nonumber \\
&=\Sigma_{\textrm{H}}^{\textrm{el}}(1,2)+\Sigma_{\textrm{xc}}^{\textrm{el}}(1,2),\label{eq:Sigmasplit}
\end{align}
where
\begin{align}
\Sigma_{\textrm{H}}^{\textrm{el}}(1,2)=-i\hbar\delta(1,2)v(1,\bar 3)G(\bar 3,\bar 3^+).
\label{eq:SigmaH}
\end{align}

\begin{figure}[htbp]
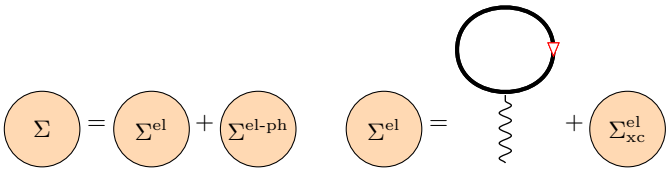

  \centering
  \[
  \begin{aligned}
  \fSigma =\fSigmael+\fSigmaelph \quad \quad 
  \fSigmael =  \HartreeEl + \Sigmaelxc
  \end{aligned}
  \]
  \caption{Electronic self-energy splitting, Eqs. \eqref{eq:fullSigma} and \eqref{eq:Sigmasplit}.}
  \label{fig:dyson-Sigma}
\end{figure}

\section{Electronic and phononic self-energies in terms of a common vertex}
\label{sec:selfenergies}

In this section, we show how the two functional derivatives in
Fig.~\ref{fig:theory-workflow} generate the electronic and phononic
self-energies via functional derivatives. We first define the electronic Green's function and the associated Bethe--Salpeter ladder response. We then use this response to derive the phonon and electronic electron-phonon self-energies in terms of the same dressed vertex, thereby establishing the basis for detailed balance.

\subsection{Exchange-correlation electronic self-energy}
Since $\Phi^{\textrm{el}}[G,v]$ has been kept formally exact in Eq. \eqref{eq:tildePhidecomp}, the clamped-nuclei electronic self-energy is exact. Its expression can be found in many reviews, such as Refs. \onlinecite{https://doi.org/10.1002/wcms.1344,RevModPhys.74.601,LarsHedin_1999,FAryasetiawan_1998,PhysRevX.13.031026}. It contains the usual GW term plus vertex corrections, as
\begin{align}
\Sigma^{\textrm{el}}_{\textrm{xc}}(1,2)=i\hbar G(1,\bar 1)W_{\textrm{el}}(1^+,\bar 2)\Gamma'(\bar 1,2;\bar 2),
\label{eq:sigmaelxcexact}
\end{align}
where $\Gamma'$ is the vertex function, which definition will be given in Eq. \eqref{eq:dysonexactvertex}.

\subsection{Kernel, vertex and polarization}
The electron-phonon self-energies derived below depend on the correlated propagation of an electron-hole pair. We therefore introduce the electronic two-particle quantities underlying the response functions defined in Eqs.~\eqref{eq:chi_el_from_Phi}--\eqref{eq:WPW2}. Although the Green's function \(G\) entering \(\Phi_{\textrm{xc}}^{\textrm{el}}[G,v]\) may depend implicitly on the phonon propagator through the full Dyson equation, it has the same functional form as in the clamped-nuclei problem. At fixed \(G\), its two-particle structure can therefore be obtained through the corresponding Bethe-Salpeter (BSE) construction. In particular, the functional derivative of the electronic exchange-correlation self-energy defines the kernel \(K_{\textrm{xc}}\) (not to be confused with the true self-consistent kernel of $\widetilde \Phi[G,D,v,g]$, which would further entail the variation of the electron-phonon self-energy). The ladder resummation generated by \(K_{\textrm{xc}}\) gives an electron-hole propagator \(L_{\textrm{xc}}\) that is irreducible with respect to Hartree Coulomb lines. Its closure defines the irreducible polarization \(P_{\textrm{el}}\) entering Eqs.~\eqref{eq:WPW1} and \eqref{eq:WPW2}. The full reducible response \(\chi_{\textrm{el}}\) defined in Eq.~\eqref{eq:chi_el_from_Phi} is subsequently recovered by restoring the Hartree contribution. We first introduce the exchange-correlation kernel, which measures how the electronic exchange-correlation self-energy changes under a variation of the one-particle propagator,
\begin{align}
K_{\textrm{xc}}(1,2;3,4)=-\frac{\delta \Sigma^{\textrm{el}}_{\textrm{xc}}(1,3)}{\delta G(4,2)}.
\end{align}
Since it is the second functional derivative of \(\Phi^{\textrm{el}}_{\textrm{xc}}[G,v]\), it obeys the symmetry
\cite{fabrizio2022course}
\begin{align}
K_{\textrm{xc}}(1,2;3,4)=K_{\textrm{xc}}(2,1;4,3).
\label{eq:Ksymm}
\end{align}
This symmetry will ensure that the ladder can be attached consistently to either side of the electron-hole propagator. In the graphical representation of the four-point kernel, the coordinates are ordered as follows: (1) corresponds to the upper-left point, (2) to the lower-right point, (3) to the lower-left point, and (4) to the upper-right point. The same convention is used for all the four-point quantities. 

In the absence of an explicit electron-hole interaction, the two particles propagate according to
\begin{align}
   L_0(1,2;3,4)
   &=
   G(1,4)G(2,3).
\end{align}
Repeated scattering through \(K_{\textrm{xc}}\) transforms this independent propagation into the correlated two-particle response \(L_{\textrm{xc}}\). The corresponding Bethe-Salpeter equation is 
\begin{align}
    L_{\textrm{xc}}(1,2;3,4)
    &=
    L_0(1,2;3,4)\nonumber\\
    &-L_0(1,\bar 3;3,\bar 1)
    K_{\textrm{xc}}(\bar 1, \bar 2;\bar 3,\bar 4)
    L_{\textrm{xc}}(\bar 4,2;\bar 2,4).\label{eq:LBSEexact}
\end{align}
Thus, \(L_{\textrm{xc}}\) contains the complete ladder resummation
of exchange-correlation interactions between the electron and the hole \cite{Reining_16}. To make the ordering of these ladder insertions explicit, we introduce an associative super-matrix product for four-point functions. For any two four-point functions \(A(1,2;3,4)\) and \(B(1,2;3,4)\) we define
\begin{equation}
    (A \star B)(1,2;3,4)
    =
    \int d2' d4'\,
    A(1,2';3,4')
    B(4',2;2',4).
    \label{eq:mult_def}
\end{equation}
The identity matrix associated to the super-matrix multiplication is
\begin{align}
I(1,2;3,4)=\delta(1,4)\delta(3,2).
\end{align}
The Bethe-Salpeter equation then takes the compact form
\begin{equation}
    L_{\textrm{xc}}
    =
    L_0-L_0 \star K_{\textrm{xc}} \star L_{\textrm{xc}},
    \quad
    L_{\textrm{xc}}
    =
    (I+L_0\star K_{\textrm{xc}})^{-1}\star L_0 .
\end{equation}

This notation allows us to transfer the complete ladder dressing from the two-particle propagator to a vertex. We define the right- and left-acting four-point vertices as
\begin{align}
    \Lambda
    &=
    (I+K_{\textrm{xc}}\star L_0)^{-1}
    =
    I-K_{\textrm{xc}}\star L_{\textrm{xc}},\\
    \widetilde{\Lambda}&=(I+L_0\star K_{\textrm{xc}})^{-1}=I-L_{\textrm{xc}}\star K_{\textrm{xc}}.
\end{align}
They provide the following two equivalent factorizations of the correlated response
\begin{align}
    L_{\textrm{xc}}&=L_0\star (I-K_{\textrm{xc}}\star L_{\textrm{xc}})=L_0\star \Lambda, \label{eq:lambdadef}\\
    L_{\textrm{xc}}&=(I+L_0\star K_{\textrm{xc}})^{-1}\star L_0 =\widetilde{\Lambda} \star L_0. \label{eq:tildelambdadef}
\end{align}
The symmetry of \(L_0\) and \(K_{\textrm{xc}}\) implies
\begin{align}
L_{\textrm{xc}}(1,2;3,4)
=
L_{\textrm{xc}}(2,1;4,3),
\end{align}
and consequently
\begin{align}
\Lambda(2,1;4,3)=\widetilde{\Lambda}(1,2;3,4).
\end{align}
The corresponding Dyson's vertex equations are
\begin{align}
\Lambda&=I-K_{\textrm{xc}} \star L_0 \star \Lambda,\label{eq:Lambda_def}\\
\widetilde{\Lambda}
&=
I-\widetilde{\Lambda}\star L_0\star K_{\textrm{xc}}.\label{eq:Lambda_tilde_def}
\end{align}
The nonlocal vertex is obtained by identifying one incoming and one outgoing coordinate of the four-point vertices as
\begin{align}
    \Gamma'(1,2;3)
    &\equiv
    \Lambda(1,3;2,3^+),
    \label{eq:three_point_ladder_vertices}\\
    \widetilde{\Gamma}'(1,2;3)
    &\equiv
    \widetilde{\Lambda}(3,1;3^+,2).
    \label{eq:three_point_ladder_vertices_tilde}
\end{align}
The symmetry established above gives
\begin{align}
\Gamma'(1,2;3)=\widetilde{\Gamma}'(1,2;3).
\end{align}
The same nonlocal vertex can therefore be used independently of the side from which the ladder is attached. This equivalence will be essential when comparing the reciprocal electronic and phononic scattering processes. The vertex satisfies the Dyson's equation
\begin{align}
\Gamma'(1,2;3)=\delta(1,3^+)\delta(2,3)-K_{\textrm{xc}}(1,\bar 5; 2,\bar 4)G(\bar 4, \bar 6)G(\bar 7, \bar 5)\Gamma'(\bar 6, \bar 7;3).
\label{eq:dysonexactvertex}
\end{align}
This is precisely the vertex entering the exact electronic self-energy in  Eq.~\eqref{eq:sigmaelxcexact}, as follows from the Hedin equations; see, for example, Eqs.~(11.11)--(11.15) of Ref.~\onlinecite{Reining_16}.

For later use, we now introduce the reducible response \(L\), which contains both Hartree and exchange-correlation contributions,
\begin{align}
K_{\textrm{tot}}(1,2;3,4) &= K_{\textrm H}(1,2;3,4) + K_\textrm{xc}(1,2;3,4),
\\
L &= L_0 - L_0\star  K_{\textrm{tot}}\star L.
\label{eq:Ltot_BSE}
\end{align}
The Hartree kernel follows from the functional derivative of Eq.~\eqref{eq:SigmaH},
\begin{align}
K_{\textrm H}(1,2;3,4)
&= i\hbar
v(1,4)\delta(1,3)\delta(2,4^+).
\label{eq:KH_def}
\end{align}
We can now connect this Bethe-Salpeter construction with the functional response introduced in Eq.~\eqref{eq:chi_el_from_Phi}. The closure of the irreducible electron-hole propagator defines
\begin{align}
P_{\textrm{el}}(1,2)
=
-i\hbar
L_{\textrm{xc}}(1,2;1^+,2^+),
\label{eq:PelLxc}
\end{align}
while restoring the Hartree kernel gives the reducible two-particle propagator
\begin{align}
L
=
L_{\textrm{xc}}
-
L_{\textrm{xc}}\star K_{\textrm H}\star L.
\end{align}
Closing the external electron-hole coordinates of this equation and using Eq.~\eqref{eq:KH_def} yields
\begin{align}
-i\hbar L(1,2;1^+,2^+)
&=
P_{\textrm{el}}(1,2)
\nonumber\\
&\quad+
P_{\textrm{el}}(1,\bar 1)
v(\bar 1,\bar 2)
\left[
-i\hbar L(\bar 2,2;\bar 2^+,2^+)
\right],
\end{align}
which has precisely the Dyson structure of Eq.~\eqref{eq:WPW1}. The Bethe--Salpeter construction therefore provides the explicit two-particle representation of the polarization and reducible response introduced above. In particular,
\begin{align}
\chi_{\textrm{el}}(1,2)
=
-i\hbar L(1,2;1^+,2^+).
\label{eq:chi_from_L}
\end{align}
The equivalence between this representation and the functional definition of Eq.~\eqref{eq:chi_el_from_Phi} follows from the fact that the kernel entering the Bethe-Salpeter equation is generated from the same functional \(\Phi_{\rm xc}^{\rm el}[G,v]\) [see discussion between Eq. (12.17) and (12.18) of Ref. \onlinecite{stefanucci2013nonequilibrium}]. In other words, the identification between Eqs. \eqref{eq:chi_from_L} and \eqref{eq:chi_el_from_Phi} is valid only when the generating functional $\Phi^{\textrm{el}}_{\textrm{xc}}[G,v]$ is taken as exact, as in our case. Since the equivalence between the functional and Bethe--Salpeter representations is an identity between skeleton functionals evaluated at fixed \(G\), it does not require \(G\) to be the clamped-nuclei propagator: the electronic lines may include electron-phonon self-energy dressing.

In the next subsection, the ladder vertex will be combined with the screened electron-phonon perturbation to define the nonlocal electron-phonon coupling \(\mathcal G^{\textrm{s}}\).
\subsection{Phonon self-energy}
\label{sec:Pi}
We now derive the phonon self-energy and reorganize it into forms that progressively expose the physical processes contained in the electronic response. The central result of this paragraph is that electronic screening and electron-hole correlations can be absorbed into dressed electron-phonon vertices. This allows the self-energy to be expressed in terms of independent electron-hole propagation, while retaining the many-body interaction effects in
the vertices.

Our starting point is the definition of the electron-phonon sector of the functional, Eq. \eqref{eq:phiephdef}. Using the relation between the interacting electronic response and the functional, Eq. \eqref{eq:chi_el_from_Phi}, it follows
\begin{align}
\widetilde \Phi^{\textrm{el-ph}}[G,D,v,g]=-\frac{1}{2}U_D(\bar 1,\bar 2)\chi_{\textrm{el}}(\bar 1, \bar 2).
\label{eq:phielphexplicit}
\end{align}
The definition of phonon self-energy as derivative of the functional, Eq. \eqref{eq:Piderphitilde}, together with the definition of $U_D$ Eq. \eqref{eq:UDdef}, implies
\begin{align}
\Pi(a,b)=g(\bar 1,a)\chi_{\textrm{el}}(\bar 1,\bar 2) g(\bar 2, b),
\end{align}
where we have used that, at fixed $G$, $\chi_{\textrm{el}}[G,v]$ has no explicit dependence on $D$ (the dependence on $D$ is implicit and appears only after Dyson's equations are solved).
The above expression entails two bare electron-phonon coupling vertices, and all screening and electron-hole correlation effects are contained in the interacting response \(\chi_{\textrm{el}}\). We can reabsorb screening effects in the vertices by combining the relations between the interacting electronic response, the polarization and $W_{\textrm{el}}$ (Eqs. \eqref{eq:chi_el_from_Phi}, \eqref{eq:WPW1} and \eqref{eq:WPW2}). Indeed, $\chi_{\textrm{el}}$ can be conveniently rewritten as
\begin{align}
\chi_{\textrm{el}}=P_{\textrm{el}}+P_{\textrm{el}}W_{\textrm{el}}P_{\textrm{el}}.
\label{eq:chidefPwP}
\end{align}
It is therefore natural to introduce the right- and left-acting
screened electron-phonon vertices as
\begin{align}
g^{\textrm{s}}(1,a)
&=
\left[
\delta(1,\bar 2)
+
W_{\textrm{el}}(1,\bar 3)
P_{\textrm{el}}(\bar 3,\bar 2)
\right]
g(\bar 2,a), \label{eq:gsdef}
\\
\widetilde g^{\textrm{s}}(1,a)
&=
g(\bar 2,a)
\left[
\delta(\bar 2,1)
+
P_{\textrm{el}}(\bar 2,\bar 3)
W_{\textrm{el}}(\bar 3,1)
\right]. \label{eq:tildegsdef}
\end{align}
These vertices represent the perturbation generated by an atomic displacement dressed by the potential induced by the electronic screening cloud. Indeed, they could be equivalently rewritten as
\begin{align}
g^{\textrm{s}}(1,a)
&=
\epsilon^{-1}(1,\bar 2)\,
g(\bar 2,a),  \quad
\epsilon^{-1}
\equiv
I+W_{\textrm{el}}P_{\textrm{el}}
\label{eq:epsilon_inv_def}\\
\widetilde g^{\textrm{s}}(1,a)
&=
g(\bar 2,a)\,
\widetilde\epsilon^{-1}(\bar 2,1), \quad
\widetilde\epsilon^{-1}
\equiv
I+P_{\textrm{el}}W_{\textrm{el}}.
\label{eq:epsilon_inv_left_def}
\end{align}
In terms of these vertices, the phonon self-energy acquires the form
\begin{equation}
\Pi(a,b)
=
 g(\bar 1,a) P_{\rm el}(\bar 1, \bar 2) g^{\textrm{s}}(\bar 2, b)
=
\widetilde g^{\textrm{s}}(\bar 1,a)P_{\rm el}(\bar 1, \bar 2)g(\bar 2, b).
\label{eq:Pi_sym}
\end{equation}
Thus, the full density response can be replaced by the irreducible polarization provided that the electrostatic screening is attached to one of the two electron-phonon vertices. The two expressions of Eq.~\eqref{eq:Pi_sym} are equivalent and correspond to assigning the screening cloud to either side of the electronic response.

The polarization \(P_{\textrm{el}}\) still contains the electron-hole correlations generated by the exchange-correlation kernel. To separate these correlations from the propagation of an independent electron-hole pair, we absorb the ladder vertex into
the electron-phonon coupling and define the nonlocal vertices
\begin{align}
\mathcal G^{\textrm{s}}(1,2;b)
\equiv
\Gamma'(1,2;\bar 3)\,g^{\textrm{s}}(\bar 3,b),\label{eq:def-Gcal-right}\\
\widetilde{\mathcal G}^{\textrm{s}}(1,2;a)
\equiv
\widetilde g^{\textrm{s}}(\bar 3,a)\,
\Gamma'(1,2;\bar 3).
\label{eq:def-Gcal-left}
\end{align}
The symmetry properties $P_{\textrm{el}}(1,2)=P_{\textrm{el}}(2,1)$ and $W_{\textrm{el}}(1,2)=W_{\textrm{el}}(2,1)$ imply
\begin{align}
\widetilde{g}^{\textrm{s}}(1;a)=g^{\textrm{s}}(1;a), \quad  \widetilde{\mathcal G}^{\textrm{s}}(1,2;a)=\mathcal G^{\textrm{s}}(1,2;a), \label{eq:tildeGeqG}
\end{align}
but we will keep the tilde sign in the following to explicit the different graphical orientation of the vertices.
Unlike the local screened vertex \(g^{\textrm{s}}\), the vertices \(\mathcal G^{\textrm{s}}\) and \(\widetilde{\mathcal G}^{\textrm{s}}\) encode the nonlocal correlations generated by repeated electron-hole interactions.

Using Eq.~\eqref{eq:PelLxc} and the factorization of the correlated
response in Eqs.~\eqref{eq:lambdadef} and
\eqref{eq:tildelambdadef}, the phonon self-energy becomes
\begin{align}
\Pi(a,b)
=&
-i\hbar
g(\bar 1,a)
L_0(\bar 1,\bar 3;\bar 1^+,\bar 4)
\mathcal G^{\textrm{s}}(\bar 4,\bar 3;b)\nonumber \\
=&-i\hbar
\widetilde{\mathcal G}^{\textrm{s}}(\bar 3,\bar 4;a)
L_0(\bar 4,\bar 2;\bar 3,\bar 2^+)
g(\bar 2,b).
\label{eq:Pi-L0-Gcal}
\end{align}
The above representation is analogous to the one introduced in Ref. \onlinecite{y1dn-m6pc}. \(L_0\) describes the propagation of an independent electron-hole pair, while the complete ladder resummation is contained in the nonlocal vertex, that include possible interactions between phonons and bounded excitons.

At this point, equation~\eqref{eq:Pi-L0-Gcal} contains the complete electron-hole ladder dressing, but only one of the two electron-phonon vertices is written in its fully dressed form. We now reorganize the same self-energy so that the independent electron-hole propagator \(L_0\) is coupled to dressed vertices on both sides.

To identify the corresponding correction, we define the electronic density matrix induced by the bare electron-phonon perturbation as
\begin{equation}
n_{\textrm{ind}}(1,2;b)
\equiv -i\hbar 
L(1,\bar 3;2,\bar 3^+)\,
g(\bar 3,b).
\label{eq:nind_def}
\end{equation}
As shown in App.~\ref{app:Gggs}, the right-acting dressed vertex can then be written as
\begin{align}
\mathcal G^{\textrm{s}}(1,2;b)
=
\delta(1,2^+)g(2,b)
-
\frac{i}{\hbar}
K_{\textrm{tot}}(1,\bar 3;2,\bar 4)
n_{\textrm{ind}}(\bar 4,\bar 3;b).
\label{eq:Gcal_g_Kn_coord}
\end{align}
This identity separates the dressed vertex into the bare electron-phonon perturbation and the electronic feedback generated through the full interaction kernel. It can therefore be used to replace the bare vertex appearing on the left-hand side of
Eq.~\eqref{eq:Pi-L0-Gcal}. Analogously, we introduce
\begin{align}
\widetilde n_{\textrm{ind}}(1,2;a)
&\equiv
-i\hbar\,
g(\bar 3,a)
L(\bar 3,2;\bar 3^+,1),
\nonumber\\
\widetilde{\mathcal G}^{\textrm{s}}(1,2;a)
&=
\delta(1,2^+)g(2,a)
-
\frac{i}{\hbar}
\widetilde n_{\textrm{ind}}(\bar 3,\bar 4;a)
K_{\textrm{tot}}(\bar 4,2;\bar 3,1).
\label{eq:Gcal_left_g_Kn_coord}
\end{align}
Substituting these relations into
Eq.~\eqref{eq:Pi-L0-Gcal}, the phonon self-energy becomes
\begin{align}
\Pi(a,b)
&=
-i\hbar\,
\widetilde{\mathcal G}^{\textrm{s}}(\bar 3,\bar 4;a)
L_0(\bar 4,\bar 5;\bar 3,\bar 6)
\mathcal G^{\textrm{s}}(\bar 6,\bar 5;b)
\nonumber\\
&\quad
+
\frac{i}{\hbar}\,
\widetilde n_{\textrm{ind}}(\bar 3,\bar 4;a)
K_{\textrm{tot}}(\bar 4,\bar 5;\bar 3,\bar 6)
n_{\textrm{ind}}(\bar 6,\bar 5;b).
\label{eq:Pi_SX_Gcal_nind}
\end{align}
The first term has the desired symmetric structure: the independent electron-hole propagator \(L_0\) is coupled to fully dressed electron-phonon vertices on both sides. The second term compensates for the interaction processes that would otherwise be counted in both vertices.

\subsection{Electron-phonon contribution to the electronic self-energy}
We now consider the reciprocal electronic process generated by the same electron-phonon functional. Whereas differentiation with respect to \(D\) opens the phonon line and produces the phonon self-energy, differentiation with respect to \(G\) opens an electronic line inside the density response and produces the electron-phonon contribution to the electronic self-energy. 

Starting from Eq.~\eqref{eq:phielphexplicit}, we obtain
\begin{align}
\Sigma^{\textrm{el-ph}}(1,2)=-\frac{1}{2}U_D(\bar 1,\bar 2)\frac{\delta \chi_{\textrm{el}}(\bar 1, \bar 2)}{\delta G(2,1^+)}.
\end{align}
The problem is thus reduced to determining how the interacting electronic response changes when one electronic propagator is varied. In particular, the variation of $\chi_{\textrm{el}}$ is related to the variation of $P_{\textrm{el}}$. The polarization definition in terms of the correlated two-particle response $L_{\textrm{xc}}$, Eq. \eqref{eq:PelLxc}, and the expression of the latter in terms of the irreducible kernel, \eqref{eq:LBSEexact}, imply that the polarization has a dependence on $G$ of the form
\begin{align}
P_{\textrm{el}}=P_{\textrm{el}}[G,v]=P_{\textrm{el}}[G,K_{\textrm{xc}}[G,v]].
\end{align}
To clarify functional dependencies, we write the total derivative of the polarization with respect to $G$ as the sum of partial derivatives
\begin{align}
\frac{\delta P_{\textrm{el}}}{\delta G}=\left. \frac{\delta P_{\textrm{el}}}{\delta G} \right|_{K_{\textrm{xc}}}+ \frac{\delta P_{\textrm{el}}}{\delta K_{\textrm{xc}}} \frac{\delta K_{\textrm{xc}}}{\delta G},
\label{eq:dPdG}
\end{align}
where $|_{K_{\textrm{xc}}}$ means that the irreducible kernel is kept fixed in the variation. As shown in App. \ref{app:variation_chi}, the variation of the interacting electronic response is expressed as
\begin{align}
\frac{\delta \chi_{\textrm{el}}}{\delta G}= \left[I+
P_{\textrm{el}}W_{\textrm{el}}\right]   \frac{\delta P_{\textrm{el}}}{\delta G} \left[I+
W_{\textrm{el}}P_{\textrm{el}}\right].
\end{align}
The two factors surrounding \(\delta P_{\textrm{el}}/\delta G\) are precisely those already absorbed into the left- and right-acting screened electron-phonon vertices. Using Eqs.~\eqref{eq:gsdef} and \eqref{eq:tildegsdef}, the self-energy can therefore be written as
\begin{align}
\Sigma^{\textrm{el-ph}}(1,2)
=
-\frac{1}{2}\,
\widetilde g^{\textrm{s}}(\bar 3,\bar a)
D(\bar a,\bar b)
g^{\textrm{s}}(\bar 4,\bar b)
\frac{\delta P_{\textrm{el}}(\bar 3,\bar 4)}
{\delta G(2,1^+)}.
\end{align}

To obtain a closed expression, we evaluate the variation of the polarization at fixed irreducible kernel \cite{PhysRevB.29.5718}
\begin{align}
\frac{\delta P_{\textrm{el}}}{\delta G}
\simeq
\left.
\frac{\delta P_{\textrm{el}}}{\delta G}
\right|_{K_{\textrm{xc}}}.
\label{eq:fixed_kernel_approx}
\end{align}
This approximation retains the complete variation of the electron-hole propagators inside the ladder while neglecting the additional variation of the irreducible interaction kernel itself. The self-energy then becomes
\begin{align}
\Sigma^{\substack{\textrm{el-ph}}}(1,2)
=
-\frac12\,
\widetilde g^{\textrm{s}}(\bar 3,\bar a)
D(\bar a,\bar b)
g^{\textrm{s}}(\bar 4,\bar b)
 \left. \frac{\delta P_{\textrm{el}}(\bar 3,\bar 4)}
{\delta G(2,1^+)}\right|_{K_{\textrm{xc}}}.
\label{eq:Sigma_eph_explicit}
\end{align}
Using the polarization definition Eq.~\eqref{eq:PelLxc}, the remaining derivative is
\begin{equation}
\left. \frac{\delta P_{\textrm{el}}(\bar 3,\bar 4)}
{\delta G(2,1^+)}\right|_{K_{\textrm{xc}}}=
-i\hbar\,
\left. \frac{\delta L_{\textrm{xc}}(\bar 3,\bar 4;\bar 3^+,\bar 4^+)}
{\delta G(2,1^+)}\right|_{K_{\textrm{xc}}}.
\end{equation}
Using the relations Eqs. \eqref{eq:Lambda_def} and \eqref{eq:Lambda_tilde_def}, as detailed in App.~\ref{app:derivKL}, we have
\begin{align}
\left. \frac{\delta L_{\textrm{xc}}(3,4;3^+,4^+)}
{\delta G(2,1^+)}\right|_{K_{\textrm{xc}}}
=
\widetilde{\Lambda}(3,\bar 1;3^+,\bar 2)
&\frac{\delta L_0(\bar 2,\bar 3;\bar 1,\bar 4)}
{\delta G(2,1^+)}
\nonumber\\
&\times \Lambda(\bar 4,4;\bar 3,4^+).
\label{eq:deltaLL}
\end{align}
Since \(L_0\) is the product of two electronic propagators, its derivative contains two terms, corresponding to opening either line of the electron-hole pair, i.e.
\begin{align}
\frac{\delta L_0(\bar 2,\bar 3;\bar 1, \bar 4)}{\delta G(2,1^+)}=\delta(\bar 2,2)\delta(\bar 4,1^+) G(\bar 3, \bar 1)+\delta(\bar 3,2)\delta(\bar 1, 1^+)G(\bar2, \bar4) 
\end{align}
Explicitly,
\begin{align}
\left. \frac{\delta L_{\textrm{xc}}(3,4;3^+,4^+)}
{\delta G(2,1^+)}\right|_{K_{\textrm{xc}}}
&=
\widetilde{\Gamma}'(\bar 1,2;3)
G(\bar 2,\bar 1)
\Gamma'(1^+,\bar 2;4)
\nonumber\\
&\quad
+
\widetilde{\Gamma}'(1^+,\bar 1;3)
G(\bar 1,\bar 2)
\Gamma'(\bar 2,2;4).
\label{eq:dLLdG_static}
\end{align}
The two contributions are equivalent after using the symmetry of the left- and right-acting ladder vertices and relabeling the dummy variables. Their multiplicity cancels the factor \(1/2\) in Eq.~\eqref{eq:Sigma_eph_explicit}. Introducing the dressed electron-phonon vertices defined in Eqs.~\eqref{eq:def-Gcal-right} and \eqref{eq:def-Gcal-left}, we finally obtain
\begin{align}
\Sigma^{\textrm{el-ph}}(1,2)
=
i\hbar\,
\widetilde{\mathcal G}^{\textrm{s}}(1^+,\bar 1;\bar a)
D(\bar a,\bar b)
G(\bar 1,\bar 2)
\mathcal G^{\textrm{s}}(\bar 2,2;\bar b).
\label{eq:Sigma_eph_W_Gcal}
\end{align}
Eq.~\eqref{eq:Sigma_eph_W_Gcal} is written in terms of fully dressed propagators. In particular, the electronic propagator \(G\), both in this equation and in the response functions entering the dressed vertices, may contain the electron--phonon dressing generated self-consistently through the  electronic Dyson equation. Such self-energy insertions are included implicitly through the dressed propagator and must not be added explicitly to the skeleton diagrams. In the SX quasiparticle reduction introduced below, this fully dressed propagator is approximated by \(G^{\rm SX}\), and the electron-phonon dressing of the internal electronic lines is consequently neglected. More advanced approximations keeping track on the electron-phonon self-energy could be implemented, and we leave it to future works.

Our derivation highlights a central advantage of the functional construction. Within a direct Hedin--Baym approach, one would have to identify the relevant vertex-corrected contributions to the electronic self-energy and combine them so as to recover the appropriate left-right structure. Here, this structure follows automatically from the derivative of the common functional
\(\widetilde{\Phi}\). Indeed, the variation of the density response first produces the screened electron-phonon vertices in the balanced form \(\widetilde g^{\textrm{s}}\) and \(g^{\textrm{s}}\), acting on opposite sides of \(\delta P_{\textrm{el}}/\delta G\). The subsequent differentiation of the polarization opens either electronic line of \(L_0=GG\), while the ladder corrections contained in the response generate naturally the corresponding left- and right-acting vertices \(\widetilde{\mathcal G}^{\textrm{s}}\) and \(\mathcal G^{\textrm{s}}\). The final two-sided dressing of the electronic self-energy therefore requires neither a diagram-by-diagram reconstruction from the Hedin-Baym equations nor an additional symmetrization prescription: it is inherited directly from the functional derivative of the electronic response.

\section{Quasiparticle linewidths and detailed balance}
\label{sec:fgr}
In this section, we finally perform the screened-exchange (SX) quasiparticle reduction represented by the
last arrow in Fig.~\ref{fig:theory-workflow}. We specialize the self-energies derived above to the SX approximation and use them to obtain Fermi golden rule expressions for the phononic and electronic
linewidths. The purpose of the SX construction is twofold: it provides a static and computationally tractable description of the electronic interaction, while retaining the static electron-hole ladder corrections that dress the electron-phonon coupling. We will show that the two linewidths are governed by the same matrix element \(\mathcal G^{\textrm{s,SX}}\), thereby making their detailed-balance relation explicit.

\begin{figure}[h]
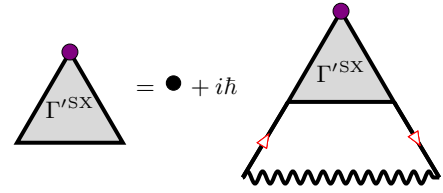

  \centering
  \[
  \begin{aligned}
  \GammaPrime
  \,
  =
  \,
  \DoubleDeltaGammaV
  \,
  +
  i\hbar
  \,
  \WelGGGammaPrime
  \end{aligned}
  \]
  \caption{
  Dyson equation for the SX electronic vertex
  \(\Gamma'{}^{\textrm{SX}}\),
  Eq.~\eqref{eq:DysonvertexSX}.
  The purple dot denotes the closure of two electronic
  coordinates.
  }
  \label{fig:dysongamma}
\end{figure}

\begin{figure}[htbp]
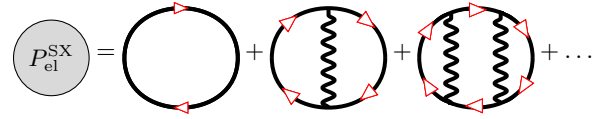

  \centering
  \[
  \PelBubble
  =
  \PolBubble
  +
  \PolBubbleOneW
  +
  \PolBubbleTwoW
  +\ldots
  \]
  \caption{
  Ladder expansion of the electronic polarization in the SX
  approximation, Eq.~\eqref{eq:P_el_static_LL}.
  For graphical convenience, the same interaction line is used
  to represent \(K_L^{\textrm{SX}}\) and
  \(W_{\textrm{el}}^{\textrm{SX}}\), although the former also
  contains the factor \(-i\hbar\).}
  \label{fig:polarization-bubble-expansion}
\end{figure}

The passage from the general formulation of Sec.~\ref{sec:selfenergies} to the SX approximation is obtained by consistently replacing all electronic building blocks by their SX counterparts. We first replace the dynamically screened interaction by a static one,
\begin{align}
W_{\textrm{el}}(1,2)
\longrightarrow
W_{\textrm{el}}^{\textrm{SX}}(1,2)
&=
W_{\textrm{el}}^{\textrm{SX}}
(\mathbf r_1,\mathbf r_2)
\delta(t_1-t_2).
\end{align}
The interaction \(W_{\textrm{el}}^{\textrm{SX}}\) is treated as a fixed effective two-body interaction. It is therefore held fixed in all functional variations and response kernels. For the moment, we regard it as a given quantity; its explicit  construction will be specified in the applications. Freezing the frequency dependence of the interaction corresponds to replacing the exact exchange-correlation kernel by the static SX ladder kernel,
\begin{align}
K_{\textrm{xc}}
&\longrightarrow
K_L^{\textrm{SX}},
\\
K_L^{\textrm{SX}}(1,2;3,4)
&=
-i\hbar\,
W_{\textrm{el}}^{\textrm{SX}}(1,2)
\delta(1,4)
\delta(3,2).
\label{eq:replacement}
\end{align}
The subscript \(L\) emphasizes that this kernel is used to resum the repeated ladder interaction between an electron and a hole. Accordingly, we denote by \(L_L^{\textrm{SX}}\) the SX ladder approximation to the correlated response
\(L_{\textrm{xc}}\). It obeys
\begin{align}
L_L^{\textrm{SX}}
&=
L_0^{\textrm{SX}}
-
L_0^{\textrm{SX}}
\star K_L^{\textrm{SX}}
\star L_L^{\textrm{SX}},
\\
L_0^{\textrm{SX}}(1,2;3,4)
&=
G^{\textrm{SX}}(1,4)
G^{\textrm{SX}}(2,3).
\end{align}
Thus, \(L_0^{\textrm{SX}}\) describes the propagation of an independent electron-hole pair built from the SX electronic propagator, whereas \(L_L^{\textrm{SX}}\) contains its complete ladder dressing by \(W_{\textrm{el}}^{\textrm{SX}}\). The same resummation can be transferred from the two-particle response to the associated nonlocal vertex. Its Dyson equation reads
\begin{align}
\Gamma'{}^{\textrm{SX}}(1,2;3)
&=
\delta(1,3^+)
\delta(2,3)
\nonumber\\
&\quad
+
i\hbar\,
W_{\textrm{el}}^{\textrm{SX}}(1,2)
G^{\textrm{SX}}(1,\bar 1)
G^{\textrm{SX}}(\bar 2,2)
\Gamma'{}^{\textrm{SX}}
(\bar 1,\bar 2;3),
\label{eq:DysonvertexSX}
\end{align}
and it is represented graphically in Fig. \ref{fig:dysongamma}.
The SX polarization is obtained by closing the electronic coordinates of the ladder response, as represented in Fig. \ref{fig:polarization-bubble-expansion}:
\begin{align}
P_{\textrm{el}}^{\textrm{SX}}(1,2)
&=
-i\hbar\,
L_L^{\textrm{SX}}(1,2;1^+,2^+).
\label{eq:P_el_static_LL}
\end{align}
The full reducible SX response is instead obtained by closing
\(L^{\textrm{SX}}\),
\begin{align}
\chi_{\textrm{el}}^{\textrm{SX}}(1,2)
&=
-i\hbar\,
L^{\textrm{SX}}(1,2;1^+,2^+),
\label{eq:chi_el_SX}
\end{align}
where \(L^{\textrm{SX}}\) satisfies
\begin{align}
L^{\textrm{SX}}
&=
L_0^{\textrm{SX}}
-
L_0^{\textrm{SX}}
\star K_{\textrm{tot}}^{\textrm{SX}}
\star L^{\textrm{SX}},
\label{eq:Ltot_BSE-SX}\\
K_{\textrm{tot}}^{\textrm{SX}}(1,2;3,4)
&=
K_{\textrm H}(1,2;3,4)
+
K_L^{\textrm{SX}}(1,2;3,4).
\label{eq:Ktot_BSE-SX}
\end{align}
In the above kernel equation, the Hartree contribution is formally left unchanged, while the exchange-correlation part is replaced by the SX ladder interaction.

Finally, the fixed-kernel derivative used in the electronic self-energy maps onto a derivative at fixed static screened interaction,
\begin{align}
\left.
\frac{\delta P_{\textrm{el}}}
{\delta G}
\right|_{K_{\textrm{xc}}}
\longrightarrow
\left.
\frac{\delta P_{\textrm{el}}^{\textrm{SX}}}
{\delta G^{\textrm{SX}}}
\right|_{W_{\textrm{el}}^{\textrm{SX}}}.
\label{eq:dPdG-SX}
\end{align}
The notation
\(\left.\vphantom{\delta P}\right|
_{W_{\textrm{el}}^{\textrm{SX}}}\)
indicates that the screened interaction is not varied when the polarization is differentiated with respect to \(G^{\textrm{SX}}\).

These substitutions provide the common SX dictionary required in all the following subsections. The phononic and electronic self-energies can now be obtained directly from their general expressions in Sec.~\ref{sec:selfenergies}, without repeating the functional derivation.

\subsection{Exchange-correlation electronic self-energy}
We first construct the electronic reference system underlying all subsequent SX response functions and self-energies. Applying the SX prescription to the exact exchange-correlation self-energy of Eq.~\eqref{eq:sigmaelxcexact}, it is customary to set the Hedin vertex to unity and retain only the static screened interaction introduced above. The exchange-correlation self-energy then becomes
\begin{align}
\Sigma_{\textrm{xc}}^{\textrm{el,SX}}(1,2)
&=
i\hbar\,
W_{\textrm{el}}^{\textrm{SX}}(1,2)
G^{\textrm{SX}}(1,2^+)
\nonumber\\
&=
\delta(t_1-t_2)
\Sigma_{\textrm{xc}}^{\textrm{el,SX}}
(\mathbf r_1,\mathbf r_2).
\label{eq:Sigma_xc_SX}
\end{align}
Here, \(1\) and \(2\) are external variables and are not integrated. Because \(W_{\textrm{el}}^{\textrm{SX}}\) is instantaneous, the self-energy is static, although it remains nonlocal in the electronic coordinates. Its spatial part can be written as
\begin{align}
\Sigma_{\textrm{xc}}^{\textrm{el,SX}}
(\mathbf r_1,\mathbf r_2)
&=
i\hbar\,
W_{\textrm{el}}^{\textrm{SX}}
(\mathbf r_1,\mathbf r_2)
G^{\textrm{SX}}
(\mathbf r_1t_1,\mathbf r_2t_1^+)
\nonumber\\
&=
-
W_{\textrm{el}}^{\textrm{SX}}
(\mathbf r_1,\mathbf r_2)
\rho^{\textrm{SX}}
(\mathbf r_1,\mathbf r_2),
\end{align}
where
\begin{align}
\rho^{\textrm{SX}}(\mathbf r_1,\mathbf r_2)
=
-i\hbar\,
G^{\textrm{SX}}
(\mathbf r_1t_1,\mathbf r_2t_1^+)
\end{align}
is the nonlocal one-particle density matrix. At equilibrium, time-translation invariance makes this equal-time quantity independent of \(t_1\).

The static self-energy can therefore be incorporated into an effective one-particle Hamiltonian,
\begin{align}
h^{\textrm{SX}}(\mathbf r_1,\mathbf r_2)
&=
h_0(\mathbf r_1)
\delta(\mathbf r_1-\mathbf r_2)
\nonumber\\
&\quad
+
\Sigma_{\textrm H}^{\textrm{el,SX}}
(\mathbf r_1,\mathbf r_2)
+
\Sigma_{\textrm{xc}}^{\textrm{el,SX}}
(\mathbf r_1,\mathbf r_2).
\label{eq:hSX_def}
\end{align}
Here, \(h_0\) contains the electron--nucleus contribution, while \(\Sigma_{\textrm H}^{\textrm{el,SX}}\) is evaluated from the self-consistent SX density. The corresponding states and energies satisfy
\begin{align}
h^{\textrm{SX}}
(\mathbf r_1,\bar{\mathbf r}_2)
\psi_i^{\textrm{SX}}(\bar{\mathbf r}_2)
=
\varepsilon_i^{\textrm{SX}}
\psi_i^{\textrm{SX}}(\mathbf r_1).
\label{eq:hSX_eigenproblem}
\end{align}
Since \(h^{\textrm{SX}}\) is static and Hermitian, the eigenvalues \(\varepsilon_i^{\textrm{SX}}\) are real and its eigenstates form the natural basis for the subsequent SX construction. In this basis, the electronic Dyson equation is solved directly. The frequency-space Green's function reads
\begin{align}
G^{\textrm{SX}}
(\mathbf r_1,\mathbf r_2;\omega)
&=
\frac{1}{2\pi}
\sum_i
\psi_i^{\textrm{SX}}(\mathbf r_1)
\psi_i^{\textrm{SX}*}(\mathbf r_2)
\nonumber\\
&\quad\times
\left[
\frac{1-f_i^{\textrm{SX}}}
{\hbar\omega-\varepsilon_i^{\textrm{SX}}+i\eta}
+
\frac{f_i^{\textrm{SX}}}
{\hbar\omega-\varepsilon_i^{\textrm{SX}}-i\eta}
\right].
\label{eq:G_SX_spectral_frequency}
\end{align}
Here, \(f_i^{\textrm{SX}}=
f(\varepsilon_i^{\textrm{SX}})\) are the equilibrium Fermi occupation factors, and the chemical potential is set to zero for notational simplicity. This provides the electronic propagator entering both the electron-hole ladder and the phononic and electronic scattering processes derived below.

\subsection{Phonon self-energy and Fermi golden rule for phonons}

Having defined the SX electronic reference system, we now determine how an atomic perturbation is renormalized by the medium. As in Sec. \ref{sec:Pi}, the construction proceeds in three steps. The bare electron-phonon perturbation is first screened by the electronic polarization, then dressed by the electron-hole ladder, and finally inserted into the phonon self-energy. Its retarded, on-shell imaginary part will provide the phonon linewidth.

The SX screened electron-phonon vertices follow directly from Eqs.~\eqref{eq:gsdef} and \eqref{eq:tildegsdef},
\begin{align}
g^{\textrm{s,SX}}(1,a)
&=
\left[
\delta(1,\bar 2)
+
W^{\textrm{SX}}_{\textrm{el}}(1,\bar 3)
P^{\textrm{SX}}_{\textrm{el}}(\bar 3,\bar 2)
\right]
g(\bar 2,a),
\label{eq:gsdefSX}
\\
\widetilde g^{\textrm{s,SX}}(1,a)
&=
g(\bar 2,a)
\left[
\delta(\bar 2,1)
+
P^{\textrm{SX}}_{\textrm{el}}(\bar 2,\bar 3)
W^{\textrm{SX}}_{\textrm{el}}(\bar 3,1)
\right].
\label{eq:tildegsdefSX}
\end{align}
The phonon self-energy of Eq.~\eqref{eq:Pi_sym} consequently becomes
\begin{align}
\Pi^{\textrm{SX}}(a,b)
&=
g(\bar 1,a)
P^{\textrm{SX}}_{\textrm{el}}(\bar 1,\bar 2)
g^{\textrm{s,SX}}(\bar 2,b)
\nonumber\\
&=
\widetilde g^{\textrm{s,SX}}(\bar 1,a)
P^{\textrm{SX}}_{\textrm{el}}(\bar 1,\bar 2)
g(\bar 2,b).
\label{eq:Pi_ep_LL_final-SX}
\end{align}

The electron-hole ladder is incorporated through the nonlocal vertices
\begin{align}
\mathcal G^{\textrm{s,SX}}(1,2;b)
&\equiv
\Gamma'{}^{\textrm{SX}}(1,2;\bar 3)
g^{\textrm{s,SX}}(\bar 3,b),
\label{eq:def-Gcal-right-SX}
\\
\widetilde{\mathcal G}^{\textrm{s,SX}}(1,2;a)
&\equiv
\widetilde g^{\textrm{s,SX}}(\bar 3,a)
\Gamma'{}^{\textrm{SX}}(1,2;\bar 3).
\label{eq:def-Gcal-left-SX}
\end{align}
The general result of Eq.~\eqref{eq:Pi-L0-Gcal} therefore maps onto
\begin{align}
\Pi^{\textrm{SX}}(a,b)
&=
-i\hbar\,
g(\bar 1,a)
L^{\textrm{SX}}_0
(\bar 1,\bar 3;\bar 1^+,\bar 4)
\mathcal G^{\textrm{s,SX}}
(\bar 4,\bar 3;b)
\nonumber\\
&=
-i\hbar\,
\widetilde{\mathcal G}^{\textrm{s,SX}}
(\bar 3,\bar 4;a)
L^{\textrm{SX}}_0
(\bar 4,\bar 2;\bar 3,\bar 2^+)
g(\bar 2,b).
\label{eq:Pi-L0-Gcal-SX}
\end{align}
Thus, the full SX electron-hole correlation is transferred from
the two-particle response to a dressed electron-phonon vertex,
leaving \(L_0^{\textrm{SX}}\) as the propagation of an independent
SX electron-hole pair.

\begin{figure}[htbp]
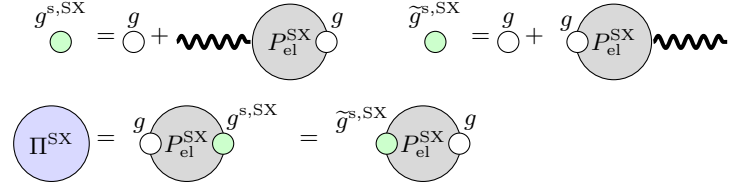

  \centering
  \[
  \begin{aligned}
  \gsVertex
  &=
  \gVertex
  +
  \WelPelgTerm
  \qquad
  \gtildeSVertex
  =
  \gVertex
  +
  \gPelWelTerm
  \\
  \PiBubbleSX
  &=
  \gPelGsTerm
  =
  \gtildeSPelgTerm
  \end{aligned}
  \]
  \caption{
  SX screened electron-phonon vertices
  \(g^{\textrm{s,SX}}\) and
  \(\widetilde g^{\textrm{s,SX}}\),
  Eqs.~\eqref{eq:gsdefSX} and
  \eqref{eq:tildegsdefSX}, and the two equivalent
  representations of the phonon self-energy in
  Eq.~\eqref{eq:Pi_ep_LL_final-SX}.
  }
  \label{fig:Pi-sym}
\end{figure}

\begin{figure}[htbp]
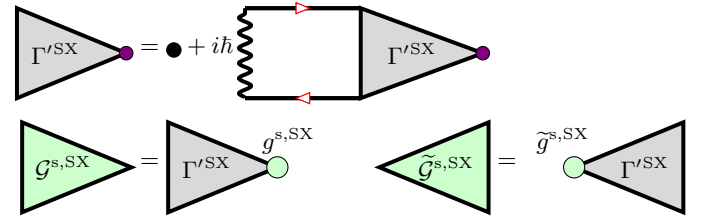

  \centering
  \[
  \begin{aligned}
  \GammaPrimeHorizontal
  &=
  \DoubleDeltaGammaH
  +
  i\hbar\,
  \WelVerticalGGGammaPrime
  \\
  \GcalSVertex
  &=
  \GammaPrimeGsTerm
  \qquad
  \GcalTildeSVertex
  =
  \GtildeSGammaPrimeTerm
  \end{aligned}
  \]
  \caption{
  SX ladder equation for the electronic vertex and construction
  of the composite electron-phonon vertices
  \(\mathcal G^{\textrm{s,SX}}\) and
  \(\widetilde{\mathcal G}^{\textrm{s,SX}}\),
  Eqs.~\eqref{eq:def-Gcal-right-SX} and
  \eqref{eq:def-Gcal-left-SX}.
  }
  \label{fig:Gcal-s-definition}
\end{figure}

\begin{figure}[htbp]
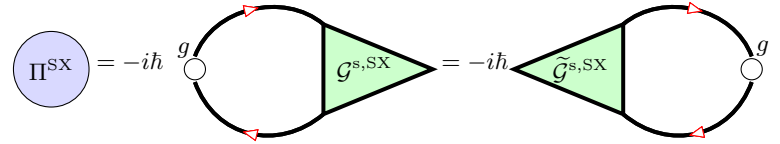

  \centering
  \[
  \PiBubbleSX
  =
  -i\hbar\,
  \gLzeroGcalSTerm
  =
  -i\hbar\,
  \GcalTildeSLzeroGTerm
  \]
  \caption{
  SX phonon self-energy expressed through the independent
  electron-hole propagator and one composite dressed vertex,
  Eq.~\eqref{eq:Pi-L0-Gcal-SX}.
  }
  \label{fig:Pi-L0-Gcal}
\end{figure}

For the derivation of the linewidth, it is useful to employ the
equivalent representation in which dressed vertices appear on both
sides of \(L_0^{\textrm{SX}}\). Applying the general result of
Eq.~\eqref{eq:Pi_SX_Gcal_nind} gives
\begin{align}
\Pi^{\textrm{SX}}(a,b)
&=
-i\hbar\,
\widetilde{\mathcal G}^{\textrm{s,SX}}
(\bar 3,\bar 4;a)
L^{\textrm{SX}}_0
(\bar 4,\bar 5;\bar 3,\bar 6)
\mathcal G^{\textrm{s,SX}}
(\bar 6,\bar 5;b)
\nonumber\\
&\quad
+
\frac{i}{\hbar}\,
\widetilde n^{\textrm{SX}}_{\textrm{ind}}
(\bar 3,\bar 4;a)
K^{\textrm{SX}}_{\textrm{tot}}
(\bar 4,\bar 5;\bar 3,\bar 6)
n^{\textrm{SX}}_{\textrm{ind}}
(\bar 6,\bar 5;b).
\label{eq:Pi_SX_Gcal_nind-SX}
\end{align}
The first term displays the desired two-sided dressed-vertex structure. 

Equation~\eqref{eq:Pi_SX_Gcal_nind-SX} was obtained in the zero-temperature time-ordered formalism. The phonon linewidth, however, is determined by the retarded self-energy. We denote the
analytic continuation of the self-energy by \(\Pi^{\textrm{R,SX}}_{\mu\mathbf q}(z)\), with
\(z=\omega+i0^+\). A quantity evaluated at \(z\) lies on the retarded side of the real-frequency axis, whereas one evaluated at \(-z\) lies at \(-\omega-i0^+\), on the advanced side. From this point onward, we adopt the linear-response normalization
of Ref.~\onlinecite{PhysRevB.111.075137}. Conventional factors of \(\pm i\hbar\) are absorbed into the definitions of the response functions. Thus, \(L_0^{\textrm{SX}}(z)\) denotes the retarded independent electron-hole response obtained by analytic continuation of \(-i\hbar L_0^{\textrm{SX}}\), and the same normalization is used for the full response and the induced
densities. Correspondingly, the factor \(-i/\hbar\) associated with the kernel term in Eq.~\eqref{eq:Pi_SX_Gcal_nind-SX} is absorbed into
\(K_{\textrm{tot}}^{\textrm{SX}}\). With these conventions, the retarded phonon self-energy reads
\begin{widetext}
\begin{align}
\Pi^{\textrm{R,SX}}_{I,\alpha;J,\beta}(z)
&=
\widetilde{\mathcal G}^{\textrm{s,SX}}
(\bar{\mathbf r}_3,\bar{\mathbf r}_4;
 I,\alpha;-z)L_0^{\textrm{SX}}
(\bar{\mathbf r}_4,\bar{\mathbf r}_5;
 \bar{\mathbf r}_3,\bar{\mathbf r}_6;z)
\mathcal G^{\textrm{s,SX}}
(\bar{\mathbf r}_6,\bar{\mathbf r}_5;
 J,\beta;z)
\nonumber\\
&\quad
-
\widetilde n_{\textrm{ind}}^{\textrm{SX}}
(\bar{\mathbf r}_3,\bar{\mathbf r}_4;
 I,\alpha;-z)
K_{\textrm{tot}}^{\textrm{SX}}
(\bar{\mathbf r}_4,\bar{\mathbf r}_5;
 \bar{\mathbf r}_3,\bar{\mathbf r}_6)
n_{\textrm{ind}}^{\textrm{SX}}
(\bar{\mathbf r}_6,\bar{\mathbf r}_5;
 J,\beta;z).
\label{eq:Pi_R_SX_Gcal_nind_atomic}
\end{align}
\end{widetext}
At finite temperature, the same retarded expression is obtained equivalently either by analytic continuation from Matsubara frequencies, or may be obtained by applying the finite-temperature Langreth rules \cite{Hyrkas2019}. Notice that the electron-phonon vertex depends on a single frequency. In general, a nonlocal vertex contains three time arguments, of which one is removed by equilibrium time-translation invariance. In the static SX approximation, every interaction rung of the ladder is instantaneous, and this property is preserved by the complete ladder resummation: the two electronic time arguments of the dressed vertex remain constrained to coincide. Only one independent time (difference) is therefore left, corresponding in frequency space to the transferred frequency \(z\). 

Eq. ~\eqref{eq:Pi_R_SX_Gcal_nind_atomic} is the ``screen-screen'' form of
Ref.~\onlinecite{PhysRevB.111.075137}, in which the left-acting vertex and induced
density are evaluated at $-z$. Its imaginary part is most directly extracted after
passing to the equivalent ``screen$^{*}$-screen'' form, obtained by replacing
$-z\to-z^{*}$ in the left-acting quantities: the two expressions define the same
holomorphic function of $z$, although each of their two terms separately is not
holomorphic. Since the phonon perturbation obeys $g_{-\mathbf q}=g_{\mathbf q}^{*}$,
in this representation the left-acting objects are the Hermitian conjugates of the
right-acting ones,
\begin{align}
\mathcal G^{\textrm{s,SX}}(\mathbf r_2;I,\alpha;\mathbf r_1;-z^{*})
&=\left[\mathcal G^{\textrm{s,SX}}(\mathbf r_1,\mathbf r_2;I,\alpha;z)\right]^{*},
\nonumber\\
n^{\textrm{SX}}_{\textrm{ind}}(\mathbf r_2,\mathbf r_1;I,\alpha;-z^{*})
&=\left[n^{\textrm{SX}}_{\textrm{ind}}(\mathbf r_1,\mathbf r_2;I,\alpha;z)\right]^{*}.
\end{align}
The kernel term then reduces to the Hermitian quadratic form
$-n^{*}K^{\textrm{SX}}_{\textrm{tot}}n$, which is real because
$K^{\textrm{SX}}_{\textrm{tot}}$ is real and obeys
Eq.~\eqref{eq:app_K_exchange}, and therefore does not contribute to the imaginary part of the self-energy.
The entire imaginary part originates from $L^{\textrm{SX}}_{0}$ sandwiched between
two complex-conjugate vertices. Further, the static quadratic contribution \(\mathcal B\) does not introduce dissipation.  Indeed, Eq. ~(B7) of Ref.~\onlinecite{PhysRevX.13.031026}) implies
\begin{align}
\mathcal B_{I,\alpha;J,\beta}
=
-\Pi^{\textrm{R}}_{I,\alpha;J,\beta}(0).
\end{align}
In short, the result follows from using the static linear response relation $\left.
\frac{\partial n^{\rm BO}}{\partial R_J}
\right|_{\mathbf R^{0,\rm BO}}
=
\chi_{\rm el}^{\rm R}(0)g_J$ in Eq. \eqref{eq:Bdef}, 
which is valid because, only for the exact xc functional, the electronic
polarizability defined through the derivative of
\(\Phi_{\rm xc}^{\rm el}\) in Eq.~\eqref{eq:chi_el_from_Phi} (and used to compute the phonon self-energy) is
identical to the polarizability generated by the corresponding Bethe-Salpeter kernel in Eq.~\eqref{eq:chi_from_L}. 
The nonadiabatic self-energy is therefore the dynamical variation
relative to the static limit,
\begin{align}
\Pi^{\textrm{NA,R}}_{I,\alpha;J,\beta}(z)
&=
\Pi^{\textrm{R}}_{I,\alpha;J,\beta}(z)
-
\Pi^{\textrm{R}}_{I,\alpha;J,\beta}(0).
\label{eq:Pi_NA_dynamic_minus_static_atomic}
\end{align}
The static phonon self-energy therefore reduces to the translationally invariant BO Hessian \(C\), preserving the acoustic sum rule. Consequently,
\begin{align}
\textrm{Im}
\left[
\Pi^{\textrm{NA,R,SX}}_{I,\alpha;J,\beta}(z)
\right]
=
\textrm{Im}
\left[
\Pi^{\textrm{R,SX}}_{I,\alpha;J,\beta}(z)
\right].
\label{eq:ImPi}
\end{align}
Thus, the phonon linewidth can be extracted directly from the
imaginary part of the retarded self-energy
\(\Pi^{\textrm{R,SX}}\).

\begin{figure}[htbp]
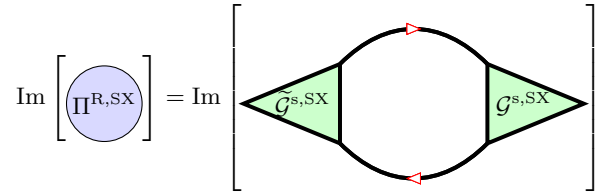

  \centering
  \[
  \begin{aligned}
  \textrm{Im}
  \left[
  \PiBubbleR
  \right]
  =
  \textrm{Im}
  \left[
  \GcalTildeRLzeroRGcalRTerm
  \right]
  \end{aligned}
  \]
  \caption{
  Imaginary part of the retarded SX phonon self-energy entering
  the linewidth, Eq.~\eqref{eq:ImPi}.
  }
  \label{fig:ImPi}
\end{figure}

We now connect this real-space formulation to the linewidth of a phonon in a periodic solid--- the molecular case can be recovered simply by setting the reciprocal wavevectors to zero in the following expressions. We denote a phonon mode by \((\mu,\mathbf q)\), where \(\mu\) is the branch index and \(\mathbf q\) its crystal momentum. The atomic index is decomposed as \(I=(s,\mathbf T)\), where \(s\) labels an atom in the unit cell and \(\mathbf T\) a Bravais-lattice vector. Electronic states are denoted by $N=(n,\mathbf k)$, $
M=(m,\mathbf k+\mathbf q)$. The Fourier and Bloch-transform conventions are provided in the Supplemental Material \cite{SI}.

In the quasi-particle approximation, the linewidth of mode \((\mu,\mathbf q)\) is defined by
\begin{align}
\Gamma^{\textrm{ph,SX}}_{\mu\mathbf q}
&\equiv
-\frac{2}{\hbar}
\lim_{\eta\to0^+}
\textrm{Im}\,
\Pi^{\textrm{R,SX}}_{\mu\mathbf q}
\left(
\omega^{\textrm{SX}}_{\mu\mathbf q}
+i\eta
\right).
\label{eq:linewidth_def}
\end{align}
Here,
\(\omega^{\textrm{SX}}_{\mu\mathbf q}\) is the (real part of) on-shell phonon frequency obtained consistently from the SX phonon Dyson equation, Eq.~\eqref{eq:DDyson}. To express the interaction in the same basis, we first project the nonlocal vertex onto the phonon mode \((\mu,\mathbf q)\),
\begin{align}
\mathcal G_{\mu\mathbf q}^{\textrm{s,SX}}
(\mathbf r_1,\mathbf r_2;z)
&\equiv
\sum_{s\alpha\mathbf T}
e^{i\mathbf q\cdot
\mathbf R^{0,\textrm{BO}}_{s\mathbf T}}
e^{s\alpha}_{\mu\mathbf q}\sqrt{\frac{\hbar
}{
2M_s
\omega^{\textrm{SX}}_{\mu\mathbf q}
}}
\nonumber\\
&\quad\times
\mathcal G^{\textrm{s,SX}}
(\mathbf r_1,\mathbf r_2;
 s\alpha\mathbf T;z).
\label{eq:Gcal_mode_projected}
\end{align}
The mode-projected vertex remains a nonlocal one-particle operator in the electronic Hilbert space. Its SX matrix elements are
\begin{align}
\left[
\mathcal G^{\textrm{s,SX}}_{\mu}
\right]^M_N(\mathbf q,z)
&\equiv
\int_{NV}
\frac{d\mathbf r_1}{NV}
\frac{d\mathbf r_2}{NV}\,
\left[
\psi_M^{\textrm{SX}}(\mathbf r_1)
\right]^*
\nonumber\\
&\quad\times
\mathcal G_{\mu\mathbf q}^{\textrm{s,SX}}
(\mathbf r_1,\mathbf r_2;z)
\psi_N^{\textrm{SX}}(\mathbf r_2),
\label{eq:Gcal_matrix_IN}
\\
\left[
\mathcal G^{\textrm{s,SX}}_{\mu}
\right]^N_M(-\mathbf q,z)
&\equiv
\int_{NV}
\frac{d\mathbf r_1}{NV}
\frac{d\mathbf r_2}{NV}\,
\left[
\psi_N^{\textrm{SX}}(\mathbf r_1)
\right]^*
\nonumber\\
&\quad\times
\mathcal G_{\mu,-\mathbf q}^{\textrm{s,SX}}
(\mathbf r_1,\mathbf r_2;z)
\psi_M^{\textrm{SX}}(\mathbf r_2).
\label{eq:Gcal_matrix_NI}
\end{align}

The generalized Fermi golden rule of derived in Ref.~\onlinecite{PhysRevB.111.075137} for the screen*-screen formulation, here reduces to
\begin{align}
\Gamma^{\textrm{ph,SX}}_{\mu\mathbf q}
&=
\frac{2\pi}{\hbar}
\sum_{N,M}
\left(
f_N^{\textrm{SX}}
-
f_M^{\textrm{SX}}
\right)
\left|
\left[
\mathcal G^{\textrm{s,SX}}_{\mu}
\right]^M_N
\left(
\mathbf q,
\omega_{\mu\mathbf q}^{\textrm{SX}}
\right)
\right|^2
\nonumber\\
&\quad\times
\delta
\left(
\varepsilon_M^{\textrm{SX}}
-
\varepsilon_N^{\textrm{SX}}
-
\hbar
\omega_{\mu\mathbf q}^{\textrm{SX}}
\right).
\label{eq:linewidth_FGR}
\end{align}
As for
the standard Fermi golden rule, this requires the electron-hole states to form a
continuum at $\omega^{\textrm{SX}}_{\mu\mathbf q}$: for undamped collective modes
such as bound excitons the matrix elements of $\mathcal G^{\textrm{s,SX}}$ diverge
as $\eta\to0^{+}$, and Eq.~\eqref{eq:Pi_R_SX_Gcal_nind_atomic} must be used directly. Equivalently, resolving the composite electronic indices,
\begin{align}
\Gamma^{\textrm{ph,SX}}_{\mu\mathbf q}
&=
\frac{2\pi}{\hbar}
\sum_{nm\mathbf k}
\left(
f_{n\mathbf k}^{\textrm{SX}}
-
f_{m\mathbf k+\mathbf q}^{\textrm{SX}}
\right) \left|
\left[
\mathcal G^{\textrm{s,SX}}_{\mu}
\right]^{
m\mathbf k+\mathbf q
}_{
n\mathbf k
}
\left(
\mathbf q,
\omega_{\mu\mathbf q}^{\textrm{SX}}
\right)
\right|^2
\nonumber\\
&\quad\times
\delta
\left(
\varepsilon_{m\mathbf k+\mathbf q}^{\textrm{SX}}
-
\varepsilon_{n\mathbf k}^{\textrm{SX}}
-
\hbar
\omega_{\mu\mathbf q}^{\textrm{SX}}
\right).
\label{eq:linewidth_FGR_explicit}
\end{align}

The occupation difference selects the net decay of the phonon into an electron-hole excitation, while the delta function enforces energy conservation. All screening and ladder corrections are contained in the single transition amplitude \(\mathcal G^{\textrm{s,SX}}\). The special case when $\mathcal{G}^{\textrm{s,SX}}$ is approximated to its static value reads
\begin{align}
\Gamma^{\textrm{ph,SX}}_{\mu\mathbf q}
&\simeq
\frac{2\pi}{\hbar}
\sum_{nm\mathbf k}
\left(
f_{n\mathbf k}^{\textrm{SX}}
-
f_{m\mathbf k+\mathbf q}^{\textrm{SX}}
\right) \left|
\left[
\mathcal G^{\textrm{s,SX}}_{\mu}
\right]^{
m\mathbf k+\mathbf q
}_{
n\mathbf k
}
\left(\mathbf q,0\right)
\right|^2
\nonumber\\
&\quad\times
\delta
\left(
\varepsilon_{m\mathbf k+\mathbf q}^{\textrm{SX}}
-
\varepsilon_{n\mathbf k}^{\textrm{SX}}
-
\hbar
\omega_{\mu\mathbf q}^{\textrm{SX}}
\right).
\label{eq:linewidth_FGR_explicit_static}
\end{align}

The next subsection shows that the same amplitude governs the reciprocal process in the electronic
linewidth.

\subsection{Electron-phonon contribution to the electronic self-energy and Fermi golden rule for electrons}

In the phonon self-energy, an electronic excitation propagates between two electron-phonon vertices. In the electronic self-energy, the internal propagators instead describe how an electron changes its state by emitting or absorbing a phonon. Since both self-energies originate from the same functional construction, the interaction amplitude is expected to be the same if no dissipation happens in the vertices.

Specializing Eq.~\eqref{eq:Sigma_eph_W_Gcal} to the SX
approximation gives
\begin{align}
\Sigma^{\textrm{el-ph,SX}}(1,2)
&=
i\hbar\,
\widetilde{\mathcal G}^{\textrm{s,SX}}
(1^+,\bar 1;\bar a)
D^{\textrm{SX}}(\bar a,\bar b)
\nonumber\\
&\quad\times
G^{\textrm{SX}}(\bar 1,\bar 2)
\mathcal G^{\textrm{s,SX}}
(\bar 2,2;\bar b).
\label{eq:Sigma_eph_W_Gcal-SX}
\end{align}
The electronic propagator is therefore connected to the phonon propagator by the same left- and right-acting vertices introduced in the previous subsection. Moreover, \(D^{\textrm{SX}}\) is the phonon propagator obtained from the Dyson equation containing \(\Pi^{\textrm{SX}}\). 

\begin{figure}[htbp]
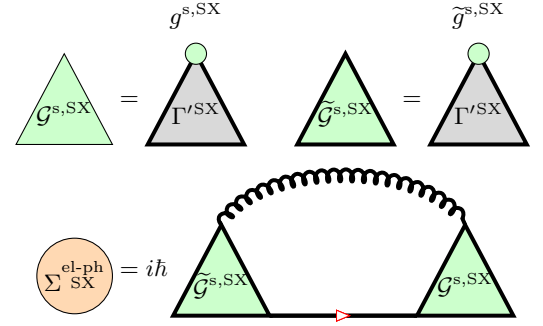

  \centering
  \[
  \begin{aligned}
  \GcalSVertexVertical
  &=
  \GammaPrimeGsTermVertical
  \qquad
  \GcalTildeSVertexVertical
  =
  \GtildeSGammaPrimeTermVertical
  \\
  \fSigmaelphSX
  &=
  i\hbar\,
  \SigmaElphRHS
  \end{aligned}
  \]
  \caption{
  SX electron-phonon self-energy, Eq.~\eqref{eq:Sigma_eph_W_Gcal-SX}.
  The internal electron and phonon propagators are connected to the vertices \(\widetilde{\mathcal G}^{\textrm{s,SX}}\) and \(\mathcal G^{\textrm{s,SX}}\).
  }
  \label{fig:Sigma-elph}
\end{figure}

We evaluate the self-energy in the basis of the SX one-particle states introduced for the phonon self-energy, where off-diagonal mixing between electronic states in $G$ is absent. A convenient route to its retarded form is to first write the convolution in Matsubara space,
\begin{align}
\Sigma^{\textrm{el-ph,SX}}_{N}(i\omega_n)
&=
-\frac{1}{\beta}
\sum_{i\nu_m}
\sum_{M,\mu\mathbf q}
\left[
\widetilde{\mathcal G}^{\textrm{s,SX}}_{\mu}
\right]^{M}_{N}
(\mathbf q,-i\nu_m)
\nonumber\\
&\quad\times
D^{\textrm{SX}}_{\mu\mathbf q}(i\nu_m)
G^{\textrm{SX}}_{M}
(i\omega_n-i\nu_m)
\nonumber\\
&\quad\times
\left[
\mathcal G^{\textrm{s,SX}}_{\mu}
\right]^{N}_{M}
(-\mathbf q,i\nu_m).
\label{eq:Sigma_eph_Matsubara_Gcal}
\end{align}
The bosonic frequency \(i\nu_m\) is transferred through the phonon line and through both dressed vertices. In the quasi-particle approximation, the electronic linewidth is obtained from the diagonal, on-shell matrix element of the retarded self-energy, obtained from analytic continuation:
\begin{align}
\Gamma^{\textrm{el,SX}}_{N}
&\equiv
-\frac{2}{\hbar}\lim_{\eta \rightarrow 0}
\operatorname{Im}
\Sigma^{\textrm{el-ph,SX,R}}_{N}
\left(
\frac{\varepsilon_N^{\textrm{SX}}}{\hbar}+i\eta
\right).
\label{eq:electron_linewidth_def}
\end{align}
In the above expression, in the argument of the self-energy we have neglected the self-consistent real part renormalization of the electronic energies due to the electron-phonon interaction, consistently with previous sections. Equation~\eqref{eq:Sigma_eph_Matsubara_Gcal} still contains the complete frequency dependence of the electron-phonon vertices and of the internal propagators. A Fermi-golden-rule form follows only under three additional conditions: both electrons and phonons must be well-defined quasiparticles, the dressed electron-phonon vertices must be approximated by their static values i.e. no dissipation happens in the interaction vertices, and they not contain excitonic bound states that invalidate the description in terms of independent asymptotic electron-hole transitions. In this static limit, the left- and right-acting matrix elements
satisfy (see App. \ref{app:reciprocity})

\begin{align}
\left[
\mathcal G^{\textrm{s,SX}}_{\mu}
\right]^{N}_{M}
(-\mathbf q,0)
&=
\left(
\left[
\mathcal G^{\textrm{s,SX}}_{\mu}
\right]^{M}_{N}
(\mathbf q,0)
\right)^*,\label{eq:GNMmqGMNq}\\
\left[
\widetilde{\mathcal G}^{\textrm{s,SX}}_{\mu}
\right]^{M}_{N}
(\mathbf q,0)&=
\left[
\mathcal G^{\textrm{s,SX}}_{\mu}
\right]^{M}_{N}
(\mathbf q,0),
\label{eq:Gcal_static_conjugation}
\end{align}
where the last equation follows trivially from Eq. \eqref{eq:tildeGeqG}. The two vertices in the self-energy therefore combine into the positive transition probability \(\left|
\mathcal G^{\textrm{s,SX}}_{\mu} \right|^2\). Performing the Matsubara sum, analytically continuing to the retarded axis, and resolving the electronic indices as
\(N=(n,\mathbf k)\) and
\(M=(m,\mathbf k+\mathbf q)\), we obtain
\begin{align}
&\Gamma^{\textrm{el,SX}}_{n\mathbf k}
\simeq
\frac{2\pi}{\hbar}
\sum_{m,\mu\mathbf q}
\left|
\left[
\mathcal G^{\textrm{s,SX}}_{\mu}
\right]^{
m\mathbf k+\mathbf q
}_{
n\mathbf k
}
(\mathbf q,0)
\right|^2
\nonumber\\
&\quad\times
\Bigg[
\left(
N_{\mu\mathbf q}^{\textrm{SX}}
+1
-
f_{m\mathbf k+\mathbf q}^{\textrm{SX}}
\right)
\delta
\left(
\varepsilon_{n\mathbf k}^{\textrm{SX}}
-
\varepsilon_{m\mathbf k+\mathbf q}^{\textrm{SX}}
-
\hbar\omega_{\mu\mathbf q}^{\textrm{SX}}
\right)
\nonumber\\
&\qquad+
\left(
N_{\mu\mathbf q}^{\textrm{SX}}
+
f_{m\mathbf k+\mathbf q}^{\textrm{SX}}
\right)
\delta
\left(
\varepsilon_{n\mathbf k}^{\textrm{SX}}
-
\varepsilon_{m\mathbf k+\mathbf q}^{\textrm{SX}}
+
\hbar\omega_{\mu\mathbf q}^{\textrm{SX}}
\right)
\Bigg].
\label{eq:electron_linewidth_FGR_SX}
\end{align}
Here,
\(N_{\mu\mathbf q}^{\textrm{SX}}\) is the Bose--Einstein occupation of the phonon mode. The first contribution describes phonon emission, whereas the second describes phonon absorption.

\begin{figure}[htbp]
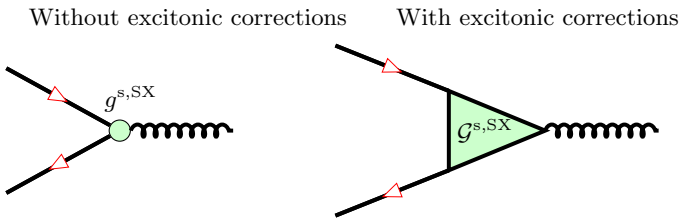

  \centering
  \[
  \begin{aligned}
  &\quad
  \textrm{Without excitonic corrections}
  \qquad
  \textrm{With excitonic corrections}
  \\
  &\FermionPhononGsVertex
  \qquad\qquad
  \FermionPhononGcalVertex
  \end{aligned}
  \]
  \caption{
  Elementary electron-phonon scattering process without and with
  excitonic vertex corrections. The phonon can be either emitted
  or absorbed. Both reciprocal processes
  are governed by the same dressed matrix element
  \(\mathcal G^{\textrm{s,SX}}\) that entails excitonic corrections.
  }
  \label{fig:epcscatt}
\end{figure}

Equations~\eqref{eq:linewidth_FGR_explicit_static} and
\eqref{eq:electron_linewidth_FGR_SX} expose the central result of
the construction. Once the static-vertex approximation is applied
consistently, the phononic and electronic linewidths contain the
same transition amplitude,
\begin{align}
\left|
\left[
\mathcal G^{\textrm{s,SX}}_{\mu}
\right]^{
m\mathbf k+\mathbf q
}_{
n\mathbf k
}
(\mathbf q,0)
\right|^2.
\end{align}
The two expressions differ only in which excitation is regarded as the decaying quasiparticle and in the corresponding equilibrium occupation factors. The detailed-balance relation can be seen explicitly by considering two electronic states satisfying
\begin{align}
\varepsilon_{m\mathbf k+\mathbf q}^{\textrm{SX}}
-
\varepsilon_{n\mathbf k}^{\textrm{SX}}
=
\hbar\omega_{\mu\mathbf q}^{\textrm{SX}}.
\end{align}
At thermal equilibrium, their Fermi and Bose occupations obey
\begin{align}
f_{n\mathbf k}^{\textrm{SX}}
\left(
1-f_{m\mathbf k+\mathbf q}^{\textrm{SX}}
\right)
N_{\mu\mathbf q}^{\textrm{SX}}
=
f_{m\mathbf k+\mathbf q}^{\textrm{SX}}
\left(
1-f_{n\mathbf k}^{\textrm{SX}}
\right)
\left(
N_{\mu\mathbf q}^{\textrm{SX}}+1
\right).
\label{eq:detailed_balance_ep}
\end{align}
The rate for phonon absorption is therefore exactly balanced by
the rate of the inverse emission process. Because the same dressed
matrix element multiplies both rates, the inclusion of excitonic
vertex corrections preserves this microscopic balance rather than
modifying the electronic and phononic processes independently, as graphically depicted in Fig. \ref{fig:epcscatt}.

\providecommand{\SmallDiagram}[1]{\scalebox{0.8}{#1}}

\newcommand{\TableRowSpace}{4.0em}
\newcommand{\TableRowSpaceA}{0.4em}
\newcommand{\TableRowSpaceB}{5.5em}

\begin{table*}[htbp]
\caption{
Summary of the diagrammatic structure leading to phonon and electron linewidths in the SX approximation.
}
\label{tab:linewidth-summary}

\centering
\scriptsize
\renewcommand{\arraystretch}{1.65}
\setlength{\tabcolsep}{3pt}

\begin{tabular}{ccc}
\hline\hline
Object & Electrons & Phonons \\

\hline

\\[\TableRowSpaceA]

Free Hamiltonian
&
\begin{tabular}{@{}c@{}}
$\displaystyle
h_0(\mathbf{x}_1)
=
-\frac{\nabla^2}{2}
+
V^{\textrm{BO}}(\mathbf{r}_1)
$
\\[0.8em]
$\displaystyle
G_0^{-1}(1,2)
=
\left[
i\hbar\partial_{t_1}
-
h_0(\mathbf x_1)
\right]
\delta(1,2)
$
\end{tabular}
&
\begin{tabular}{@{}c@{}}
$\displaystyle
H_{0,\mathrm{ph}}
=
\sum_I
\frac{\mathbf P_I^2}{2M_I}
+
\frac12
\sum_{I\alpha,J\beta}
U_{I\alpha}
C_{I\alpha,J\beta}
U_{J\beta}
$
\\[0.8em]
$\displaystyle
D_0^{-1}(a,b)
=
-
\left[
M_I\delta_{IJ}\delta_{\alpha\beta}
\partial_{t_a}^2
+
C_{I\alpha,J\beta}
\right]
\delta(t_a-t_b)
$
\end{tabular}
\\[\TableRowSpace]

\hline

\\[\TableRowSpaceA]

Green's functions Dyson equations
&
\begin{tabular}{@{}c@{}}
\SmallDiagram{\(\displaystyle
\G
=
\Gzero
+
\GzeroSigmaG
\)}\\[1.5em]
\SmallDiagram{\(\displaystyle
\fSigma =\fSigmael+\fSigmaelph 
\)
}
\end{tabular}

&
\begin{tabular}{@{}c@{}}
\SmallDiagram{\(\displaystyle
\Dphonon
=
\Dzero
+
\DzeroPiD
\)}
\\[1.5em]
\SmallDiagram{\(\displaystyle
\PiNABubble
=
\PiBubble
+
\BBubble
\)}
\end{tabular}
\\[\TableRowSpace]

\hline

\\[\TableRowSpaceA]

$\Gamma'$ Dyson equation
&
\begin{tabular}{@{}c@{}}
\SmallDiagram{\(\displaystyle
\GammaPrime
=
\DoubleDeltaGammaV
+
i\hbar  \, \WelGGGammaPrime
\)}
\end{tabular}
&
\begin{tabular}{@{}c@{}}
\SmallDiagram{\(\displaystyle
\GammaPrimeHorizontal
=
\DoubleDeltaGammaH
+
 i\hbar  \, \WelVerticalGGGammaPrime
\)}
\end{tabular}
\\[\TableRowSpace]

\hline

\\[\TableRowSpaceA]

Electron-phonon vertex $\mathcal G$
&
\begin{tabular}{@{}c@{}}
\SmallDiagram{\(\displaystyle
\gsVertex
=
\gVertex
+
\WelPelgTerm
\)}
\\[1.6em]
\SmallDiagram{\(\displaystyle
\GcalSVertexVertical
=
\GammaPrimeGsTermVertical
\)}
\end{tabular}
&
\begin{tabular}{@{}c@{}}
\SmallDiagram{\(\displaystyle
\gsVertex
=
\gVertex
+
\WelPelgTerm
\)}
\\[3.8em]
\SmallDiagram{\(\displaystyle
\GcalSVertex
=
\GammaPrimeGsTerm
\)}
\end{tabular}
\\[\TableRowSpaceB]

\hline

\\[\TableRowSpaceA]

Self-energy
&
\begin{tabular}{@{}c@{}}
\SmallDiagram{\(\displaystyle
\fSigmaelphSX
=
i\hbar\,
\SigmaElphRHS
\)}
\end{tabular}
&
\begin{tabular}{@{}c@{}}
\SmallDiagram{\(\displaystyle
  \PiBubbleSX
  =
  -i\hbar
  \gLzeroGcalSTerm
  = -i\hbar \,
  \GcalTildeSLzeroGTerm
\)}
\end{tabular}
\\[\TableRowSpace]

\hline

\\[\TableRowSpaceA]

Linewidth definition (SX approximation)
&
\begin{tabular}{@{}c@{}}
$\displaystyle
\Gamma^{\mathrm{el,SX}}_{n\mathbf k}
=
-\frac{2}{\hbar}
\mathrm{Im}\,
\Sigma^{\mathrm{el-ph,SX,R}}_{n\mathbf k}
\left(
\epsilon^{\mathrm{SX}}_{n\mathbf k}
\right)
$
\end{tabular}
&
\begin{tabular}{@{}c@{}}
$\displaystyle
\Gamma^{\mathrm{ph,SX}}_{\mu\mathbf q}
=
-\frac{1}{\hbar}
2\,
\mathrm{Im}\,
\Pi^{\mathrm{SX,R}}_{\mu\mathbf q}
\left(
\omega^{\mathrm{SX}}_{\mu\mathbf q}\right)
$
\end{tabular}
\\[\TableRowSpace]

 & & \begin{tabular}{@{}c@{}}
 \SmallDiagram{\(\displaystyle
\mathrm{Im}\!\left[
\PiBubbleR
\right]
=
\mathrm{Im}\!\left[
\GcalTildeRLzeroRGcalRTerm
\right]
\)}
\end{tabular}
\\[\TableRowSpace]

\hline

\\[\TableRowSpaceA]

Fermi golden-rule linewidth \\ [-2.0em]
(SX approximation with static vertices)
&
\begin{tabular}{@{}c@{}}
$\displaystyle
\Gamma^{\mathrm{el,SX}}_{n\mathbf k}
\simeq
\frac{2\pi}{\hbar}
\sum_{m,\mu\mathbf q}
\left|
\left[
\mathcal G^{s,\mathrm{SX}}_\mu
\right]^{m,\mathbf k+\mathbf q}_{n,\mathbf k}
(\mathbf q,0)
\right|^2
$
\\[0.8em]
$\displaystyle
\times
\Big[
\left(
N^{\mathrm{SX}}_{\mu\mathbf q}
+
1
-
f^{\mathrm{SX}}_{m,\mathbf k+\mathbf q}
\right)
\delta
\left(
\epsilon^{\mathrm{SX}}_{n\mathbf k}
-
\epsilon^{\mathrm{SX}}_{m,\mathbf k+\mathbf q}
-
\hbar\omega^{\mathrm{SX}}_{\mu\mathbf q}
\right)
$
\\[0.8em]
$\displaystyle
+
\left(
N^{\mathrm{SX}}_{\mu\mathbf q}
+
f^{\mathrm{SX}}_{m,\mathbf k+\mathbf q}
\right)
\delta
\left(
\epsilon^{\mathrm{SX}}_{n\mathbf k}
-
\epsilon^{\mathrm{SX}}_{m,\mathbf k+\mathbf q}
+
\hbar\omega^{\mathrm{SX}}_{\mu\mathbf q}
\right)
\Big]
$
\end{tabular}
&
\begin{tabular}{@{}c@{}}
$\displaystyle
\Gamma^{\mathrm{ph,SX}}_{\mu\mathbf q}
\simeq
\frac{2\pi}{\hbar}
\sum_{n m\mathbf k}
\left(
f^{\mathrm{SX}}_{n\mathbf k}
-
f^{\mathrm{SX}}_{m,\mathbf k+\mathbf q}
\right)
$
\\[0.8em]
$\displaystyle
\times
\left|
\left[
\mathcal G^{s,\mathrm{SX}}_\mu
\right]^{m,\mathbf k+\mathbf q}_{n,\mathbf k}
\left(\mathbf q,0\right)
\right|^2
$
\\[0.8em]
$\displaystyle
\times
\delta
\left(
\epsilon^{\mathrm{SX}}_{m,\mathbf k+\mathbf q}
-
\epsilon^{\mathrm{SX}}_{n\mathbf k}
-
\hbar\omega^{\mathrm{SX}}_{\mu\mathbf q}
\right)
$
\end{tabular}
\\[\TableRowSpace]

\hline\hline

Object
&
$\displaystyle W^{\textrm{SX}}_{\mathrm{el}}$
&
$\displaystyle P^{\textrm{SX}}_{\mathrm{el}}$
\\[1.0em]

\hline 

\\[\TableRowSpaceA]

Screened Coulomb and polarization
&
\begin{tabular}{@{}c@{}}
$\displaystyle
W_{\textrm{el}}^{\textrm{SX}}(1,2)
=
W_{\textrm{el}}^{\textrm{SX}}(\mathbf r_1,\mathbf r_2)
\delta(t_1-t_2)
$
\end{tabular}
&
\begin{tabular}{@{}c@{}}
\SmallDiagram{\(\displaystyle
\PelBubble
=
\PolBubble
+
\PolBubbleOneW
+
\PolBubbleTwoW
+
\ldots
\)}
\end{tabular}
\\[\TableRowSpace]

\hline\hline
\end{tabular}

\end{table*}

\section{Application to graphene: electronic and phononic linewidths}
\label{sec:graphene}

We now apply the formalism to freestanding graphene and evaluate the electronic and phononic linewidths including electron-hole ladder corrections. The calculation requires three many-body ingredients: the quasiparticle energies \(\varepsilon^{\textrm{SX}}\), the phonon frequencies \(\omega^{\textrm{SX}}\), and the excitonically dressed electron-phonon vertex \(\mathcal G^{\textrm{s,SX}}\). Splitting the effect of MBPT on all these ingredients is physically relevant because their renormalizations partially cancel out. The SX electronic dispersion reduces the phase space available for scattering, whereas the Bethe-Salpeter ladder strongly enhances the scattering amplitude. The resulting linewidth cannot therefore be inferred from the renormalization of either the bands or the vertex alone.

Our computational strategy follows Refs.~\onlinecite{y1dn-m6pc,PhysRevB.111.075137,PhysRevB.111.075118}. Here we describe the physical construction and the resulting hierarchy of approximations; numerical parameters and convergence tests are reported in the Supplemental Material~\cite{SI}.

\subsection{Quasiparticle dispersions and reduction of scattering phase space}

The graphene band structure, and the effect of many body corrections, has been extensively studied \cite{Bostwick2007,PhysRevLett.101.226405,doi:10.1073/pnas.1100242108,
Elias2011,PhysRevLett.109.116802,PhysRevB.87.045425,
PhysRevB.109.075120}. We here describe the low-energy \(\pi/\pi^\ast\) electronic structure of graphene using a tight-binding (TB) model that accurately reproduces the DFT bands and response functions. Within this model, the SX Hamiltonian of Eq.~\eqref{eq:hSX_def} is solved self-consistently. We start from the screened Coulomb interaction evaluated in the random-phase approximation using DFT energies, solve the electronic problem with the corresponding screened-exchange self-energy, and update the screening using the resulting quasiparticle energies. The procedure is iterated until convergence of \(h^{\textrm{SX}}\) and \(\varepsilon^{\textrm{SX}}\).

Computing \(W_{\textrm{el}}\) from the self-consistent SX energies corresponds to the diagrammatic construction of Fig.~\ref{fig:dyson-Pel}, with the irreducible polarization in Fig.~\ref{fig:polarization-bubble-expansion} truncated at the independent electron-hole bubble. We verified that adding the next-order diagrams changes the screened interaction by only a few percents and does not alter the hierarchy of linewidths discussed below. We also disregard self-energy insertions of \(\Sigma^{\textrm{el-ph}}\) on the electronic Green's function and therefore determine the electronic quasiparticles at clamped nuclei.

\begin{figure*}[t]
    \centering
        \centering
        \includegraphics[width=\linewidth]{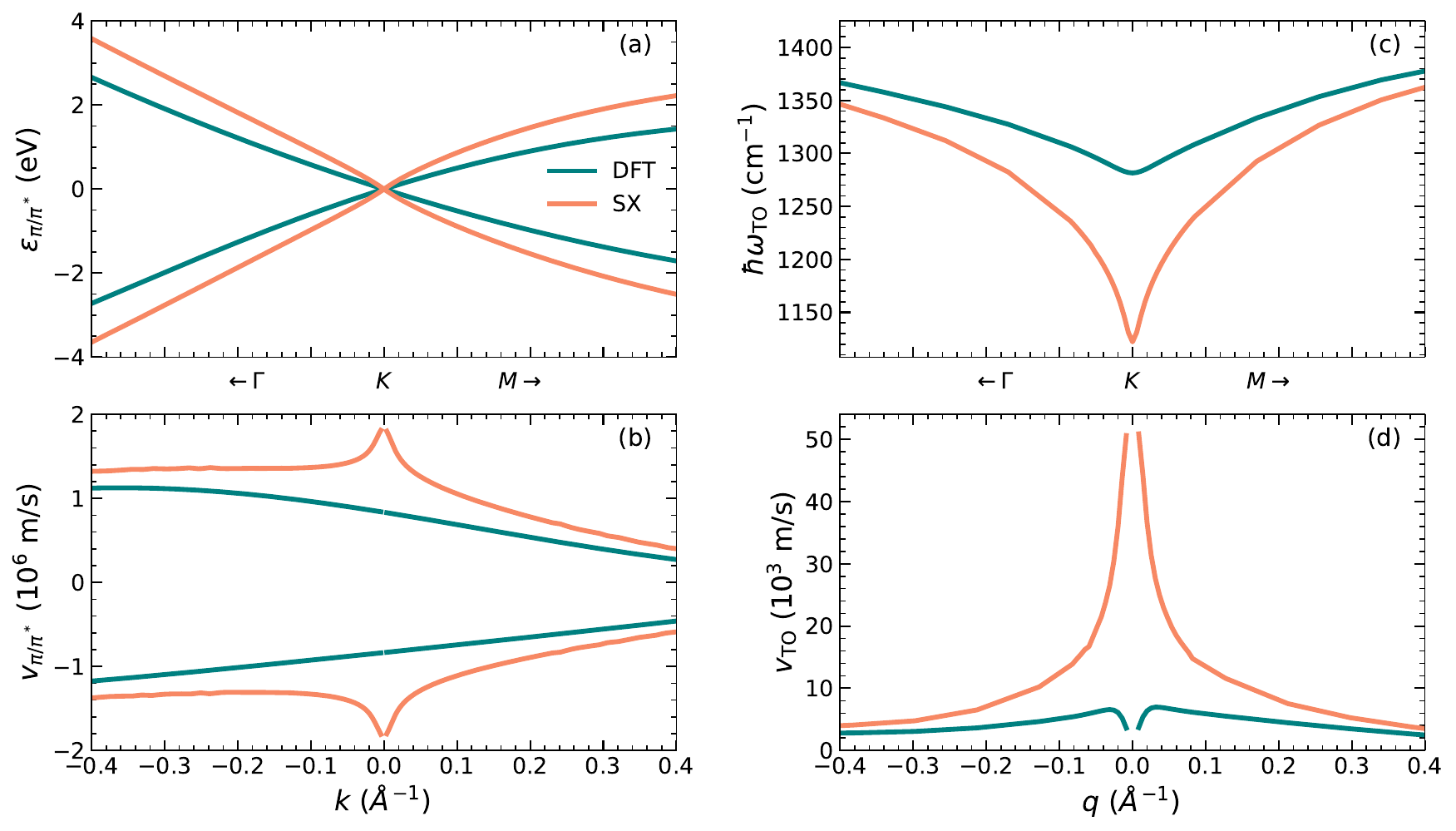}
    \caption{
    Many-body renormalization of the electronic and phononic dispersions of graphene. $k$ and $q$ points magnitudes are referred to the $K$ point. (a-b) Electronic \(\pi/\pi^\ast\) bands and group velocities along selected high-symmetry directions at the DFT and SX levels. The SX dispersion is steeper close to the Dirac point, reducing the electronic density of states and the scattering phase space. (c-d) Static TO-phonon dispersion and group velocity near the \(K\) point, calculated without and with electron-hole ladder corrections. The correlated response deepens the Kohn anomaly, redshifts the TO frequency near \(K\), and strongly enhances its group velocity.
    }
    \label{fig:bandvel}
\end{figure*}

The resulting electronic bands and group velocities are shown in Fig.~\ref{fig:bandvel} (a) and (b). Screened exchange steepens the Dirac cone and produces the characteristic logarithmic increase of the quasiparticle velocity close to the Fermi level \cite{doi:10.1073/pnas.1100242108}. The corresponding density of states per unit energy is reduced. This band renormalization therefore suppresses both kinds of scattering considered here: it decreases the number of final electronic states available to an electron emitting or absorbing a phonon and, at fixed phonon energy, reduces the joint density of electron-hole states into which a phonon can decay.

The phonon dispersion of graphene is well known at the DFT level \cite{PhysRevB.67.035401,WirtzRubio2004,PhysRevLett.92.075501,PhysRevLett.93.185503,PhysRevB.71.205214,PhysRevLett.97.266407,PhysRevB.78.165421}, and the effects of many-body corrections have been recently studied in Ref. \cite{y1dn-m6pc}, which we follow here. We compute the real part of the phonon spectrum from the static force constants of Eq.~\eqref{eq:Cdef} at the SX level. The adiabatic approximation is applied only to the real part of the phonon frequency and to the electron-phonon vertex. The phonon linewidth remains an intrinsically dynamical quantity, obtained from the imaginary part of the retarded self-energy evaluated at the static phonon frequency. It is therefore nonzero whenever this finite frequency lies within the electron-hole continuum, so that decay into an electron-hole pair is allowed; in gapless graphene this condition can be satisfied even at charge neutrality by interband transitions. For the present purpose, the phonon frequency primarily determines the energy-conservation surface and the thresholds for phonon emission and absorption. Its precise approximation has a smaller effect on the linewidth than the electronic phase space and the electron-phonon vertex. Nevertheless, electron-hole correlations strongly reshape the static TO branch near the \(K\) point, as shown in Fig.~\ref{fig:bandvel} (c) and (d). The many-body result deepens the zone-boundary Kohn anomaly, producing a redshift of about \(150~\textrm{cm}^{-1}\) near \(K\) and an approximately tenfold increase of the directional group velocity relative to DFT \cite{y1dn-m6pc}.

\subsection{Excitonic enhancement of the electron-phonon vertex}

We next determine the interaction matrix element entering the scattering rates. The statically screened coupling \(g^{\textrm{s}}\) is represented in the TB model through a real-space bond-stretching interaction fitted to DFPT results \cite{PhysRevB.90.125414,PhysRevB.84.035433,PhysRevB.110.L201405}. We denote this reference coupling by \(g^{\textrm{s,DFT}}\). In the present implementation, the change of the local vertex produced solely by replacing the DFT dielectric screening with the SX screening is neglected. Our tests indicate that this correction is subleading relative to the electron-hole ladder dressing that transforms \(g^{\textrm{s,DFT}}\) into the nonlocal vertex \(\mathcal G^{\textrm{s,SX}}\).

\begin{figure*}[t]
    \centering
    \includegraphics[width=\linewidth]
    {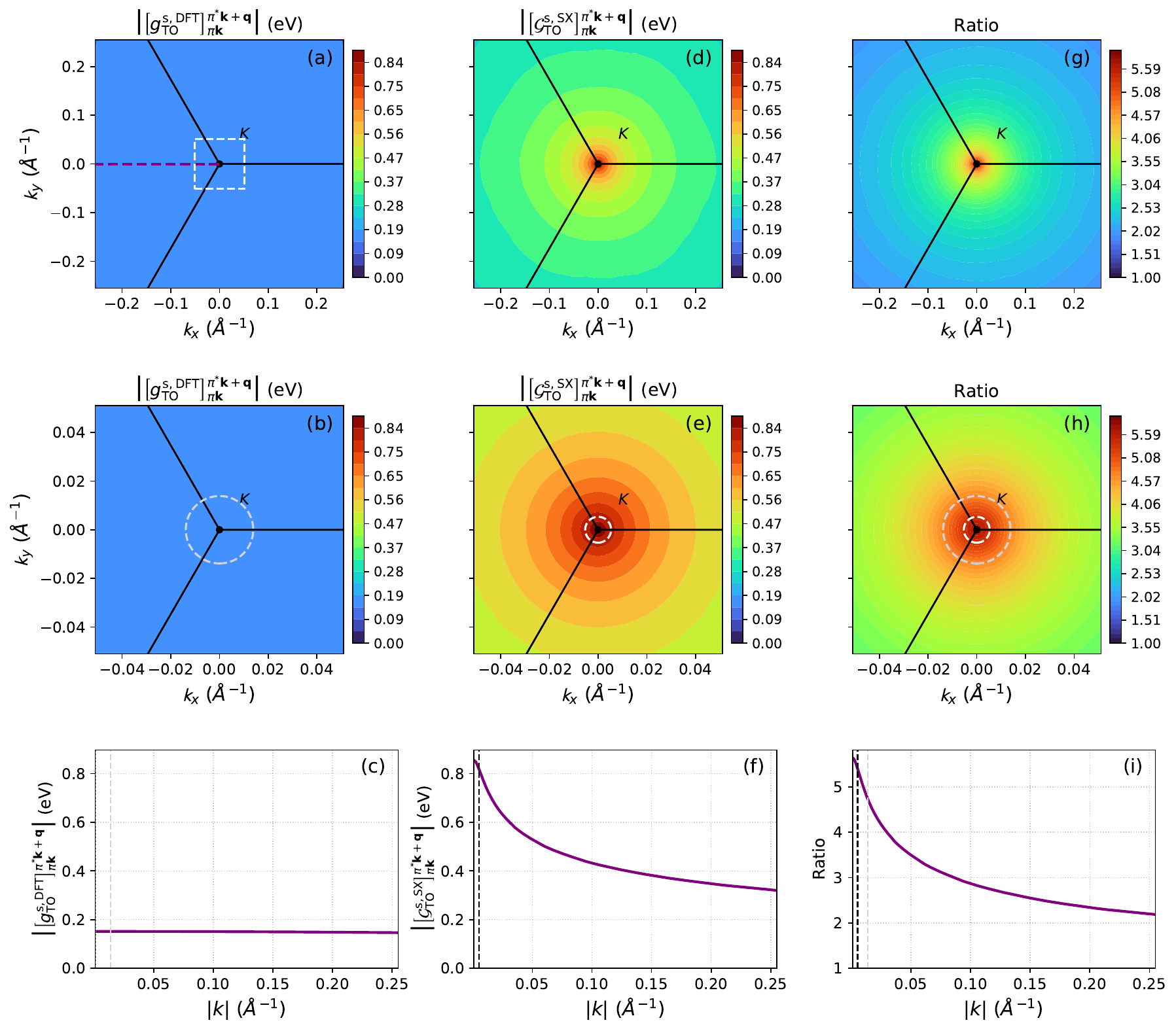}
    \caption{
    Excitonic enhancement of the intervalley electron-phonon vertex for the TO phonon at \(\mathbf q=\mathbf K\). The $k$ value has been shifted so to be zero at the Dirac point. The left (a-b-c) and central (d-e-f) columns show the interband matrix elements of the local DFPT coupling \(g^{\textrm{s,DFT}}\) and of the nonlocal dressed vertex \(\mathcal G^{\textrm{s,SX}}\), respectively; the right column (g-h-i) shows their ratio. The second row display the smaller momentum regions around \(K\) indicated by the white rectangular box of the upper left panel, while the last row shows the cut indicated in the upper left panel by a dashed purple line. The gray and white contours in the second row represent the energy-conserving electron-hole pairs obtained with the DFT and SX quasiparticle spectra, obtained from the equalities $\varepsilon^{\textrm{DFT/SX}}_{\pi^*\mathbf{k+q}}-\varepsilon^{\textrm{DFT/SX}}_{\pi\mathbf{k}}=\hbar\omega^{\textrm{DFT/SX}}_{\textrm{TO} \mathbf{K}}$ respectively, assuming isotropy for the band structure. The vertical markers in the bottom row delimit the SX energy-conserving contour. The dressed vertex is enhanced by approximately a factor six at \(K\) and by approximately \(5.5\) over the momentum region directly entering the linewidth. 
    }
    \label{fig:epc}
\end{figure*}

The dressed vertex \(\mathcal G^{\textrm{s,SX}}\) is obtained by solving the Sternheimer form of the Bethe-Salpeter equation within the TB model, following Ref.~\onlinecite{PhysRevB.111.075137}. This is equivalent to the vertex Dyson equation represented in Fig.~\ref{fig:Gcal-s-definition}. Starting from \(g^{\textrm{s,DFT}}\), we evaluate the potential induced through the exchange-correlation kernel,  propagate the corresponding electron-hole response with \(L_0^{\textrm{SX}}\), and iterate until \(\mathcal G^{\textrm{s,SX}}\) is converged. Fig. ~\ref{fig:epc} shows the resulting interband matrix elements for an intervalley phonon with \(\mathbf q=\mathbf K\) (corresponding to an electron scattering of the following type: $\pi\mathbf{k}\rightarrow \pi^*\mathbf{k+K}$). The DFPT vertex is nearly structureless on the momentum scale relevant to the decay process. The excitonic vertex instead grows rapidly as the electronic wave vector $\mathbf{k}$ approaches $\mathbf{K}$. Its magnitude is enhanced by a factor of approximately six at K and by approximately 5.5 on the energy-conserving contour sampled by the SX quasiparticles. The contours in Fig.~\ref{fig:epc} also show that the band and vertex corrections are not completely independent. Because the SX Dirac cone is steeper, energy conservation for a phonon of fixed energy selects electron-hole pairs closer to the valley center. This reduces the phase-space measure, but it simultaneously makes the scattering process sample the region where the vertex enhancement is largest. The final rate results from the competition between these two effects. In particular, an amplitude enhancement of about \(5\) corresponds to an enhancement of the transition probability of order \(25\), before the reduced SX phase space is included.

\subsection{Electronic and phononic linewidths}

Having determined the quasiparticle spectra and the corresponding vertex,
we evaluate the electronic linewidth from
Eq.~\eqref{eq:electron_linewidth_FGR_SX} and the phononic linewidth
from Eq.~\eqref{eq:linewidth_FGR_explicit_static}. Schematically, the
fully consistent SX result has the structure
\begin{align}
\Gamma^{\textrm{el/ph,SX}}
=
\Gamma^{\textrm{el/ph}}
\left[
\mathcal G^{\textrm{s,SX}},
\varepsilon^{\textrm{SX}},
\omega^{\textrm{SX}}
\right].
\end{align}
To isolate the effects of quasiparticle and vertex renormalization, we
compare three calculation levels:
\begin{align}
\Gamma^{\textrm{el/ph,SX+BSE}}
&=
\Gamma^{\textrm{el/ph}}
\left[
\mathcal G^{\textrm{s,SX}},
\varepsilon^{\textrm{SX}},
\omega^{\textrm{SX}}
\right],
\label{eq:FGRappr0}
\\
\Gamma^{\textrm{el/ph,SX+DFPT}}
&=
\Gamma^{\textrm{el/ph}}
\left[
g^{\textrm{s,DFT}},
\varepsilon^{\textrm{SX}},
\omega^{\textrm{DFT}}
\right],
\label{eq:FGRappr1}
\\
\Gamma^{\textrm{el/ph,DFT+DFPT}}
&=
\Gamma^{\textrm{el/ph}}
\left[
g^{\textrm{s,DFT}},
\varepsilon^{\textrm{DFT}},
\omega^{\textrm{DFT}}
\right].
\label{eq:FGRappr2}
\end{align}
In this notation, the first label specifies the electronic band-structure level, SX or DFT, whereas the second specifies the linear-response treatment of the electron-phonon vertex: BSE for the many-body dressed vertex \(\mathcal G^{\textrm{s,SX}}\) and DFPT for the vertex \(g^{\textrm{s,DFT}}\). The complete expressions are summarized in  Table~\ref{tab:linewidth-approximations}.

%

\begin{table*}[t]
\caption{
Electronic and phononic linewidths used in the comparison.
For each approximation, the electronic linewidth is shown first and
the phononic linewidth immediately below it. The SX+BSE expressions
coincide with Eqs.~\eqref{eq:electron_linewidth_FGR_SX} and
\eqref{eq:linewidth_FGR_explicit_static}.
}
\label{tab:linewidth-approximations}

\centering
\renewcommand{\arraystretch}{1.55}
\setlength{\tabcolsep}{4pt}

\resizebox{\textwidth}{!}{%
\begin{tabular}{c}
\hline\hline

SX+BSE
\\[1.0em]

$\displaystyle
\Gamma^{\textrm{el,SX+BSE}}_{n\mathbf k}
=
\frac{2\pi}{\hbar}
\sum_{m,\mu\mathbf q}
\left|
\left[
\mathcal G^{\mu,\textrm{SX}}
\right]^{m,\mathbf k+\mathbf q}_{n,\mathbf k}
(\mathbf q,0)
\right|^2
\left[
\left(
N^{\textrm{SX}}_{\mu\mathbf q}
+
1
-
f^{\textrm{SX}}_{m,\mathbf k+\mathbf q}
\right)
\delta\!\left(
\varepsilon^{\textrm{SX}}_{n\mathbf k}
-
\varepsilon^{\textrm{SX}}_{m,\mathbf k+\mathbf q}
-
\hbar\omega^{\textrm{SX}}_{\mu\mathbf q}
\right)
+
\left(
N^{\textrm{SX}}_{\mu\mathbf q}
+
f^{\textrm{SX}}_{m,\mathbf k+\mathbf q}
\right)
\delta\!\left(
\varepsilon^{\textrm{SX}}_{n\mathbf k}
-
\varepsilon^{\textrm{SX}}_{m,\mathbf k+\mathbf q}
+
\hbar\omega^{\textrm{SX}}_{\mu\mathbf q}
\right)
\right]
$
\\[1.3em]

$\displaystyle
\Gamma^{\textrm{ph,SX+BSE}}_{\nu\mathbf q}
=
\frac{2\pi}{\hbar}
\sum_{nm\mathbf k}
\left(
f^{\textrm{SX}}_{n\mathbf k}
-
f^{\textrm{SX}}_{m,\mathbf k+\mathbf q}
\right)
\left|
\left[
\mathcal G^{\nu,\textrm{R,SX}}
\right]^{m,\mathbf k+\mathbf q}_{n,\mathbf k}
(\mathbf q,0)
\right|^2
\delta\!\left(
\varepsilon^{\textrm{SX}}_{m,\mathbf k+\mathbf q}
-
\varepsilon^{\textrm{SX}}_{n\mathbf k}
-
\hbar\omega^{\textrm{SX}}_{\nu\mathbf q}
\right)
$
\\[1.5em]

\hline

SX+DFPT
\\[1.0em]

$\displaystyle
\Gamma^{\textrm{el,SX+DFPT}}_{n\mathbf k}
=
\frac{2\pi}{\hbar}
\sum_{m,\mu\mathbf q}
\left|
\left[
g^{\textrm{s,DFT}}_{\mu}
\right]^{m\mathbf k+\mathbf q}_{n\mathbf k}
(\mathbf q,0)
\right|^2
\left[
\left(
N_{\mu\mathbf q}^{\textrm{DFT}}
+
1
-
f_{m\mathbf k+\mathbf q}^{\textrm{SX}}
\right)
\delta\!\left(
\varepsilon_{n\mathbf k}^{\textrm{SX}}
-
\varepsilon_{m\mathbf k+\mathbf q}^{\textrm{SX}}
-
\hbar\omega_{\mu\mathbf q}^{\textrm{DFT}}
\right)
+
\left(
N_{\mu\mathbf q}^{\textrm{DFT}}
+
f_{m\mathbf k+\mathbf q}^{\textrm{SX}}
\right)
\delta\!\left(
\varepsilon_{n\mathbf k}^{\textrm{SX}}
-
\varepsilon_{m\mathbf k+\mathbf q}^{\textrm{SX}}
+
\hbar\omega_{\mu\mathbf q}^{\textrm{DFT}}
\right)
\right]
$
\\[1.3em]

$\displaystyle
\Gamma^{\textrm{ph,SX+DFPT}}_{\mu\mathbf q}
=
\frac{2\pi}{\hbar}
\sum_{nm\mathbf k}
\left(
f_{n\mathbf k}^{\textrm{SX}}
-
f_{m\mathbf k+\mathbf q}^{\textrm{SX}}
\right)
\left|
\left[
g^{\textrm{s,DFT}}_{\mu}
\right]^{m\mathbf k+\mathbf q}_{n\mathbf k}
(\mathbf q,0)
\right|^2
\delta\!\left(
\varepsilon_{m\mathbf k+\mathbf q}^{\textrm{SX}}
-
\varepsilon_{n\mathbf k}^{\textrm{SX}}
-
\hbar\omega_{\mu\mathbf q}^{\textrm{DFT}}
\right)
$
\\[1.5em]

\hline

DFT+DFPT
\\[1.0em]

$\displaystyle
\Gamma^{\textrm{el,DFT+DFPT}}_{n\mathbf k}
=
\frac{2\pi}{\hbar}
\sum_{m,\mu\mathbf q}
\left|
\left[
g^{\textrm{s,DFT}}_{\mu}
\right]^{m\mathbf k+\mathbf q}_{n\mathbf k}
(\mathbf q,0)
\right|^2
\left[
\left(
N_{\mu\mathbf q}^{\textrm{DFT}}
+
1
-
f_{m\mathbf k+\mathbf q}^{\textrm{DFT}}
\right)
\delta\!\left(
\varepsilon_{n\mathbf k}^{\textrm{DFT}}
-
\varepsilon_{m\mathbf k+\mathbf q}^{\textrm{DFT}}
-
\hbar\omega_{\mu\mathbf q}^{\textrm{DFT}}
\right)
+
\left(
N_{\mu\mathbf q}^{\textrm{DFT}}
+
f_{m\mathbf k+\mathbf q}^{\textrm{DFT}}
\right)
\delta\!\left(
\varepsilon_{n\mathbf k}^{\textrm{DFT}}
-
\varepsilon_{m\mathbf k+\mathbf q}^{\textrm{DFT}}
+
\hbar\omega_{\mu\mathbf q}^{\textrm{DFT}}
\right)
\right]
$
\\[1.3em]

$\displaystyle
\Gamma^{\textrm{ph,DFT+DFPT}}_{\mu\mathbf q}
=
\frac{2\pi}{\hbar}
\sum_{nm\mathbf k}
\left(
f_{n\mathbf k}^{\textrm{DFT}}
-
f_{m\mathbf k+\mathbf q}^{\textrm{DFT}}
\right)
\left|
\left[
g^{\textrm{s,DFT}}_{\mu}
\right]^{m\mathbf k+\mathbf q}_{n\mathbf k}
(\mathbf q,0)
\right|^2
\delta\!\left(
\varepsilon_{m\mathbf k+\mathbf q}^{\textrm{DFT}}
-
\varepsilon_{n\mathbf k}^{\textrm{DFT}}
-
\hbar\omega_{\mu\mathbf q}^{\textrm{DFT}}
\right)
$
\\[1.5em]

\hline\hline
\end{tabular}%
}

\end{table*}

The intermediate SX+DFPT calculation is essential for interpreting the results. Relative to DFT+DFPT, it changes the electronic spectrum while retaining the DFPT vertex. The difference between these two calculations therefore predominantly measures the phase-space suppression caused by the steeper SX bands. The subsequent change from SX+DFPT to SX+BSE instead isolates the effect of the electron-hole ladder on the scattering amplitude.

\begin{figure*}[t]
        \centering
        \includegraphics[width=\linewidth]
        {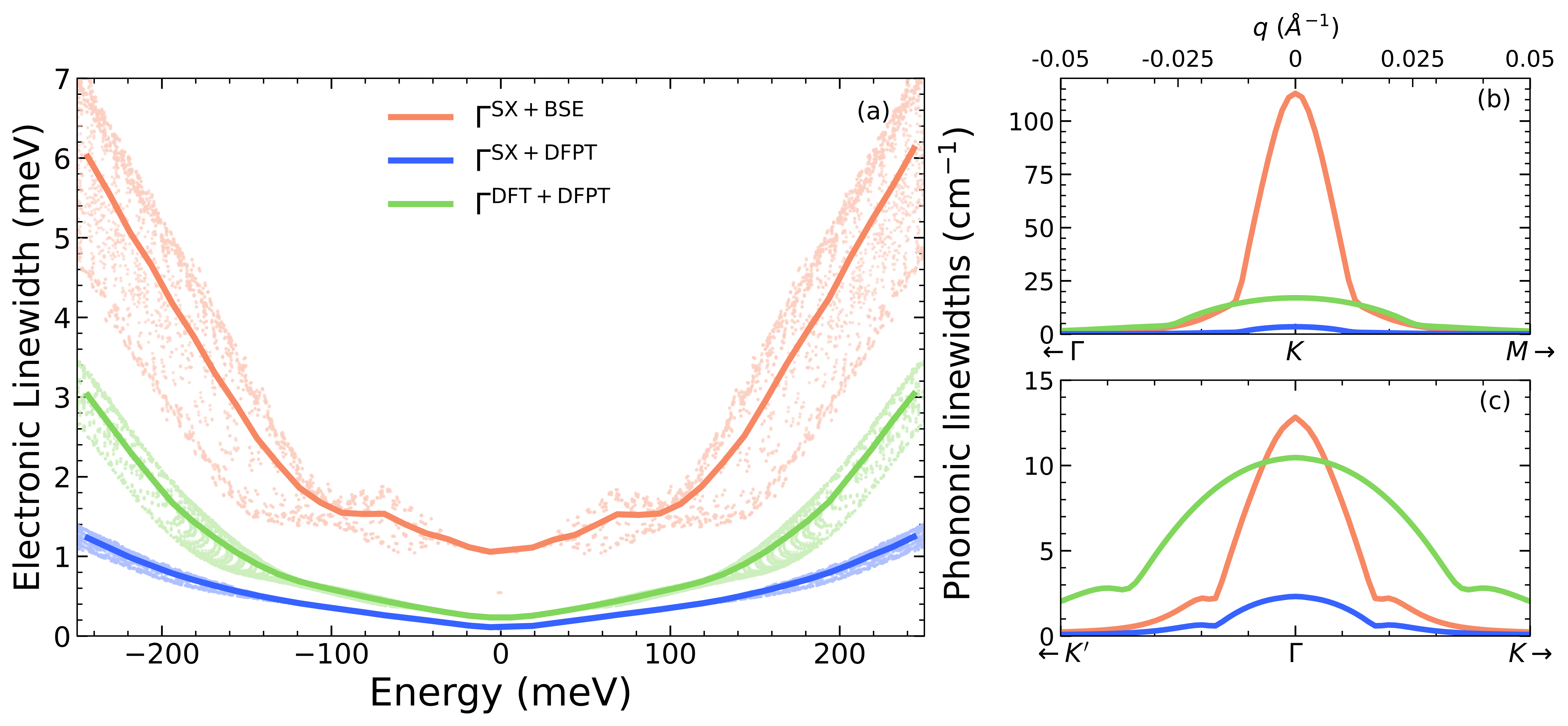}
    \caption{ Electron-phonon induced electronic and phononic linewidths obtained with the three levels of approximation defined in Eqs.~\eqref{eq:FGRappr0}--\eqref{eq:FGRappr2}.  The SX+BSE calculation uses \(\mathcal G^{\textrm{s,SX}}\), \(\varepsilon^{\textrm{SX}}\), and \(\omega^{\textrm{SX}}\); SX+DFPT retains the SX electronic energies but replaces the correlated vertex with \(g^{\textrm{s,DFT}}\); DFT+DFPT uses the DFT bands, frequencies, and coupling. 
    (a) Electronic linewidth as a function of quasiparticle energy. Symbols represent the values calculated for individual \(\mathbf k\) points, and continuous curves are energy-bin averages. The acoustic-phonon contribution is included using the model of Ref.~\onlinecite{PhysRevB.90.125414}. (b-c) Phonon linewidth, reported as the full width at half maximum, for (b) the TO phonon near \(K\) and (c) the LO phonon near $\Gamma$. $q$ points magnitudes are referred to the $K$ point or the $\Gamma$ point respectively. In both panels, using the SX spectrum alone (SX+DFPT) drastically suppresses the linewidth, whereas the electron-hole ladder produces a larger enhancement that yields a final linewidth above the DFT+DFPT result, except for the LO phonon near $\Gamma$ where the effects compensate almost exactly due to the gauge-coupling nature of the coupling.
    }
    \label{fig:linewidth}
\end{figure*}

The left panel of Fig.~\ref{fig:linewidth} shows the electronic linewidth for every \(\mathbf k\) point of the fine mesh as a function of the quasiparticle energy. The solid curves are obtained by binning the individual values in energy. The linear acoustic-phonon contribution, absent from the optical-phonon TB model, is included using the analytical DFT model of Ref.~\onlinecite{PhysRevB.90.125414}; results
without this contribution are given in the Supplemental
Material~\cite{SI}.

The electronic linewidth has a minimum close to the Dirac point, where the electronic density of states and the phase space for inelastic scattering are smallest. It increases with \(\lvert\varepsilon_{n\mathbf k}\rvert\) as additional final states and phonon emission channels become available. The approximate electron-hole symmetry of the result reflects the corresponding symmetry of the low-energy \(\pi/\pi^\ast\) spectrum.

The comparison among the three curves demonstrates that the band and vertex corrections act in opposite directions. Keeping the DFPT vertex while replacing the DFT bands by the SX bands lowers the
electronic linewidth by approximately a factor of three. The increased quasiparticle velocity therefore suppresses scattering substantially even though the same interaction vertex coupling is used. When the electron-hole ladder is included, the linewidth increases from the SX+DFPT curve to the full SX+BSE result. The latter exceeds the standard DFT+DFPT linewidth by at least a factor of two over the energy range shown. The excitonic enhancement of the matrix element thus not only compensates the loss of phase space but reverses its effect. The change is not a simple rigid rescaling of the DFT result. Different electronic energies sample different phonon branches, momentum transfers, and regions of the nonlocal vertex. Consequently, the momentum dependence of \(\mathcal G^{\textrm{s,SX}}\) modifies both the magnitude and the energy dependence of the linewidth.

The right panel of Fig.~\ref{fig:linewidth} shows the linewidth of the TO phonon near \(K\) (upper) and of the LO phonon near \(\Gamma\) (lower), computed directly on a reciprocal space line. The DFT+DFPT result provides the conventional baseline. For both modes, replacing the DFT spectrum with the SX quasiparticle bands while retaining \(g^{\textrm{s,DFT}}\) lowers the linewidth, because the steeper Dirac cone reduces the joint density of states on the decay surface. The SX+DFPT curve therefore provides the phononic counterpart of the phase-space suppression observed for the electronic linewidth. The full SX+BSE result displays the opposite behavior. Near \(K\), the TO linewidth increases from approximately \(20~\textrm{cm}^{-1}\) at the DFT+DFPT level to approximately \(100~\textrm{cm}^{-1}\), corresponding to a net enhancement of about a factor five \cite{y1dn-m6pc}. This increase remains substantially smaller than the enhancement of the squared vertex, which is of order \(25\), because the larger scattering amplitude is partially compensated by the reduced electron-hole phase space associated with the SX bands, consistently with the behavior found for the electronic linewidth. The redshift of the TO branch modifies the precise energy-conservation contour but is not the main origin of the linewidth enhancement. The situation is qualitatively different for the long-wavelength optical mode near \(\Gamma\). There, the electron-phonon coupling is constrained by the small-wavevector gauge-field cancellation \cite{PhysRevB.76.045430,Bistoni_2019,Rostami2018,PhysRevB.110.L201405}, so that quasiparticle and ladder corrections largely compensate each other \cite{y1dn-m6pc}. Accordingly, the SX+BSE linewidth is brought back close to its DFT+DFPT value rather than being strongly enhanced.

This distinction between the \(K\) and \(\Gamma\) phonon modes behaviour also clarifies the energy dependence of the electronic linewidth. Electronic scattering samples contributions from both regions of the phonon spectrum: close to charge neutrality, the lowest-energy optical scattering channel is dominated by the \(K\)-point phonons, for which the vertex enhancement is large, and the many-body enhancement of the electronic linewidth is therefore most pronounced. At larger carrier energies, scattering by the \(\Gamma\)-point optical modes becomes available as well; because their many-body correction is much weaker, their contribution dilutes the relative enhancement.

The parallel hierarchy of the electronic and phononic results is a direct manifestation of the common-vertex construction developed in Secs.~\ref{sec:selfenergies} and~\ref{sec:fgr}. The same \(\lvert\mathcal G^{\textrm{s,SX}}\rvert^2\) that enhances the decay of the TO phonon into an electron-hole pair also enhances the reciprocal electronic processes involving absorption or emission of that phonon. Doping and dielectric screening provide additional control over this competition and will be discussed in a future work. At low carrier densities, the excitonic enhancement persists. At higher doping, the increased electronic screening weakens the ladder correction. Encapsulation in hexagonal boron nitride similarly reduces the effective Coulomb interaction and therefore decreases the vertex enhancement \cite{y1dn-m6pc}. These effects may contribute to the dependence of the apparent electron-phonon strength inferred from angle-resolved photoemission experiments \cite{PhysRevB.76.205411,PhysRevB.81.041403,Bostwick2007}. Finally, the static quasiparticle vertex obtained here is directly suited for Boltzmann transport calculations. Because \(\mathcal G^{\textrm{s,SX}}\) is expressed between electronic quasiparticle states, it can replace the conventional DFPT matrix element in electron and phonon collision integrals without requiring explicit propagation of excitonic quasiparticles. For graphene, the strong enhancement of intervalley TO scattering suggests that vertex corrections can substantially increase the room temperature contribution of zone-boundary optical phonons to intrinsic resistivity, hot-carrier relaxation, and coupled electron-phonon dynamics, despite the simultaneous increase of the quasiparticle velocity. 

\section{Conclusions}
\label{sec:conclusions}

We have developed a many-body framework for incorporating electron-hole correlations into electron-phonon scattering while preserving the detailed-balance relation between electronic and phononic processes. Rather than introducing a corrected electron-phonon coupling phenomenologically, we started from the microscopic Hamiltonian of interacting electrons and nuclei and expanded it around the Born-Oppenheimer equilibrium geometry. This construction makes the reference electronic and phononic problems explicit and provides a controlled starting point for introducing screening and vertex corrections without ambiguity or double counting. We then formulated the problem through a common effective action for the electronic and phononic propagators. The approximation adopted here selects, in a controlled way, the part of the interaction functional that contains a single phonon propagator. It therefore retains the single-phonon processes relevant for linewidths while preserving the diagrammatic structure and symmetry factors inherited from the electronic exchange-correlation functional. Higher-order multiphonon processes, not treated in this work, could in principle be included using a similar logic.

A central result of this formulation is that the electronic and phononic self-energies are generated from the same approximate functional. Electronic screening first transforms the bare electron-phonon interaction into a screened local vertex, while electron-hole ladder corrections further dress it into a nonlocal vertex. The same correlated vertex enters both the electron and phonon self-energies. Vertex corrections are therefore treated coherently in the two subsystems, rather than being introduced independently for electronic and phononic scattering. In the quasiparticle and static-vertex limit considered here, we include electron-hole correlations effects inside the electron-phonon vertices. The intermediate electron-hole pair is then described in terms of independent electronic quasiparticles, while excitonic effects remain encoded in the scattering amplitude. This organization is complementary to approaches in which excitons are introduced as explicit propagating quasiparticles and is particularly convenient for calculating electron and phonon linewidths. It leads to Fermi golden rule expressions containing the same correlated matrix element and makes the detailed-balance relation explicit.

We applied the formalism to graphene using many-body-renormalized electronic bands and phonon frequencies together with a Bethe-Salpeter correction to the electron-phonon vertex. The calculation reveals two competing effects. The renormalization of the electronic dispersion increases the carrier velocity and reduces the available scattering phase space, thereby suppressing the linewidth. The vertex correction acts in the opposite direction by enhancing the electron-phonon matrix element. For both electronic and phononic linewidth, the enhancement of the interaction overcompensates for the reduction of phase space, producing a final linewidth larger than that obtained from a standard density-functional calculation. The opposite effects of band and vertex renormalization are therefore common to both electronic and phononic linewidths. Accounting for only one of these contributions can consequently lead to an incomplete or even qualitatively misleading estimate of the linewidth.

The present work provides a controlled extension of many-body perturbation theory from the real parts of electronic and phononic self-energies to their imaginary parts. It also supplies static electron-phonon vertices containing excitonic effects that can be used within conventional quasiparticle calculations without introducing explicit exciton propagation. This makes the approach suitable for future studies of coupled electron-phonon transport, Raman and infrared spectra, hot-carrier relaxation, and superconductivity in weakly screened materials.

\section{Acknowledgments}
We thank J. Fiore, M. Udina, A. Ferretti, D. Varsano and J. M. Lihm for useful discussions.
We acknowledge the EuroHPC Joint Undertaking for awarding this project access to the EuroHPC supercomputer LUMI, hosted by CSC (Finland) and the LUMI consortium through a EuroHPC Regular Access call. This project has received funding from the European Research Council (ERC) under the European Union’s Horizon 2020 research and innovation program (MORE-TEM ERC-SYN Project, Grant Agreement No. 951215).

\appendix

\section{Variation of the reducible electronic response}
\label{app:variation_chi}

In this appendix we derive the variation of the reducible electronic
response \(\chi_{\textrm{el}}\). We start from the Dyson
equation of Eq.~\eqref{eq:WPW1}, i.e. :
\begin{equation}
\chi_{\textrm{el}}
=
P_{\textrm{el}}
+
P_{\textrm{el}}
v
\chi_{\textrm{el}}.
\end{equation}
Therefore, differentiating with respect to $G$ we find
\begin{align}
 \frac{\delta\chi_{\textrm{el}}}{\delta G}
&=
 \frac{\delta P_{\textrm{el}}}{\delta G}
+
 \frac{\delta P_{\textrm{el}}}{\delta G}
v
\chi_{\textrm{el}}
+
P_{\textrm{el}}
v
\frac{\delta\chi_{\textrm{el}}}{\delta G}.
\label{eq:app_delta_chi_1}
\end{align}
Collecting the terms proportional to
\(\delta\chi_{\textrm{el}}/\delta G\) on the left-hand side gives
\begin{align}
\left(
I-P_{\textrm{el}}v
\right)
\frac{\delta\chi_{\textrm{el}}}{\delta G}
&=
\frac{\delta P_{\textrm{el}}}{\delta G}
\left(
I+v\chi_{\textrm{el}}
\right).
\label{eq:app_delta_chi_2}
\end{align}
Multiplying from the left by
\(\left(I-P_{\textrm{el}}v\right)^{-1}\), we obtain
\begin{align}
\frac{\delta\chi_{\textrm{el}}}{\delta G}
&=
\left(
I-P_{\textrm{el}}v
\right)^{-1}
 \frac{\delta P_{\textrm{el}}}{\delta G}
\left(
I+v\chi_{\textrm{el}}
\right).
\label{eq:app_delta_chi_3}
\end{align}

The factors appearing in Eq.~\eqref{eq:app_delta_chi_3} can be expressed
in terms of the screened interaction
\(W_{\textrm{el}}\) using Eq.~\eqref{eq:WPW2} i.e. 
\begin{equation}
W_{\textrm{el}}
=
v
+
vP_{\textrm{el}}
W_{\textrm{el}}.
\label{eq:app_W_Dyson}
\end{equation}
Multiplying Eq.~\eqref{eq:app_W_Dyson} from the left by
\(P_{\textrm{el}}\) gives
\begin{equation}
P_{\textrm{el}}
W_{\textrm{el}}
=
P_{\textrm{el}}v
+
P_{\textrm{el}}
v
P_{\textrm{el}}
W_{\textrm{el}}.
\end{equation}
It follows that
\begin{align}
\left(
I-P_{\textrm{el}}v
\right)
\left(
I+
P_{\textrm{el}}
W_{\textrm{el}}
\right)
&=
I,
\end{align}
and hence
\begin{equation}
\left(
I-P_{\textrm{el}}v
\right)^{-1}
=
I+
P_{\textrm{el}}
W_{\textrm{el}}.
\label{eq:app_left_screening_identity}
\end{equation}

The second factor can be rewritten using
Eq.~\eqref{eq:chidefPwP},
\begin{equation}
\chi_{\textrm{el}}
=
P_{\textrm{el}}
+
P_{\textrm{el}}
W_{\textrm{el}}
P_{\textrm{el}}.
\label{eq:app_chi_PWP}
\end{equation}
Multiplication from the left by \(v\) gives
\begin{align}
v\chi_{\textrm{el}}
&=
vP_{\textrm{el}}
+
vP_{\textrm{el}}
W_{\textrm{el}}
P_{\textrm{el}}.
\end{align}
On the other hand, multiplying
Eq.~\eqref{eq:app_W_Dyson} from the right by
\(P_{\textrm{el}}\) gives
\begin{align}
W_{\textrm{el}}
P_{\textrm{el}}
&=
vP_{\textrm{el}}
+
vP_{\textrm{el}}
W_{\textrm{el}}
P_{\textrm{el}}.
\end{align}
Therefore,
\begin{equation}
v\chi_{\textrm{el}}
=
W_{\textrm{el}}
P_{\textrm{el}},
\label{eq:vchiWpel}
\end{equation}
and consequently
\begin{equation}
I+v\chi_{\textrm{el}}
=
I+
W_{\textrm{el}}
P_{\textrm{el}}.
\label{eq:app_right_screening_identity}
\end{equation}

Substituting Eqs.~\eqref{eq:app_left_screening_identity} and
\eqref{eq:app_right_screening_identity} into
Eq.~\eqref{eq:app_delta_chi_3}, we finally obtain
\begin{equation}
\frac{\delta\chi_{\textrm{el}}}{\delta G}
=
\left[
I+
P_{\textrm{el}}
W_{\textrm{el}}
\right]
 \frac{\delta P_{\textrm{el}}}{\delta G}
\left[
I+
W_{\textrm{el}}
P_{\textrm{el}}
\right].
\label{eq:delta_chi_fixed_v}
\end{equation}

\section{Proof of Eq. \eqref{eq:Gcal_g_Kn_coord}}
\label{app:Gggs}
In this appendix we derive the identities used in the Fermi golden rule
section. For a four-point object \(A\) acting on a two-leg electron-phonon vertex
\(h(1,2;b)\), we use the convention
\begin{equation}
    (A\star h)(1,2;b)
    =
    A(1,\bar 3;2,\bar 4)
    h(\bar 4,\bar 3;b),
\end{equation}
and, for a left contraction,
\begin{equation}
    (h\star A)(1,2;a)
    =
    h(\bar 3,\bar 4;a)
    A(\bar 4,2;\bar 3,1).
\end{equation}

When the local (screened or not) electron-phonon coupling is used as an object with two electronic
legs, we write
\begin{equation}
g(1,2;b)
\equiv
\delta(1,2^+)g(2,b).
\label{eq:app_g_two_leg}
\end{equation}
Therefore
\begin{equation}
(L\star g)(1,2;b)
=
L(1,\bar 3;2,\bar 3^+)
g(\bar 3,b).
\end{equation}

In this notation, we define the induced density matrix by
\begin{equation}
n_{\textrm{ind}}(1,2;b)
\equiv
-i\hbar\,
(L\star g)(1,2;b).
\label{eq:app_nind_def}
\end{equation}
Equivalently,
\begin{equation}
L\star g
=
\frac{i}{\hbar}
n_{\textrm{ind}}.
\label{eq:app_Lg_nind}
\end{equation}

We now recall the screened electron-phonon vertex used in the main text,
\begin{equation}
g^{\textrm{s}}
=
\left(
I+
W_{\textrm{el}}
P_{\textrm{el}}
\right)g .
\label{eq:app_gs_main_def}
\end{equation}
Using Eq. \eqref{eq:vchiWpel} and the kernel convention used in the main text, we obtain the equivalent self-consistent form
\begin{equation}
g^{\textrm{s}}
=
g
-
\frac{i}{\hbar}
K_{\textrm H}
P_{\textrm{el}}
g^{\textrm{s}}.
\label{eq:app_gs_KH_P}
\end{equation}

The ladder-dressed electron-phonon vertex is defined as
\begin{equation}
\mathcal G^{\textrm{s}}
=
\Lambda\star g^{\textrm{s}}.
\label{eq:app_Gcal_def}
\end{equation}
Using Eq.~\eqref{eq:Lambda_def} in Eq. \eqref{eq:app_Gcal_def}, we find the Dyson equation
\begin{equation}
\mathcal G^{\textrm{s}}
=
g^{\textrm{s}}
-
K_\textrm{xc}\star L_0\star\mathcal G^{\textrm{s}}.
\label{eq:app_Gcal_ladder}
\end{equation}

We now prove that the amputated full response coincides with the
ladder-dressed electron-phonon vertex defined in the main text. To this
end, we first separate the Hartree from the exchange and correlation in the full
Bethe-Salpeter equation. Starting from
\begin{equation}
L
=
L_0
-
L_0\star
\left(
K_{\textrm H}
+
K_\textrm{xc}
\right)
\star
L,
\end{equation}
we rewrite it as
\begin{equation}
\left(
I+
L_0\star K_\textrm{xc}
\right)
\star
L
=
L_0
-
L_0\star K_{\textrm H}\star L .
\label{eq:BSE-app}
\end{equation}
Multiplying from the left by
\((I+L_0\star K_\textrm{xc})^{-1}\), and using
\[
L_\textrm{xc}
=
(I+L_0\star K_\textrm{xc})^{-1}\star L_0,
\]
we obtain
\begin{equation}
L
=
L_\textrm{xc}
-
L_\textrm{xc}\star K_{\textrm H}\star L .
\label{eq:app_L_split_H_L}
\end{equation}
Closing Eq.~\eqref{eq:app_L_split_H_L} on the bare electron-phonon vertex
gives
\begin{equation}
L\star g
=
L_\textrm{xc}\star
\left[
g
-
K_{\textrm H}\star L\star g
\right].
\label{eq:app_Lg_split_H_L}
\end{equation}
We now define the quantity in square brackets as
\begin{equation}
h^{\textrm{s}}
\equiv
g
-
K_{\textrm H}\star L\star g .
\label{eq:app_hs_def}
\end{equation}
This can also be written as
\begin{equation}
h^{\textrm{s}}
=
g
-
\frac{i}{\hbar}
K_{\textrm H}\star n_{\textrm{ind}} .
\label{eq:app_hs_g_KHn}
\end{equation}
Equation~\eqref{eq:app_Lg_split_H_L} then becomes
\begin{equation}
L\star g
=
L_\textrm{xc}\star h^{\textrm{s}} .
\label{eq:app_Lg_LLhs}
\end{equation}

We now show that \(h^{\textrm{s}}\) is precisely the screened vertex
\(g^{\textrm{s}}\) used in the main text. Inserting
Eq.~\eqref{eq:app_Lg_LLhs} back into Eq.~\eqref{eq:app_hs_def} gives
\begin{equation}
h^{\textrm{s}}
=
g
-
K_{\textrm H}\star
L_\textrm{xc}\star
h^{\textrm{s}} .
\label{eq:app_hs_selfconsistent}
\end{equation}
Using
\[
P_{\textrm{el}}(1,2)
=
-i\hbar\,
L_\textrm{xc}(1,2;1^+,2^+),
\]
Eq.~\eqref{eq:app_hs_selfconsistent} is equivalent to
\begin{equation}
h^{\textrm{s}}
=
g
-
\frac{i}{\hbar}
K_{\textrm H}
P_{\textrm{el}}
h^{\textrm{s}} .
\label{eq:app_hs_KH_P}
\end{equation}
Therefore
\begin{equation}
h^{\textrm{s}}
=
g^{\textrm{s}} .
\label{eq:app_hs_equals_gs}
\end{equation}

Using Eq.~\eqref{eq:app_hs_equals_gs} in
Eq.~\eqref{eq:app_Lg_LLhs}, and using
\(L_\textrm{xc}=L_0\star\Lambda\), we obtain
\begin{align}
L\star g
=
L_\textrm{xc}\star g^{\textrm{s}}
=
L_0\star\Lambda\star g^{\textrm{s}}
=
L_0\star\mathcal G^{\textrm{s}} .
\label{eq:app_Lg_L0G_proved}
\end{align}
Multiplying by \(L_0^{-1}\) from the left finally gives the desired
identity
\begin{equation}
L_0^{-1}\star L\star g
=
\mathcal G^{\textrm{s}} .
\label{eq:app_amputated_equals_Gcal}
\end{equation}

We can now use the full Bethe-Salpeter equation to obtain the compact
form of this vertex. Multiplying Eq. \eqref{eq:BSE-app} by \(L_0^{-1}\) from the left and closing on \(g\), we get
\begin{equation}
L_0^{-1}\star L\star g
=
g
-
K_{\textrm{tot}}\star L\star g .
\end{equation}
Recognizing the density matrix and using Eq.~\eqref{eq:app_amputated_equals_Gcal}, we obtain
\begin{equation}
\mathcal G^{\textrm{s}}
=
g
-
\frac{i}{\hbar}
K_{\textrm{tot}}\star
n_{\textrm{ind}} .
\label{eq:app_Gcal_g_Kn}
\end{equation}
In explicit coordinates, this is Eq. \eqref{eq:Gcal_g_Kn_coord}. A similar proof can be done for the left vertex.

\section{Proof of Eq.~\eqref{eq:deltaLL}}
\label{app:derivKL}
In this appendix we derive the variation of \(L_\textrm{xc}\). All products and inverses appearing below are understood with
respect to the \(\star\)-product defined in Eq.~\eqref{eq:mult_def}.

We start from the Bethe-Salpeter equation
\begin{equation}
L_{\textrm{xc}}
=
L_0
-
L_0\star K_{\textrm{xc}}\star L_{\textrm{xc}} .
\label{eq:app_LL_BSE}
\end{equation}
The derivative with respect to $G$ reads
\begin{align}
\left. \frac{\delta L_{\textrm{xc}}}{\delta G} \right|_{K_{\textrm{xc}}}
=
\frac{\delta L_0}{\delta G}
-
\frac{\delta L_0}{\delta G}\star K_{\textrm{xc}}\star L_{\textrm{xc}}
-
\left. L_0\star K_{\textrm{xc}}\star\frac{\delta L_{\textrm{xc}}}{\delta G} \right|_{K_{\textrm{xc}}},
\label{eq:app_delta_LL_1}
\end{align}
where we have put to zero the derivative $\delta K_{\textrm{xc}}/\delta G$.
Collecting the terms proportional to \(\delta L_{\textrm{xc}}/\delta G\) on the left-hand
side, we obtain
\begin{align}
\left(
I+L_0\star K_{\textrm{xc}}
\right)\star\frac{\delta L_{\textrm{xc}}}{\delta G}
&=
\frac{\delta L_0}{\delta G}\star
\left(
I-K_{\textrm{xc}}\star L_{\textrm{xc}}
\right).
\label{eq:app_delta_LL_2}
\end{align}

Using the definitions Eq. \eqref{eq:Lambda_def} and \eqref{eq:Lambda_tilde_def} and multiplying Eq.~\eqref{eq:app_delta_LL_2} from the left by
\(\widetilde{\Lambda}\), we find
\begin{align}
\left. \frac{\delta L_{\textrm{xc}}}{\delta G} \right|_{K_{\textrm{xc}}}
&=
\widetilde{\Lambda}
\star\frac{\delta L_0}{\delta G}
\star\Lambda.
\label{eq:app_delta_LL_general}
\end{align}

\section{Conditions of validity of the static reciprocity relation}
\label{app:reciprocity}
We establish Eq.~\eqref{eq:GNMmqGMNq}, on which the reduction of both
linewidths to a single $|\mathcal G^{\textrm{s,SX}}_{\mu}|^{2}$ rests.
We assume non-degeneracy of electronic and phononic states, otherwise
the same conclusions apply considering the appropriate sums over the
degenerate subspaces. We work in the basis of the SX states of
Eq.~\eqref{eq:hSX_eigenproblem}, with the four-point index placement of
the Supplemental Material~\cite{SI}, and in the linear-response
normalization introduced below Eq.~\eqref{eq:Pi_SX_Gcal_nind-SX}, in
which all prefactors are real. The derivation is carried out at fixed
Cartesian atomic index $I\alpha$ and for generic state labels: no
crystal momentum enters before the phonon contraction, so that the
result holds for molecules as well.
Combining $L=L_{0}-L_{0}\star K_{\textrm{tot}}\star L$ with
Eq.~\eqref{eq:app_Gcal_g_Kn} gives the self-consistent-field form of the
dressed vertex,
\begin{align}
\big[\mathcal G^{\textrm{s,SX}}_{I\alpha}\big]^{M}_{N}(z)
=
\big[g_{I\alpha}\big]^{M}_{N}
-\sum_{M'N'}
\big[K^{\textrm{SX}}_{\textrm{tot}}\big]^{M'N'}_{MN}\,
\big[L^{\textrm{SX}}_{0}\big]^{M'}_{N'}(z)\,
\big[\mathcal G^{\textrm{s,SX}}_{I\alpha}\big]^{M'}_{N'}(z),
\label{eq:app_Gcal_selfconsistent}
\end{align}
where we used that $L^{\textrm{SX}}_{0}$ is diagonal in the pair index.
We solve Eq.~\eqref{eq:app_Gcal_selfconsistent} by iteration, starting
from $[\mathcal G^{(0)}]^{M}_{N}=[g_{I\alpha}]^{M}_{N}$, and show
inductively that at $z=0$ every iterate obeys
\begin{align}
\Big(\big[X\big]^{M}_{N}\Big)^{*}=\big[X\big]^{N}_{M}.
\label{eq:app_conj_property}
\end{align}
\emph{Bare vertex.} Since $\hat H$ is Hermitian and the displacement
$U_{I\alpha}$ is real, $\hat g_{I\alpha}$ is Hermitian and
$\mathcal G^{(0)}$ obeys Eq.~\eqref{eq:app_conj_property}. This is
automatic and imposes no restriction; in particular the coupling need be
neither local nor real.
\emph{Electron-hole propagator.} The Hermiticity of $h^{\textrm{SX}}$
makes $\varepsilon^{\textrm{SX}}_{N}$ and $f^{\textrm{SX}}_{N}$ real, so
that in
\begin{align}
\big[L^{\textrm{SX}}_{0}\big]^{M}_{N}(z)
=
\frac{f^{\textrm{SX}}_{N}-f^{\textrm{SX}}_{M}}
{\varepsilon^{\textrm{SX}}_{N}-\varepsilon^{\textrm{SX}}_{M}+\hbar z}
\label{eq:app_L0ME}
\end{align}
numerator and denominator change sign together under
$M\leftrightarrow N$ at $z=0$, giving
\begin{align}
\big([L^{\textrm{SX}}_{0}]^{M}_{N}(0)\big)^{*}
=[L^{\textrm{SX}}_{0}]^{N}_{M}(0).
\end{align}
The degenerate limit, given by
$\partial f/\partial\varepsilon$, is likewise symmetric. At $z\neq0$ the
two orderings differ by $\hbar z\to-\hbar z$ in the denominator, which
is the origin of the static limit; what survives in general is
$\big([L^{\textrm{SX}}_{0}]^{M}_{N}(z)\big)^{*}
=[L^{\textrm{SX}}_{0}]^{N}_{M}(-z^{*})$, whence the evaluation of the
left and right vertices at opposite frequencies in
Eqs.~\eqref{eq:Pi_R_SX_Gcal_nind_atomic} and
\eqref{eq:Sigma_eph_Matsubara_Gcal}.
\emph{Interaction kernels.} With the coordinate ordering introduced
below Eq.~\eqref{eq:Ksymm}, the matrix elements of a four-point kernel
read
\begin{align}
\big[K\big]^{M'N'}_{MN}
=
\int\frac{d1\,d2\,d3\,d4}{(NV)^{4}}\;
\psi^{*}_{M}(1)\,\psi^{*}_{N'}(2)\,\psi_{N}(3)\,\psi_{M'}(4)\;
K(1,2;3,4).
\label{eq:app_KME}
\end{align}
$K^{\textrm{SX}}_{\textrm{tot}}$
obeys~\cite{PhysRevB.111.075137}
\begin{align}
K^{\textrm{SX}}_{\textrm{tot}}(1,2;3,4)=K^{\textrm{SX}}_{\textrm{tot}}(3,4;1,2).
\label{eq:app_K_exchange}
\end{align}
Conjugating
Eq.~\eqref{eq:app_KME}, relabelling $1\leftrightarrow3$ and
$2\leftrightarrow4$, and using Eq.~\eqref{eq:app_K_exchange}, one
obtains
\begin{align}
\Big(\big[K\big]^{M'N'}_{MN}\Big)^{*}
=
\big[K\big]^{N'M'}_{NM} ,
\label{eq:app_conj_rule}
\end{align}
which is Eq.~\eqref{eq:app_conj_property} applied separately to the
incoming and to the outgoing pair.
\emph{Induction.} Assume that $\mathcal G^{(p)}$ obeys
Eq.~\eqref{eq:app_conj_property}. Conjugating the $(p+1)$-th iterate of
Eq.~\eqref{eq:app_Gcal_selfconsistent} at $z=0$, using the three results
above and relabelling the dummy pair $M'\leftrightarrow N'$,
\begin{align}
\Big(\big[\mathcal G^{(p+1)}\big]^{M}_{N}\Big)^{*}
&=
\big[g_{I\alpha}\big]^{N}_{M}
-\sum_{M'N'}
\big[K^{\textrm{SX}}_{\textrm{tot}}\big]^{M'N'}_{NM}\,
\big[L^{\textrm{SX}}_{0}\big]^{M'}_{N'}(0)\,
\big[\mathcal G^{(p)}\big]^{M'}_{N'}
\nonumber\\
&=\big[\mathcal G^{(p+1)}\big]^{N}_{M} .
\end{align}
Since $\mathcal G^{(0)}$ obeys Eq.~\eqref{eq:app_conj_property}, so does
every iterate, hence
\begin{align}
\Big(\big[\mathcal G^{\textrm{s,SX}}_{I\alpha}\big]^{M}_{N}(0)\Big)^{*}
=
\big[\mathcal G^{\textrm{s,SX}}_{I\alpha}\big]^{N}_{M}(0).
\label{eq:app_Gcal_hermitian}
\end{align}
$M$ and $N$ are arbitrary
SX states and no relation between states at $\mathbf k$ and $-\mathbf k$
appears anywhere, so that time-reversal symmetry plays no role.
Conversely, Eq.~\eqref{eq:app_Gcal_hermitian} fails if a finite
electron-phonon lifetime is inserted in $G^{\textrm{SX}}$, the resulting
complex energies spoiling Eq.~\eqref{eq:app_L0ME}, consistently with the
clamped-nuclei prescription of Sec.~\ref{sec:graphene}.
\emph{Contraction with the phonon.} Crystal momentum is generated only
now. Writing $I=(s,\mathbf T)$ and performing the lattice Fourier
transform contained in Eq.~\eqref{eq:Gcal_mode_projected},
\begin{align}
\big[\mathcal G^{\textrm{s,SX}}_{s\alpha}\big]^{M}_{N}(\mathbf q,0)
\equiv
\sum_{\mathbf T}
e^{i\mathbf q\cdot\mathbf R^{0,\textrm{BO}}_{s\mathbf T}}\,
\big[\mathcal G^{\textrm{s,SX}}_{s\alpha\mathbf T}\big]^{M}_{N}(0),
\label{eq:app_Gcal_q_def}
\end{align}
which vanishes unless $\mathbf k_{M}=\mathbf k_{N}+\mathbf q$, so that
from here on $N=(n,\mathbf k)$ and $M=(m,\mathbf k+\mathbf q)$.
Conjugating Eq.~\eqref{eq:app_Gcal_q_def} term by term and using
Eq.~\eqref{eq:app_Gcal_hermitian},
\begin{align}
\Big(\big[\mathcal G^{\textrm{s,SX}}_{s\alpha}\big]^{M}_{N}
(\mathbf q,0)\Big)^{*}
=
\big[\mathcal G^{\textrm{s,SX}}_{s\alpha}\big]^{N}_{M}(-\mathbf q,0),
\label{eq:app_reciprocity_cartesian}
\end{align}
the conjugation of the phase being the only new operation involved. The
two sides involve the same two states: the label $-\mathbf q$ records
which of them is the bra, and no state at reversed crystal momentum
appears.
\emph{Projection onto phonon branches.} The remaining factors in
Eq.~\eqref{eq:Gcal_mode_projected} transfer
Eq.~\eqref{eq:app_reciprocity_cartesian} to
Eq.~\eqref{eq:GNMmqGMNq} if and only if the polarization vectors and
frequencies obey
\begin{align}
e^{s\alpha}_{\mu,-\mathbf q}=\left[e^{s\alpha}_{\mu\mathbf q}\right]^{*},
\qquad
\omega_{\mu,-\mathbf q}=\omega_{\mu\mathbf q} ,
\label{eq:app_pol_conjugation}
\end{align}
that is, if and only if the matrix whose eigenvectors define the modes
is the Bloch transform of a real real-space matrix. This is always the
case for the static force constants of Eq.~\eqref{eq:Cdef} used here,
since these are second derivatives of a real function of the real
nuclear coordinates. It may fail for modes obtained from the
non-adiabatic dynamical matrix on shell when time-reversal symmetry is
broken, the phonon self-energy then acquiring an antisymmetric
imaginary part that splits circularly polarized modes and endows them
with a finite angular momentum, as for chiral
phonons~\cite{PhysRevLett.115.115502,PhysRevLett.112.085503,PhysRevB.110.L201405, PhysRevX.14.011041,PhysRevLett.126.225703, Fachin2025, PhysRevLett.134.206701}. In that case
Eq.~\eqref{eq:app_pol_conjugation} has no counterpart, and
Eqs.~\eqref{eq:linewidth_FGR_explicit_static} and
\eqref{eq:electron_linewidth_FGR_SX} can no longer be reduced to a
common $|\mathcal G^{\textrm{s,SX}}_{\mu}|^{2}$.
\bibliography{biblio}

@article{Giustino2017,
  ids = {Giustino_2017,Giustino₂017},
  title = {Electron-Phonon Interactions from First Principles},
  author = {Giustino, Feliciano},
  year = {2017},
  month = feb,
  journal = {Reviews of Modern Physics},
  volume = {89},
  number = {1},
  eprint = {1603.06965},
  pages = {015003},
  publisher = {{American Physical Society}},
  issn = {0034-6861},
  doi = {10.1103/RevModPhys.89.015003},
  archiveprefix = {arxiv},
  arxivid = {1603.06965},
  isbn = {0034-6861{\r 1}539-0756},
  mag_id = {2310390902},
  pmcid = {null},
  pmid = {null}
}

@article{PhysRevX.13.031026,
  title = {In and Out-of-Equilibrium Ab Initio Theory of Electrons and Phonons},
  author = {Stefanucci, Gianluca and van Leeuwen, Robert and Perfetto, Enrico},
  journal = {Phys. Rev. X},
  volume = {13},
  issue = {3},
  pages = {031026},
  numpages = {40},
  year = {2023},
  month = {Sep},
  publisher = {American Physical Society},
  doi = {10.1103/PhysRevX.13.031026},
  url = {https://link.aps.org/doi/10.1103/PhysRevX.13.031026}
}

@book{Reining_16, place={Cambridge}, title={Interacting Electrons: Theory and Computational Approaches}, publisher={Cambridge University Press}, author={Martin, Richard M. and Reining, Lucia and Ceperley, David M.}, year={2016}}

@article{PhysRev.139.A796,
  title = {New Method for Calculating the One-Particle Green's Function with Application to the Electron-Gas Problem},
  author = {Hedin, Lars},
  journal = {Phys. Rev.},
  volume = {139},
  issue = {3A},
  pages = {A796--A823},
  numpages = {0},
  year = {1965},
  month = {Aug},
  publisher = {American Physical Society},
  doi = {10.1103/PhysRev.139.A796},
  url = {https://link.aps.org/doi/10.1103/PhysRev.139.A796}
}

@book{fabrizio2022course,
  title={A Course in Quantum Many-Body Theory: From Conventional Fermi Liquids to Strongly Correlated Systems},
  author={Fabrizio, M.},
  isbn={9783031163050},
  series={Graduate Texts in Physics},
  url={https://books.google.it/books?id=eY2ZEAAAQBAJ},
  year={2022},
  publisher={Springer International Publishing}
}

@article{PhysRevB.84.035433,
  title = {Theory of double-resonant Raman spectra in graphene: Intensity and line shape of defect-induced and two-phonon bands},
  author = {Venezuela, Pedro and Lazzeri, Michele and Mauri, Francesco},
  journal = {Phys. Rev. B},
  volume = {84},
  issue = {3},
  pages = {035433},
  numpages = {25},
  year = {2011},
  month = {Jul},
  publisher = {American Physical Society},
  doi = {10.1103/PhysRevB.84.035433},
  url = {https://link.aps.org/doi/10.1103/PhysRevB.84.035433}
}

@article{PhysRevB.90.125414,
  title = {Phonon-limited resistivity of graphene by first-principles calculations: Electron-phonon interactions, strain-induced gauge field, and Boltzmann equation},
  author = {Sohier, Thibault and Calandra, Matteo and Park, Cheol-Hwan and Bonini, Nicola and Marzari, Nicola and Mauri, Francesco},
  journal = {Phys. Rev. B},
  volume = {90},
  issue = {12},
  pages = {125414},
  numpages = {18},
  year = {2014},
  month = {Sep},
  publisher = {American Physical Society},
  doi = {10.1103/PhysRevB.90.125414},
  url = {https://link.aps.org/doi/10.1103/PhysRevB.90.125414}
}

@article{PhysRevLett.93.185503,
  title = {Kohn Anomalies and Electron-Phonon Interactions in Graphite},
  author = {Piscanec, S. and Lazzeri, M. and Mauri, Francesco and Ferrari, A. C. and Robertson, J.},
  journal = {Phys. Rev. Lett.},
  volume = {93},
  issue = {18},
  pages = {185503},
  numpages = {4},
  year = {2004},
  month = {Oct},
  publisher = {American Physical Society},
  doi = {10.1103/PhysRevLett.93.185503},
  url = {https://link.aps.org/doi/10.1103/PhysRevLett.93.185503}
}

@article{PhysRevB.111.075118,
  title = {High- and low-energy many-body effects of graphene in a unified approach},
  author = {Guandalini, Alberto and Caldarelli, Giovanni and Macheda, Francesco and Mauri, Francesco},
  journal = {Phys. Rev. B},
  volume = {111},
  issue = {7},
  pages = {075118},
  numpages = {15},
  year = {2025},
  month = {Feb},
  publisher = {American Physical Society},
  doi = {10.1103/PhysRevB.111.075118},
  url = {https://link.aps.org/doi/10.1103/PhysRevB.111.075118}
}

@article{y1dn-m6pc,
  title = {Excitonic Effects in Phonons: Reshaping the Graphene Kohn Anomalies and Lifetimes},
  author = {Guandalini, Alberto and Macheda, Francesco and Caldarelli, Giovanni and Mauri, Francesco},
  journal = {Phys. Rev. Lett.},
  volume = {135},
  issue = {7},
  pages = {076401},
  numpages = {8},
  year = {2025},
  month = {Aug},
  publisher = {American Physical Society},
  doi = {10.1103/y1dn-m6pc},
  url = {https://link.aps.org/doi/10.1103/y1dn-m6pc}
}

@book{stefanucci2013nonequilibrium,
  title={Nonequilibrium Many-Body Theory of Quantum Systems: A Modern Introduction},
  author={Stefanucci, G. and van Leeuwen, R.},
  isbn={9780521766173},
  lccn={2012050475},
  url={https://books.google.it/books?id=6GsrjPFXLDYC},
  year={2013},
  publisher={Cambridge University Press}
}

@article{PhysRevB.111.075137,
  title = {Variational formulation of dynamical electronic response functions in the presence of nonlocal exchange interactions},
  author = {Caldarelli, Giovanni and Guandalini, Alberto and Macheda, Francesco and Mauri, Francesco},
  journal = {Phys. Rev. B},
  volume = {111},
  issue = {7},
  pages = {075137},
  numpages = {20},
  year = {2025},
  month = {Feb},
  publisher = {American Physical Society},
  doi = {10.1103/PhysRevB.111.075137},
  url = {https://link.aps.org/doi/10.1103/PhysRevB.111.075137}
}

@article{PhysRevB.29.5718,
  title = {Effects of dynamical screening on resonances at inner-shell thresholds in semiconductors},
  author = {Strinati, G.},
  journal = {Phys. Rev. B},
  volume = {29},
  issue = {10},
  pages = {5718--5726},
  numpages = {0},
  year = {1984},
  month = {May},
  publisher = {American Physical Society},
  doi = {10.1103/PhysRevB.29.5718},
  url = {https://link.aps.org/doi/10.1103/PhysRevB.29.5718}
}

@miscellanea{SI,
title={Supplementary information}
}

@article{Marini_1,
  title = {Many-body perturbation theory approach to the electron-phonon interaction with density-functional theory as a starting point},
  author = {Marini, Andrea and Ponc\'e, S. and Gonze, X.},
  journal = {Phys. Rev. B},
  volume = {91},
  issue = {22},
  pages = {224310},
  numpages = {22},
  year = {2015},
  month = {Jun},
  publisher = {American Physical Society},
  doi = {10.1103/PhysRevB.91.224310},
  url = {https://link.aps.org/doi/10.1103/PhysRevB.91.224310}
}

@article{PhysRevB.111.024307,
  title = {Exact formula with two dynamically screened electron-phonon couplings for positive phonon-linewidths approximations},
  author = {Stefanucci, Gianluca and Perfetto, Enrico},
  journal = {Phys. Rev. B},
  volume = {111},
  issue = {2},
  pages = {024307},
  numpages = {5},
  year = {2025},
  month = {Jan},
  publisher = {American Physical Society},
  doi = {10.1103/PhysRevB.111.024307},
  url = {https://link.aps.org/doi/10.1103/PhysRevB.111.024307}
}

@article{SciPostPhys.18.1.009,
	title = {Excitonic Bloch equations from first principles},
	pages = {009},
	author = {Stefanucci, Gianluca and Perfetto, Enrico},
	journal = {SciPost Phys.},
	volume = {18},
	year = {2025},
	publisher = {SciPost},
	doi = {10.21468/SciPostPhys.18.1.009},
	url = {https://scipost.org/10.21468/SciPostPhys.18.1.009}
}

@article{PhysRevLett.134.186401,
  title = {Nonperturbative Self-Consistent Electron-Phonon Spectral Functions and Transport},
  author = {Lihm, Jae-Mo and Ponc\'e, Samuel},
  journal = {Phys. Rev. Lett.},
  volume = {134},
  issue = {18},
  pages = {186401},
  numpages = {11},
  year = {2025},
  month = {May},
  publisher = {American Physical Society},
  doi = {10.1103/PhysRevLett.134.186401},
  url = {https://link.aps.org/doi/10.1103/PhysRevLett.134.186401}
}

@article{mtjf-ylf7,
  title = {Beyond-Quasiparticle Transport with Vertex Correction: Self-Consistent Ladder Formalism for Electron-Phonon Interactions},
  author = {Lihm, Jae-Mo and Ponc\'e, Samuel},
  journal = {Phys. Rev. X},
  volume = {16},
  issue = {1},
  pages = {011008},
  numpages = {47},
  year = {2026},
  month = {Jan},
  publisher = {American Physical Society},
  doi = {10.1103/mtjf-ylf7},
  url = {https://link.aps.org/doi/10.1103/mtjf-ylf7}
}

@article{Mishra_2025,
   title={Electron–phonon vertex correction effect in superconducting H3S},
   volume={11},
   ISSN={2057-3960},
   url={http://dx.doi.org/10.1038/s41524-025-01818-9},
   DOI={10.1038/s41524-025-01818-9},
   number={1},
   journal={npj Computational Materials},
   publisher={Springer Science and Business Media LLC},
   author={Mishra, Shashi B. and Mori, Hitoshi and Margine, Elena R.},
   year={2025},
   month=Nov
}

@article{Venanzi2023,
  title = {Probing {{Enhanced Electron-Phonon Coupling}} in {{Graphene}} by {{Infrared Resonance Raman Spectroscopy}}},
  author = {Venanzi, Tommaso and Graziotto, Lorenzo and Macheda, Francesco and Sotgiu, Simone and Ouaj, Taoufiq and Stellino, Elena and Fasolato, Claudia and Postorino, Paolo and Mi{\v s}eikis, Vaidotas and Metzelaars, Marvin and K{\"o}gerler, Paul and Beschoten, Bernd and Coletti, Camilla and Roddaro, Stefano and Calandra, Matteo and Ortolani, Michele and Stampfer, Christoph and Mauri, Francesco and Baldassarre, Leonetta},
  year = {2023},
  month = jun,
  journal = {Physical Review Letters},
  volume = {130},
  number = {25},
  pages = {256901},
  issn = {0031-9007, 1079-7114},
  doi = {10.1103/PhysRevLett.130.256901},
  urldate = {2023-11-20},
  langid = {english}
}

@article{Gruneis2009,
  title = {Phonon Surface Mapping of Graphite: {{Disentangling}} Quasi-Degenerate Phonon Dispersions},
  author = {Gr{\"u}neis, A. and Serrano, J. and Bosak, A. and Lazzeri, M. and Molodtsov, S. L. and Wirtz, L. and Attaccalite, C. and Krisch, M. and Rubio, A. and Mauri, F. and Pichler, T.},
  year = {2009},
  journal = {Physical Review B - Condensed Matter and Materials Physics},
  eprint = {0904.3205},
  issn = {10980121},
  doi = {10.1103/PhysRevB.80.085423},
  archiveprefix = {arxiv},
  arxivid = {0904.3205},
  isbn = {1098-0121}
}

@article{Marini2021,
  title = {Lattice Dynamics of Photoexcited Insulators from Constrained Density-Functional Perturbation Theory},
  author = {Marini, Giovanni and Calandra, Matteo},
  year = {2021},
  month = oct,
  journal = {Physical Review B},
  volume = {104},
  number = {14},
  pages = {144103},
  issn = {2469-9950, 2469-9969},
  doi = {10.1103/PhysRevB.104.144103},
  urldate = {2023-11-20},
  langid = {english}
}

@article{Harb2006,
  title = {Carrier {{Relaxation}} and {{Lattice Heating Dynamics}} in {{Silicon Revealed}} by {{Femtosecond Electron Diffraction}}},
  author = {Harb, Maher and Ernstorfer, Ralph and Dartigalongue, Thibault and Hebeisen, Christoph T. and Jordan, Robert E. and Miller, R. J. Dwayne},
  year = {2006},
  month = dec,
  journal = {The Journal of Physical Chemistry B},
  volume = {110},
  number = {50},
  pages = {25308--25313},
  issn = {1520-6106, 1520-5207},
  doi = {10.1021/jp064649n},
  urldate = {2023-11-20},
  langid = {english}
}

@article{Bernardi2014,
  title = {{\emph{Ab }}{{{\emph{Initio}}}} {{Study}} of {{Hot Carriers}} in the {{First Picosecond}} after {{Sunlight Absorption}} in {{Silicon}}},
  author = {Bernardi, Marco and {Vigil-Fowler}, Derek and Lischner, Johannes and Neaton, Jeffrey B. and Louie, Steven G.},
  year = {2014},
  month = jun,
  journal = {Physical Review Letters},
  volume = {112},
  number = {25},
  pages = {257402},
  issn = {0031-9007, 1079-7114},
  doi = {10.1103/PhysRevLett.112.257402},
  urldate = {2023-11-20},
  langid = {english}
}

@article{Tielrooij2018,
  title = {Out-of-Plane Heat Transfer in van Der {{Waals}} Stacks through Electron-Hyperbolic Phonon Coupling},
  author = {Tielrooij, Klaas Jan and Hesp, Niels C.H. and Principi, Alessandro and Lundeberg, Mark B. and Pogna, Eva A.A. and Banszerus, Luca and Mics, Zolt{\'a}n and Massicotte, Mathieu and Schmidt, Peter and Davydovskaya, Diana and Purdie, David G. and Goykhman, Ilya and Soavi, Giancarlo and Lombardo, Antonio and Watanabe, Kenji and Taniguchi, Takashi and Bonn, Mischa and Turchinovich, Dmitry and Stampfer, Christoph and Ferrari, Andrea C. and Cerullo, Giulio and Polini, Marco and Koppens, Frank H.L.},
  year = {2018},
  journal = {Nature Nanotechnology},
  volume = {13},
  number = {1},
  eprint = {1702.03766},
  pages = {41--46},
  issn = {17483395},
  doi = {10.1038/s41565-017-0008-8},
  archiveprefix = {arxiv},
  arxivid = {1702.03766},
  isbn = {1748-3395},
  pmid = {29180742}
}

@Article{Mocatti2026,
author={Mocatti, Stefano
and Marini, Giovanni
and Volpato, Giulio
and Cudazzo, Pierluigi
and Calandra, Matteo},
title={Nonequilibrium photocarrier and phonon dynamics from first principles: a unified treatment of carrier-carrier, carrier-phonon, and phonon-phonon scattering},
journal={npj Computational Materials},
year={2026},
month={May},
day={17},
issn={2057-3960},
doi={10.1038/s41524-026-02104-y},
url={https://doi.org/10.1038/s41524-026-02104-y}
}

@article{Bardeen1957,
  title = {Theory of {{Superconductivity}}},
  author = {Bardeen, J. and Cooper, L. N. and Schrieffer, J. R.},
  year = {1957},
  month = dec,
  journal = {Physical Review},
  volume = {108},
  number = {5},
  pages = {1175--1204},
  issn = {0031-899X},
  doi = {10.1103/PhysRev.108.1175},
  urldate = {2023-11-20},
  langid = {english}
}

@book{Schrieffer1999,
  title = {Theory of Superconductivity},
  author = {Schrieffer, John R.},
  year = {1999},
  series = {Advanced Book Classics},
  edition = {Rev. printing, 4. print},
  publisher = {{Westview Press}},
  address = {{Bolder, Colo}},
  isbn = {978-0-7382-0120-7},
  langid = {english}
}

@article{Marsiglio2020,
  title = {Eliashberg Theory: {{A}} Short Review},
  shorttitle = {Eliashberg Theory},
  author = {Marsiglio, F.},
  year = {2020},
  month = jun,
  journal = {Annals of Physics},
  volume = {417},
  pages = {168102},
  issn = {00034916},
  doi = {10.1016/j.aop.2020.168102},
  urldate = {2023-11-20},
  langid = {english}
}

@book{Grimvall1981TheEI,
  title={The electron-phonon interaction in metals},
  author={G{\"o}ran Grimvall},
  year={1981},
  publisher={North-Holland, Amsterdam}
}

@book{ziman2001electrons,
  title={Electrons and Phonons: The Theory of Transport Phenomena in Solids},
  author={Ziman, J.M.},
  isbn={9780198507796},
  lccn={2001268325},
  series={International series of monographs on physics},
  url={https://books.google.it/books?id=UtEy63pjngsC},
  year={2001},
  publisher={OUP Oxford}
}

@article{RevModPhys.73.515,
  title = {Phonons and related crystal properties from density-functional perturbation theory},
  author = {Baroni, Stefano and de Gironcoli, Stefano and Dal Corso, Andrea and Giannozzi, Paolo},
  journal = {Rev. Mod. Phys.},
  volume = {73},
  issue = {2},
  pages = {515--562},
  numpages = {0},
  year = {2001},
  month = {Jul},
  publisher = {American Physical Society},
  doi = {10.1103/RevModPhys.73.515},
  url = {https://link.aps.org/doi/10.1103/RevModPhys.73.515}
}

@article{PhysRevB.98.201201,
  title = {Magnetotransport phenomena in $p$-doped diamond from first principles},
  author = {Macheda, Francesco and Bonini, Nicola},
  journal = {Phys. Rev. B},
  volume = {98},
  issue = {20},
  pages = {201201},
  numpages = {5},
  year = {2018},
  month = {Nov},
  publisher = {American Physical Society},
  doi = {10.1103/PhysRevB.98.201201},
  url = {https://link.aps.org/doi/10.1103/PhysRevB.98.201201}
}

@Article{Macheda2020,
author={Macheda, Francesco
and Ponc{\'e}, Samuel
and Giustino, Feliciano
and Bonini, Nicola},
title={Theory and Computation of Hall Scattering Factor in Graphene},
journal={Nano Letters},
year={2020},
month={Dec},
day={09},
publisher={American Chemical Society},
volume={20},
number={12},
pages={8861-8865},
issn={1530-6984},
doi={10.1021/acs.nanolett.0c03874},
url={https://doi.org/10.1021/acs.nanolett.0c03874}
}

@article{PhysRevResearch.3.043022,
  title = {First-principles predictions of Hall and drift mobilities in semiconductors},
  author = {Ponc\'e, Samuel and Macheda, Francesco and Margine, Elena Roxana and Marzari, Nicola and Bonini, Nicola and Giustino, Feliciano},
  journal = {Phys. Rev. Research},
  volume = {3},
  issue = {4},
  pages = {043022},
  numpages = {36},
  year = {2021},
  month = {Oct},
  publisher = {American Physical Society},
  doi = {10.1103/PhysRevResearch.3.043022},
  url = {https://link.aps.org/doi/10.1103/PhysRevResearch.3.043022}
}

@Article{Ponce2020,
author={Ponc{\'e}, Samuel
and Li, Wenbin
and Reichardt, Sven
and Giustino, Feliciano},
title={First-principles calculations of charge carrier mobility and conductivity in bulk semiconductors and two-dimensional materials},
journal={Reports on Progress in Physics},
year={2020},
month={Feb},
day={04},
publisher={IOP Publishing},
volume={83},
number={3},
pages={036501},
issn={0034-4885},
doi={10.1088/1361-6633/ab6a43},
url={https://doi.org/10.1088/1361-6633/ab6a43}
}

@article{Pizzi2020,
	doi = {10.1088/1361-648x/ab51ff},
	url = {https://doi.org/10.1088%2F1361-648x%2Fab51ff},
	year = 2020,
	month = {jan},
	publisher = {{IOP} Publishing},
	volume = {32},
	number = {16},
	pages = {165902},
	author = {Giovanni Pizzi and Valerio Vitale and Ryotaro Arita and Stefan Blügel and Frank Freimuth and Guillaume G{\'{e}}ranton and Marco Gibertini and Dominik Gresch and Charles Johnson and Takashi Koretsune and Julen Iba{\~{n}}ez-Azpiroz and Hyungjun Lee and Jae-Mo Lihm and Daniel Marchand and Antimo Marrazzo and Yuriy Mokrousov and Jamal I Mustafa and Yoshiro Nohara and Yusuke Nomura and Lorenzo Paulatto and Samuel Ponc{\'{e}} and Thomas Ponweiser and Junfeng Qiao and Florian Thöle and Stepan S Tsirkin and Ma{\l}gorzata Wierzbowska and Nicola Marzari and David Vanderbilt and Ivo Souza and Arash A Mostofi and Jonathan R Yates},
	title = {Wannier90 as a community code: new features and applications},
	journal = {Journal of Physics: Condensed Matter}
}

@article{PONCE2016116,
title = {EPW: Electron–phonon coupling, transport and superconducting properties using maximally localized Wannier functions},
journal = {Computer Physics Communications},
volume = {209},
pages = {116-133},
year = {2016},
issn = {0010-4655},
doi = {https://doi.org/10.1016/j.cpc.2016.07.028},
url = {https://www.sciencedirect.com/science/article/pii/S0010465516302260},
author = {S. Poncé and E.R. Margine and C. Verdi and F. Giustino}
}

@article{PhysRevB.76.165108,
  title = {Electron-phonon interaction using Wannier functions},
  author = {Giustino, Feliciano and Cohen, Marvin L. and Louie, Steven G.},
  journal = {Phys. Rev. B},
  volume = {76},
  issue = {16},
  pages = {165108},
  numpages = {19},
  year = {2007},
  month = {Oct},
  publisher = {American Physical Society},
  doi = {10.1103/PhysRevB.76.165108},
  url = {https://link.aps.org/doi/10.1103/PhysRevB.76.165108}
}

@article{PhysRevB.1.910,
  title = {Microscopic Theory of Force Constants in the Adiabatic Approximation},
  author = {Pick, Robert M. and Cohen, Morrel H. and Martin, Richard M.},
  journal = {Phys. Rev. B},
  volume = {1},
  issue = {2},
  pages = {910--920},
  numpages = {0},
  year = {1970},
  month = {Jan},
  publisher = {American Physical Society},
  doi = {10.1103/PhysRevB.1.910},
  url = {https://link.aps.org/doi/10.1103/PhysRevB.1.910}
}

@article{doi:10.1119/1.1934059,
author = {Born,Max  and Huang,Kun },
title = {Dynamical Theory of Crystal Lattices},
journal = {American Journal of Physics},
volume = {23},
number = {7},
pages = {474-474},
year = {1955},
doi = {10.1119/1.1934059},
URL = {https://doi.org/10.1119/1.1934059},
eprint = {https://doi.org/10.1119/1.1934059}
}

@article{PhysRevB.13.694,
  title = {Microscopic theory of electron-phonon interaction in insulators or semiconductors},
  author = {Vogl, P.},
  journal = {Phys. Rev. B},
  volume = {13},
  issue = {2},
  pages = {694--704},
  numpages = {0},
  year = {1976},
  month = {Jan},
  publisher = {American Physical Society},
  doi = {10.1103/PhysRevB.13.694},
  url = {https://link.aps.org/doi/10.1103/PhysRevB.13.694}
}

@article{PhysRevResearch.2.033055,
  title = {First-principles study of electronic transport and structural properties of ${\mathrm{Cu}}_{12}{\mathrm{Sb}}_{4}{\mathrm{S}}_{13}$ in its high-temperature phase},
  author = {Di Paola, Cono and Macheda, Francesco and Laricchia, Savio and Weber, Cedric and Bonini, Nicola},
  journal = {Phys. Rev. Research},
  volume = {2},
  issue = {3},
  pages = {033055},
  numpages = {7},
  year = {2020},
  month = {Jul},
  publisher = {American Physical Society},
  doi = {10.1103/PhysRevResearch.2.033055},
  url = {https://link.aps.org/doi/10.1103/PhysRevResearch.2.033055}
}

@Article{Betz2013,
author={Betz, A. C.
and Jhang, S. H.
and Pallecchi, E.
and Ferreira, R.
and F{\`e}ve, G.
and Berroir, J.-M.
and Pla{\c{c}}ais, B.},
title={Supercollision cooling in undoped graphene},
journal={Nature Physics},
year={2013},
month={Feb},
day={01},
volume={9},
number={2},
pages={109-112},
issn={1745-2481},
doi={10.1038/nphys2494},
url={https://doi.org/10.1038/nphys2494}
}

@article{PhysRevB.84.075449,
  title = {Relaxation of optically excited carriers in graphene},
  author = {Kim, Raseong and Perebeinos, Vasili and Avouris, Phaedon},
  journal = {Phys. Rev. B},
  volume = {84},
  issue = {7},
  pages = {075449},
  numpages = {5},
  year = {2011},
  month = {Aug},
  publisher = {American Physical Society},
  doi = {10.1103/PhysRevB.84.075449},
  url = {https://link.aps.org/doi/10.1103/PhysRevB.84.075449}
}

@article{PhysRevB.103.134304,
  title = {First-principles theory of infrared vibrational spectroscopy of metals and semimetals: Application to graphite},
  author = {Binci, Luca and Barone, Paolo and Mauri, Francesco},
  journal = {Phys. Rev. B},
  volume = {103},
  issue = {13},
  pages = {134304},
  numpages = {10},
  year = {2021},
  month = {Apr},
  publisher = {American Physical Society},
  doi = {10.1103/PhysRevB.103.134304},
  url = {https://link.aps.org/doi/10.1103/PhysRevB.103.134304}
}

@article{PhysRevB.97.045201,
  title = {First-principles study of electron and hole mobilities of Si and GaAs},
  author = {Ma, Jinlong and Nissimagoudar, Arun S. and Li, Wu},
  journal = {Phys. Rev. B},
  volume = {97},
  issue = {4},
  pages = {045201},
  numpages = {9},
  year = {2018},
  month = {Jan},
  publisher = {American Physical Society},
  doi = {10.1103/PhysRevB.97.045201},
  url = {https://link.aps.org/doi/10.1103/PhysRevB.97.045201}
}

@article{PhysRevB.94.201201,
  title = {Ab initio electron mobility and polar phonon scattering in GaAs},
  author = {Zhou, Jin-Jian and Bernardi, Marco},
  journal = {Phys. Rev. B},
  volume = {94},
  issue = {20},
  pages = {201201},
  numpages = {6},
  year = {2016},
  month = {Nov},
  publisher = {American Physical Society},
  doi = {10.1103/PhysRevB.94.201201},
  url = {https://link.aps.org/doi/10.1103/PhysRevB.94.201201}
}

@article{PhysRevB.102.094308,
  title = {Phonon-limited electron mobility in Si, GaAs, and GaP with exact treatment of dynamical quadrupoles},
  author = {Brunin, Guillaume and Miranda, Henrique Pereira Coutada and Giantomassi, Matteo and Royo, Miquel and Stengel, Massimiliano and Verstraete, Matthieu J. and Gonze, Xavier and Rignanese, Gian-Marco and Hautier, Geoffroy},
  journal = {Phys. Rev. B},
  volume = {102},
  issue = {9},
  pages = {094308},
  numpages = {16},
  year = {2020},
  month = {Sep},
  publisher = {American Physical Society},
  doi = {10.1103/PhysRevB.102.094308},
  url = {https://link.aps.org/doi/10.1103/PhysRevB.102.094308}
}

@Article{Lee2020,
author={Lee, Nien-En
and Zhou, Jin-Jian
and Chen, Hsiao-Yi
and Bernardi, Marco},
title={Ab initio electron-two-phonon scattering in GaAs from next-to-leading order perturbation theory},
journal={Nature Communications},
year={2020},
month={Mar},
day={30},
volume={11},
number={1},
pages={1607},
issn={2041-1723},
doi={10.1038/s41467-020-15339-0},
url={https://doi.org/10.1038/s41467-020-15339-0}
}

@article{PhysRevB.100.085204,
  title = {Hole mobility of strained GaN from first principles},
  author = {Ponc\'e, Samuel and Jena, Debdeep and Giustino, Feliciano},
  journal = {Phys. Rev. B},
  volume = {100},
  issue = {8},
  pages = {085204},
  numpages = {16},
  year = {2019},
  month = {Aug},
  publisher = {American Physical Society},
  doi = {10.1103/PhysRevB.100.085204},
  url = {https://link.aps.org/doi/10.1103/PhysRevB.100.085204}
}

@article{PhysRev.127.1391,
  title = {Self-Consistent Approximations in Many-Body Systems},
  author = {Baym, Gordon},
  journal = {Phys. Rev.},
  volume = {127},
  issue = {4},
  pages = {1391--1401},
  year = {1962},
  month = {Aug},
  publisher = {American Physical Society},
  doi = {10.1103/PhysRev.127.1391},
  url = {https://link.aps.org/doi/10.1103/PhysRev.127.1391}
}

@article{PhysRev.118.1417,
  title = {Ground-State Energy of a Many-Fermion System. II},
  author = {Luttinger, J. M. and Ward, J. C.},
  journal = {Phys. Rev.},
  volume = {118},
  issue = {5},
  pages = {1417--1427},
  year = {1960},
  month = {Jun},
  publisher = {American Physical Society},
  doi = {10.1103/PhysRev.118.1417},
  url = {https://link.aps.org/doi/10.1103/PhysRev.118.1417}
}

@article{PhysRevB.62.4927,
  title = {Electron-hole excitations and optical spectra from first principles},
  author = {Rohlfing, Michael and Louie, Steven G.},
  journal = {Phys. Rev. B},
  volume = {62},
  issue = {8},
  pages = {4927--4944},
  year = {2000},
  month = {Aug},
  publisher = {American Physical Society},
  doi = {10.1103/PhysRevB.62.4927},
  url = {https://link.aps.org/doi/10.1103/PhysRevB.62.4927}
}

@article{Migdal1958,
  title = {Interaction between Electrons and Lattice Vibrations in a Normal Metal},
  author = {Migdal, A. B.},
  journal = {Soviet Physics JETP},
  volume = {7},
  number = {6},
  pages = {996--1001},
  year = {1958},
  url = {https://jetp.ras.ru/cgi-bin/e/index/e/7/6/p996?a=list}
}

@article{Eliashberg1960,
  title = {Interactions between Electrons and Lattice Vibrations in a Superconductor},
  author = {Eliashberg, G. M.},
  journal = {Soviet Physics JETP},
  volume = {11},
  number = {3},
  pages = {696--702},
  year = {1960},
  url = {https://www.jetp.ras.ru/cgi-bin/e/index/e/11/3/p696?a=list}
}

@article{PhysRevB.6.2577,
  title = {Neutron Spectroscopy of Superconductors},
  author = {Allen, Philip B.},
  journal = {Phys. Rev. B},
  volume = {6},
  issue = {7},
  pages = {2577--2579},
  year = {1972},
  month = {Oct},
  publisher = {American Physical Society},
  doi = {10.1103/PhysRevB.6.2577},
  url = {https://link.aps.org/doi/10.1103/PhysRevB.6.2577}
}

@article{PhysRevB.12.905,
  title = {Transition temperature of strong-coupled superconductors reanalyzed},
  author = {Allen, P. B. and Dynes, R. C.},
  journal = {Phys. Rev. B},
  volume = {12},
  issue = {3},
  pages = {905--922},
  year = {1975},
  month = {Aug},
  publisher = {American Physical Society},
  doi = {10.1103/PhysRevB.12.905},
  url = {https://link.aps.org/doi/10.1103/PhysRevB.12.905}
}

@article{PhysRevLett.97.266407,
  title = {Nonadiabatic Kohn Anomaly in a Doped Graphene Monolayer},
  author = {Lazzeri, Michele and Mauri, Francesco},
  journal = {Phys. Rev. Lett.},
  volume = {97},
  issue = {26},
  pages = {266407},
  year = {2006},
  month = {Dec},
  publisher = {American Physical Society},
  doi = {10.1103/PhysRevLett.97.266407},
  url = {https://link.aps.org/doi/10.1103/PhysRevLett.97.266407}
}

@article{PhysRevLett.99.086804,
  title = {Velocity Renormalization and Carrier Lifetime in Graphene from the Electron-Phonon Interaction},
  author = {Park, Cheol-Hwan and Giustino, Feliciano and Cohen, Marvin L. and Louie, Steven G.},
  journal = {Phys. Rev. Lett.},
  volume = {99},
  issue = {8},
  pages = {086804},
  year = {2007},
  month = {Aug},
  publisher = {American Physical Society},
  doi = {10.1103/PhysRevLett.99.086804},
  url = {https://link.aps.org/doi/10.1103/PhysRevLett.99.086804}
}

@article{Ferrari2013,
  title = {Raman spectroscopy as a versatile tool for studying the properties of graphene},
  author = {Ferrari, Andrea C. and Basko, Denis M.},
  journal = {Nature Nanotechnology},
  volume = {8},
  pages = {235--246},
  year = {2013},
  doi = {10.1038/nnano.2013.46},
  url = {https://doi.org/10.1038/nnano.2013.46}
}

@article{RevModPhys.84.1067,
  title = {Electron-Electron Interactions in Graphene: Current Status and Perspectives},
  author = {Kotov, Valeri N. and Uchoa, Bruno and Pereira, Vitor M. and Guinea, F. and Castro Neto, A. H.},
  journal = {Rev. Mod. Phys.},
  volume = {84},
  issue = {3},
  pages = {1067--1125},
  year = {2012},
  month = {Jul},
  publisher = {American Physical Society},
  doi = {10.1103/RevModPhys.84.1067},
  url = {https://link.aps.org/doi/10.1103/RevModPhys.84.1067}
}

@article{Gonzalez1994,
  title = {Non-Fermi liquid behavior of electrons in the half-filled honeycomb lattice},
  author = {Gonz{\'a}lez, J. and Guinea, F. and Vozmediano, M. A. H.},
  journal = {Nuclear Physics B},
  volume = {424},
  number = {3},
  pages = {595--618},
  year = {1994},
  doi = {10.1016/0550-3213(94)90410-3}
}

@article{Elias2011,
  title = {Dirac cones reshaped by interaction effects in suspended graphene},
  author = {Elias, D. C. and Gorbachev, R. V. and Mayorov, A. S. and Morozov, S. V. and Zhukov, A. A. and Blake, P. and Ponomarenko, L. A. and Grigorieva, I. V. and Novoselov, K. S. and Guinea, F. and Geim, A. K.},
  journal = {Nature Physics},
  volume = {7},
  pages = {701--704},
  year = {2011},
  doi = {10.1038/nphys2049},
  url = {https://doi.org/10.1038/nphys2049}
}

@article{PhysRevLett.122.186402,
  title = {Electron-Phonon Coupling from \textit{Ab Initio} Linear-Response Theory within the \(GW\) Method: Correlation-Enhanced Interactions and Superconductivity in \(\mathrm{Ba}_{1-x}\mathrm{K}_x\mathrm{BiO}_3\)},
  author = {Li, Zhenglu and Antonius, Gabriel and Wu, Meng and da Jornada, Felipe H. and Louie, Steven G.},
  journal = {Phys. Rev. Lett.},
  volume = {122},
  issue = {18},
  pages = {186402},
  year = {2019},
  month = {May},
  publisher = {American Physical Society},
  doi = {10.1103/PhysRevLett.122.186402},
  url = {https://link.aps.org/doi/10.1103/PhysRevLett.122.186402}
}

@article{Bernardi2016,
  title = {First-principles dynamics of electrons and phonons},
  author = {Bernardi, Marco},
  journal = {The European Physical Journal B},
  volume = {89},
  pages = {239},
  year = {2016},
  doi = {10.1140/epjb/e2016-70399-4},
  url = {https://doi.org/10.1140/epjb/e2016-70399-4}
}

@article{PhysRevLett.125.107401,
  title = {Exciton-Phonon Interaction and Relaxation Times from First Principles},
  author = {Chen, Hsiao-Yi and Sangalli, Davide and Bernardi, Marco},
  journal = {Phys. Rev. Lett.},
  volume = {125},
  issue = {10},
  pages = {107401},
  year = {2020},
  month = {Sep},
  publisher = {American Physical Society},
  doi = {10.1103/PhysRevLett.125.107401},
  url = {https://link.aps.org/doi/10.1103/PhysRevLett.125.107401}
}

@article{Li2024GWPT,
  title = {Electron-phonon coupling from \(GW\) perturbation theory: Practical workflow combining BerkeleyGW, ABINIT, and EPW},
  author = {Li, Zhenglu and Antonius, Gabriel and Chan, Yang-Hao and Louie, Steven G.},
  journal = {Computer Physics Communications},
  volume = {295},
  pages = {109003},
  year = {2024},
  doi = {10.1016/j.cpc.2023.109003}
}

@article{PhysRevB.105.085111,
  title = {Theory of Exciton-Phonon Coupling},
  author = {Antonius, Gabriel and Louie, Steven G.},
  journal = {Phys. Rev. B},
  volume = {105},
  issue = {8},
  pages = {085111},
  numpages = {14},
  year = {2022},
  month = {Feb},
  publisher = {American Physical Society},
  doi = {10.1103/PhysRevB.105.085111},
  url = {https://link.aps.org/doi/10.1103/PhysRevB.105.085111}
}

@article{PhysRevB.84.085406,
  title = {Dielectric screening in two-dimensional insulators: Implications for excitonic and impurity states in graphane},
  author = {Cudazzo, Pierluigi and Tokatly, Ilya V. and Rubio, Angel},
  journal = {Phys. Rev. B},
  volume = {84},
  issue = {8},
  pages = {085406},
  numpages = {11},
  year = {2011},
  month = {Aug},
  publisher = {American Physical Society},
  doi = {10.1103/PhysRevB.84.085406},
  url = {https://link.aps.org/doi/10.1103/PhysRevB.84.085406}
}

@article{PhysRevB.88.045318,
  title = {Theory of neutral and charged excitons in monolayer transition metal dichalcogenides},
  author = {Berkelbach, Timothy C. and Hybertsen, Mark S. and Reichman, David R.},
  journal = {Phys. Rev. B},
  volume = {88},
  issue = {4},
  pages = {045318},
  numpages = {6},
  year = {2013},
  month = {Jul},
  publisher = {American Physical Society},
  doi = {10.1103/PhysRevB.88.045318},
  url = {https://link.aps.org/doi/10.1103/PhysRevB.88.045318}
}

@article{PhysRevLett.113.076802,
  title = {Exciton Binding Energy and Nonhydrogenic Rydberg Series in Monolayer \(\mathrm{WS}_2\)},
  author = {Chernikov, Alexey and Berkelbach, Timothy C. and Hill, Heather M. and Rigosi, Albert and Li, Yilei and Aslan, Ozgur Burak and Reichman, David R. and Hybertsen, Mark S. and Heinz, Tony F.},
  journal = {Phys. Rev. Lett.},
  volume = {113},
  issue = {7},
  pages = {076802},
  numpages = {5},
  year = {2014},
  month = {Aug},
  publisher = {American Physical Society},
  doi = {10.1103/PhysRevLett.113.076802},
  url = {https://link.aps.org/doi/10.1103/PhysRevLett.113.076802}
}

@article{PhysRevB.92.245123,
  title = {Excitons in van der Waals heterostructures: The important role of dielectric screening},
  author = {Latini, Simone and Olsen, Thomas and Thygesen, Kristian S.},
  journal = {Phys. Rev. B},
  volume = {92},
  issue = {24},
  pages = {245123},
  numpages = {13},
  year = {2015},
  month = {Dec},
  publisher = {American Physical Society},
  doi = {10.1103/PhysRevB.92.245123},
  url = {https://link.aps.org/doi/10.1103/PhysRevB.92.245123}
}

@article{PhysRevLett.111.216805,
  title = {Optical Spectrum of \(\mathrm{MoS}_2\): Many-Body Effects and Diversity of Exciton States},
  author = {Qiu, Diana Y. and da Jornada, Felipe H. and Louie, Steven G.},
  journal = {Phys. Rev. Lett.},
  volume = {111},
  issue = {21},
  pages = {216805},
  numpages = {5},
  year = {2013},
  month = {Nov},
  publisher = {American Physical Society},
  doi = {10.1103/PhysRevLett.111.216805},
  url = {https://link.aps.org/doi/10.1103/PhysRevLett.111.216805}
}

@article{PhysRevB.93.235435,
  title = {Screening and many-body effects in two-dimensional crystals: Monolayer \(\mathrm{MoS}_2\)},
  author = {Qiu, Diana Y. and da Jornada, Felipe H. and Louie, Steven G.},
  journal = {Phys. Rev. B},
  volume = {93},
  issue = {23},
  pages = {235435},
  numpages = {13},
  year = {2016},
  month = {Jun},
  publisher = {American Physical Society},
  doi = {10.1103/PhysRevB.93.235435},
  url = {https://link.aps.org/doi/10.1103/PhysRevB.93.235435}
}

@article{PhysRevB.78.081406,
  title = {Impact of the electron-electron correlation on phonon dispersion: Failure of LDA and GGA DFT functionals in graphene and graphite},
  author = {Lazzeri, Michele and Attaccalite, Claudio and Wirtz, Ludger and Mauri, Francesco},
  journal = {Phys. Rev. B},
  volume = {78},
  issue = {8},
  pages = {081406(R)},
  numpages = {4},
  year = {2008},
  month = {Aug},
  publisher = {American Physical Society},
  doi = {10.1103/PhysRevB.78.081406},
  url = {https://link.aps.org/doi/10.1103/PhysRevB.78.081406}
}

@Article{Graziotto2024,
author={Graziotto, Lorenzo
and Macheda, Francesco
and Venanzi, Tommaso
and Marchese, Guglielmo
and Sotgiu, Simone
and Ouaj, Taoufiq
and Stellino, Elena
and Fasolato, Claudia
and Postorino, Paolo
and Metzelaars, Marvin
and K{\"o}gerler, Paul
and Beschoten, Bernd
and Calandra, Matteo
and Ortolani, Michele
and Stampfer, Christoph
and Mauri, Francesco
and Baldassarre, Leonetta},
title={Infrared Resonance Raman of Bilayer Graphene: Signatures of Massive Fermions and Band Structure on the 2D Peak},
journal={Nano Letters},
year={2024},
month={Feb},
day={14},
publisher={American Chemical Society},
volume={24},
number={6},
pages={1867-1873},
issn={1530-6984},
doi={10.1021/acs.nanolett.3c03502},
url={https://doi.org/10.1021/acs.nanolett.3c03502}
}

@misc{macheda2026coupleddynamicalboltzmanntransport,
      title={Coupled dynamical Boltzmann transport equations with long-range electron-phonon and electron-electron interactions in 2D materials}, 
      author={Francesco Macheda and Thibault Sohier},
      year={2026},
      eprint={2604.01746},
      archivePrefix={arXiv},
      primaryClass={cond-mat.mes-hall},
      url={https://arxiv.org/abs/2604.01746}, 
}

@misc{dinar2026electronictransportbnencasulatedgraphene,
      title={Electronic transport in BN-encasulated graphene limited by remote phonon scattering}, 
      author={Khalid Dinar and Francesco Macheda and Alberto Guandalini and Matthieu Paillet and Christophe Consejo and Frederic Teppe and Benoit Jouault and Thibault Sohier and Sébastien Nanot},
      year={2026},
      eprint={2604.00678},
      archivePrefix={arXiv},
      primaryClass={cond-mat.mes-hall},
      url={https://arxiv.org/abs/2604.00678}, 
}

@misc{Laricchia2025QSGW,
      title={Electron-Phonon Coupling in Many-Body Perturbation Theory: Developments within the Quasiparticle Self-Consistent GW approximation and LMTO Formalism}, 
      author={Savio Laricchia and Casey Eichstaedt and Dimitar Pashov and Mark van Schilfgaarde},
      year={2025},
      eprint={2404.02902},
      archivePrefix={arXiv},
      primaryClass={cond-mat.mtrl-sci},
      url={https://arxiv.org/abs/2404.02902}, 
}

@article{PhysRevB.110.L201405,
  title = {Nearly quantized Born effective charges as probes for the topological phase transition in the Haldane and Kane-Mele models},
  author = {Fachin, Paolo and Macheda, Francesco and Barone, Paolo and Mauri, Francesco},
  journal = {Phys. Rev. B},
  volume = {110},
  issue = {20},
  pages = {L201405},
  numpages = {7},
  year = {2024},
  month = {Nov},
  publisher = {American Physical Society},
  doi = {10.1103/PhysRevB.110.L201405},
  url = {https://link.aps.org/doi/10.1103/PhysRevB.110.L201405}
}

@article{
doi:10.1073/pnas.1100242108,
author = {David A. Siegel  and Cheol-Hwan Park  and Choongyu Hwang  and Jack Deslippe  and Alexei V. Fedorov  and Steven G. Louie  and Alessandra Lanzara },
title = {Many-body interactions in quasi-freestanding graphene},
journal = {Proceedings of the National Academy of Sciences},
volume = {108},
number = {28},
pages = {11365-11369},
year = {2011},
doi = {10.1073/pnas.1100242108},
URL = {https://www.pnas.org/doi/abs/10.1073/pnas.1100242108},
eprint = {https://www.pnas.org/doi/pdf/10.1073/pnas.1100242108}}

@article{PhysRevB.76.205411,
  title = {Electron-phonon coupling and electron self-energy in electron-doped graphene: Calculation of angular-resolved photoemission spectra},
  author = {Calandra, Matteo and Mauri, Francesco},
  journal = {Phys. Rev. B},
  volume = {76},
  issue = {20},
  pages = {205411},
  numpages = {9},
  year = {2007},
  month = {Nov},
  publisher = {American Physical Society},
  doi = {10.1103/PhysRevB.76.205411},
  url = {https://link.aps.org/doi/10.1103/PhysRevB.76.205411}
}

@article{PhysRevB.81.041403,
  title = {Electron-phonon coupling in potassium-doped graphene: Angle-resolved photoemission spectroscopy},
  author = {Bianchi, M. and Rienks, E. D. L. and Lizzit, S. and Baraldi, A. and Balog, R. and Hornek\ae{}r, L. and Hofmann, Ph.},
  journal = {Phys. Rev. B},
  volume = {81},
  issue = {4},
  pages = {041403(R)},
  numpages = {4},
  year = {2010},
  month = {Jan},
  publisher = {American Physical Society},
  doi = {10.1103/PhysRevB.81.041403},
  url = {https://link.aps.org/doi/10.1103/PhysRevB.81.041403}
}

@article{PhysRevD.10.2428,
  title = {Effective action for composite operators},
  author = {Cornwall, John M. and Jackiw, R. and Tomboulis, E.},
  journal = {Phys. Rev. D},
  volume = {10},
  issue = {8},
  pages = {2428--2445},
  numpages = {0},
  year = {1974},
  month = {Oct},
  publisher = {American Physical Society},
  doi = {10.1103/PhysRevD.10.2428},
  url = {https://link.aps.org/doi/10.1103/PhysRevD.10.2428}
}

@article{PhysRevB.106.075119,
  title = {Ab initio self-consistent many-body theory of polarons at all couplings},
  author = {Lafuente-Bartolome, Jon and Lian, Chao and Sio, Weng Hong and Gurtubay, Idoia G. and Eiguren, Asier and Giustino, Feliciano},
  journal = {Phys. Rev. B},
  volume = {106},
  issue = {7},
  pages = {075119},
  numpages = {20},
  year = {2022},
  month = {Aug},
  publisher = {American Physical Society},
  doi = {10.1103/PhysRevB.106.075119},
  url = {https://link.aps.org/doi/10.1103/PhysRevB.106.075119}
}

@book{NegeleOrland1988,
  author    = {Negele, John W. and Orland, Henri},
  title     = {Quantum Many-Particle Systems},
  publisher = {Addison-Wesley},
  year      = {1988}
}

@article{PhysRevB.111.144309,
  title = {First-principles equation for coherent phonons: Dynamics and polaron distortions},
  author = {Stefanucci, Gianluca and Perfetto, Enrico},
  journal = {Phys. Rev. B},
  volume = {111},
  issue = {14},
  pages = {144309},
  numpages = {11},
  year = {2025},
  month = {Apr},
  publisher = {American Physical Society},
  doi = {10.1103/PhysRevB.111.144309},
  url = {https://link.aps.org/doi/10.1103/PhysRevB.111.144309}
}

@article{FAryasetiawan_1998,
doi = {10.1088/0034-4885/61/3/002},
url = {https://doi.org/10.1088/0034-4885/61/3/002},
year = {1998},
month = {mar},
publisher = {},
volume = {61},
number = {3},
pages = {237},
author = {F Aryasetiawan and O Gunnarsson},
title = {The  GW method},
journal = {Reports on Progress in Physics}
}

@article{LarsHedin_1999,
doi = {10.1088/0953-8984/11/42/201},
url = {https://doi.org/10.1088/0953-8984/11/42/201},
year = {1999},
month = {oct},
publisher = {},
volume = {11},
number = {42},
pages = {R489},
author = {Lars Hedin},
title = {On 
correlation effects in electron spectroscopies and 
the GW approximation},
journal = {Journal of Physics: Condensed Matter}
}

@article{RevModPhys.74.601,
  title = {Electronic excitations: density-functional versus many-body Green's-function approaches},
  author = {Onida, Giovanni and Reining, Lucia and Rubio, Angel},
  journal = {Rev. Mod. Phys.},
  volume = {74},
  issue = {2},
  pages = {601--659},
  numpages = {0},
  year = {2002},
  month = {Jun},
  publisher = {American Physical Society},
  doi = {10.1103/RevModPhys.74.601},
  url = {https://link.aps.org/doi/10.1103/RevModPhys.74.601}
}

@article{https://doi.org/10.1002/wcms.1344,
author = {Reining, Lucia},
title = {The GW approximation: content, successes and limitations},
journal = {WIREs Computational Molecular Science},
volume = {8},
number = {3},
pages = {e1344},
doi = {https://doi.org/10.1002/wcms.1344},
url = {https://wires.onlinelibrary.wiley.com/doi/abs/10.1002/wcms.1344},
eprint = {https://wires.onlinelibrary.wiley.com/doi/pdf/10.1002/wcms.1344},
year = {2018}
}

@article{83vn-7zyw,
  title = {First-Principles Predictions of Carrier Mobility with Record Accuracy Using GW Perturbation Theory},
  author = {Pant, Nick and Tiwari, Sabyasachi and Louie, Steven G. and Li, Zhenglu and Giustino, Feliciano},
  journal = {Phys. Rev. Lett.},
  volume = {137},
  issue = {5},
  pages = {056303},
  numpages = {7},
  year = {2026},
  month = {Jul},
  publisher = {American Physical Society},
  doi = {10.1103/83vn-7zyw},
  url = {https://link.aps.org/doi/10.1103/83vn-7zyw}
}

@article{PerfettoWuStefanucci2024, author = {Perfetto, Enrico and Wu, Kai and Stefanucci, Gianluca}, title = {Theory of coherent phonons coupled to excitons}, journal = {npj 2D Materials and Applications}, volume = {8}, pages = {40}, year = {2024}, doi = {10.1038/s41699-024-00474-9} }

@article{StefanucciPerfetto2026, author = {Stefanucci, Gianluca and Perfetto, Enrico}, title = {Unified First-Principles Formula for Time-Resolved ARPES Spectra of Coherent and Incoherent Excitons beyond the Dilute Limit}, journal = {Phys. Rev. Lett.}, volume = {137}, pages = {026402}, year = {2026}, doi = {10.1103/zhs1-rv1k} }

@Article{Lee2023,
author={Lee, Hyungjun
and Ponc{\'e}, Samuel
and Bushick, Kyle
and Hajinazar, Samad
and Lafuente-Bartolome, Jon
and Leveillee, Joshua
and Lian, Chao
and Lihm, Jae-Mo
and Macheda, Francesco
and Mori, Hitoshi
and Paudyal, Hari
and Sio, Weng Hong
and Tiwari, Sabyasachi
and Zacharias, Marios
and Zhang, Xiao
and Bonini, Nicola
and Kioupakis, Emmanouil
and Margine, Elena R.
and Giustino, Feliciano},
title={Electron--phonon physics from first principles using the EPW code},
journal={npj Computational Materials},
year={2023},
month={Aug},
day={25},
volume={9},
number={1},
pages={156},
issn={2057-3960},
doi={10.1038/s41524-023-01107-3},
url={https://doi.org/10.1038/s41524-023-01107-3}
}

@article{MARINI2024108950, title = {epiq: An open-source software for the calculation of electron-phonon interaction related properties}, journal = {Computer Physics Communications}, volume = {295}, pages = {108950}, year = {2024}, issn = {0010-4655}, doi = {https://doi.org/10.1016/j.cpc.2023.108950}, url = {https://www.sciencedirect.com/science/article/pii/S0010465523002953}, author = {Giovanni Marini and Guglielmo Marchese and Gianni Profeta and Jelena Sjakste and Francesco Macheda and Nathalie Vast and Francesco Mauri and Matteo Calandra}, }

@article{ZHOU2021107970,
title = {Perturbo: A software package for ab initio electron–phonon interactions, charge transport and ultrafast dynamics},
journal = {Computer Physics Communications},
volume = {264},
pages = {107970},
year = {2021},
issn = {0010-4655},
doi = {https://doi.org/10.1016/j.cpc.2021.107970},
url = {https://www.sciencedirect.com/science/article/pii/S0010465521000837},
author = {Jin-Jian Zhou and Jinsoo Park and I-Te Lu and Ivan Maliyov and Xiao Tong and Marco Bernardi}
}

@article{Antonius2014ZPR,
  author  = {Antonius, Gabriel and Ponc{\'e}, Samuel and Boulanger, Pierre
             and C{\^o}t{\'e}, Michel and Gonze, Xavier},
  title   = {Many-Body Effects on the Zero-Point Renormalization
             of the Band Structure},
  journal = {Phys. Rev. Lett.},
  volume  = {112},
  pages   = {215501},
  year    = {2014},
  doi     = {10.1103/PhysRevLett.112.215501}
}

@article{Monserrat2016Correlation,
  author  = {Monserrat, Bartomeu},
  title   = {Correlation Effects on Electron-Phonon Coupling in
             Semiconductors: Many-Body Theory along Thermal Lines},
  journal = {Phys. Rev. B},
  volume  = {93},
  pages   = {100301},
  year    = {2016},
  doi     = {10.1103/PhysRevB.93.100301}
}

@article{PhysRevLett.98.067005,
  title = {Electron-Phonon Renormalization in Cuprate Superconductors},
  author = {Zhang, Peihong and Louie, Steven G. and Cohen, Marvin L.},
  journal = {Phys. Rev. Lett.},
  volume = {98},
  issue = {6},
  pages = {067005},
  numpages = {4},
  year = {2007},
  month = {Feb},
  publisher = {American Physical Society},
  doi = {10.1103/PhysRevLett.98.067005},
  url = {https://link.aps.org/doi/10.1103/PhysRevLett.98.067005}
}

@article{PhysRevLett.105.265501,
  title = {Electron-Phonon Renormalization of the Direct Band Gap of Diamond},
  author = {Giustino, Feliciano and Louie, Steven G. and Cohen, Marvin L.},
  journal = {Phys. Rev. Lett.},
  volume = {105},
  issue = {26},
  pages = {265501},
  numpages = {4},
  year = {2010},
  month = {Dec},
  publisher = {American Physical Society},
  doi = {10.1103/PhysRevLett.105.265501},
  url = {https://link.aps.org/doi/10.1103/PhysRevLett.105.265501}
}

@article{6m5p-mmg3,
  title = {Large Many-Electron Effects in the Temperature-Dependent Electron-Phonon Renormalization of Semiconductor Band Gaps},
  author = {Gong, Xiaoxun and Li, Zhenglu and Louie, Steven G.},
  journal = {Phys. Rev. Lett.},
  volume = {136},
  issue = {21},
  pages = {216401},
  numpages = {9},
  year = {2026},
  month = {May},
  publisher = {American Physical Society},
  doi = {10.1103/6m5p-mmg3},
  url = {https://link.aps.org/doi/10.1103/6m5p-mmg3}
}

@article{PhysRevB.76.045430,
  title = {Symmetry-based approach to electron-phonon interactions in graphene},
  author = {Ma\~nes, J. L.},
  journal = {Phys. Rev. B},
  volume = {76},
  issue = {4},
  pages = {045430},
  numpages = {10},
  year = {2007},
  month = {Jul},
  publisher = {American Physical Society},
  doi = {10.1103/PhysRevB.76.045430},
  url = {https://link.aps.org/doi/10.1103/PhysRevB.76.045430}
}

@article{Bistoni_2019,
doi = {10.1088/2053-1583/ab2ce0},
url = {https://doi.org/10.1088/2053-1583/ab2ce0},
year = {2019},
month = {jul},
publisher = {IOP Publishing},
volume = {6},
number = {4},
pages = {045015},
author = {Bistoni, O and Barone, P and Cappelluti, E and Benfatto, L and Mauri, F},
title = {Giant effective charges and piezoelectricity in gapped graphene},
journal = {2D Materials}
}

@Article{Rostami2018,
author={Rostami, Habib
and Guinea, Francisco
and Polini, Marco
and Rold{\'a}n, Rafael},
title={Piezoelectricity and valley chern number in inhomogeneous hexagonal 2D crystals},
journal={npj 2D Materials and Applications},
year={2018},
month={May},
day={31},
volume={2},
number={1},
pages={15},
issn={2397-7132},
doi={10.1038/s41699-018-0061-7},
url={https://doi.org/10.1038/s41699-018-0061-7}
}

@article{Bostwick2007,
  author  = {Bostwick, Aaron and Ohta, Taisuke and Seyller, Thomas and Horn, Karsten and Rotenberg, Eli},
  title   = {Quasiparticle dynamics in graphene},
  journal = {Nature Physics},
  volume  = {3},
  pages   = {36--40},
  year    = {2007},
  doi     = {10.1038/nphys477}
}

@article{PhysRevLett.101.226405,
  author  = {Trevisanutto, Paolo E. and Giorgetti, Christine and Reining, Lucia and Ladisa, Massimo and Olevano, Valerio},
  title   = {Ab Initio {$GW$} Many-Body Effects in Graphene},
  journal = {Physical Review Letters},
  volume  = {101},
  number  = {22},
  pages   = {226405},
  year    = {2008},
  doi     = {10.1103/PhysRevLett.101.226405}
}

@article{PhysRevLett.109.116802,
  author  = {Chae, Jungseok and Jung, Suyong and Young, Andrea F. and Dean, Cory R. and Wang, Lei and Gao, Yuanda and Watanabe, Kenji and Taniguchi, Takashi and Hone, James and Shepard, Kenneth L. and Kim, Philip and Zhitenev, Nikolai B. and Stroscio, Joseph A.},
  title   = {Renormalization of the Graphene Dispersion Velocity Determined from Scanning Tunneling Spectroscopy},
  journal = {Physical Review Letters},
  volume  = {109},
  number  = {11},
  pages   = {116802},
  year    = {2012},
  doi     = {10.1103/PhysRevLett.109.116802}
}

@article{PhysRevB.87.045425,
  author  = {Das Sarma, S. and Hwang, E. H.},
  title   = {Velocity renormalization and anomalous quasiparticle dispersion in extrinsic graphene},
  journal = {Physical Review B},
  volume  = {87},
  number  = {4},
  pages   = {045425},
  year    = {2013},
  doi     = {10.1103/PhysRevB.87.045425}
}

@article{PhysRevB.109.075120,
  author  = {Guandalini, Alberto and Leon, Dario A. and D'Amico, Pino and Cardoso, Claudia and Ferretti, Andrea and Rontani, Massimo and Varsano, Daniele},
  title   = {Efficient {$GW$} calculations via interpolation of the screened interaction in momentum and frequency space: The case of graphene},
  journal = {Physical Review B},
  volume  = {109},
  number  = {7},
  pages   = {075120},
  year    = {2024},
  doi     = {10.1103/PhysRevB.109.075120}
}

@article{PhysRevB.67.035401,
  author = {Dubay, O. and Kresse, G.},
  title = {Accurate density functional calculations for the phonon dispersion relations of graphite layer and carbon nanotubes},
  journal = {Phys. Rev. B},
  volume = {67},
  issue = {3},
  pages = {035401},
  year = {2003},
  doi = {10.1103/PhysRevB.67.035401}
}

@article{WirtzRubio2004,
  author = {Wirtz, Ludger and Rubio, Angel},
  title = {The phonon dispersion of graphite revisited},
  journal = {Solid State Commun.},
  volume = {131},
  number = {3-4},
  pages = {141--152},
  year = {2004},
  doi = {10.1016/j.ssc.2004.04.042}
}

@article{PhysRevLett.92.075501,
  author = {Maultzsch, J. and Reich, S. and Thomsen, C. and Requardt, H. and Ordej\'on, P.},
  title = {Phonon Dispersion in Graphite},
  journal = {Phys. Rev. Lett.},
  volume = {92},
  issue = {7},
  pages = {075501},
  year = {2004},
  doi = {10.1103/PhysRevLett.92.075501}
}

@article{PhysRevB.71.205214,
  author = {Mounet, Nicolas and Marzari, Nicola},
  title = {First-principles determination of the structural, vibrational and thermodynamic properties of diamond, graphite, and derivatives},
  journal = {Phys. Rev. B},
  volume = {71},
  issue = {20},
  pages = {205214},
  year = {2005},
  doi = {10.1103/PhysRevB.71.205214}
}

@article{PhysRevB.78.165421,
  author = {Saha, Srijan Kumar and Waghmare, U. V. and Krishnamurthy, H. R. and Sood, A. K.},
  title = {Phonons in few-layer graphene and interplanar interaction: A first-principles study},
  journal = {Phys. Rev. B},
  volume = {78},
  issue = {16},
  pages = {165421},
  year = {2008},
  doi = {10.1103/PhysRevB.78.165421}
}

@article{PhysRevLett.115.115502,
title = {Chiral Phonons at High-Symmetry Points in Monolayer Hexagonal Lattices},
author = {Zhang, Lifa and Niu, Qian},
journal = {Phys. Rev. Lett.},
volume = {115},
issue = {11},
pages = {115502},
numpages = {5},
year = {2015},
month = {9},
publisher = {American Physical Society},
doi = {10.1103/PhysRevLett.115.115502},
url = {https://link.aps.org/doi/10.1103/PhysRevLett.115.115502}
}

@article{PhysRevLett.112.085503,
title = {Angular Momentum of Phonons and the Einstein--de Haas Effect},
author = {Zhang, Lifa and Niu, Qian},
journal = {Phys. Rev. Lett.},
volume = {112},
issue = {8},
pages = {085503},
numpages = {5},
year = {2014},
month = {2},
publisher = {American Physical Society},
doi = {10.1103/PhysRevLett.112.085503},
url = {https://link.aps.org/doi/10.1103/PhysRevLett.112.085503}
}

@article{PhysRevX.14.011041,
title = {Adiabatic Dynamics of Coupled Spins and Phonons in Magnetic Insulators},
author = {Ren, Shang and Bonini, John and Stengel, Massimiliano and Dreyer, Cyrus E. and Vanderbilt, David},
journal = {Phys. Rev. X},
volume = {14},
issue = {1},
pages = {011041},
numpages = {30},
year = {2024},
month = {3},
publisher = {American Physical Society},
doi = {10.1103/PhysRevX.14.011041},
url = {https://link.aps.org/doi/10.1103/PhysRevX.14.011041}
}

@article{PhysRevLett.126.225703,
title = {Intrinsic Vibrational Angular Momentum from Nonadiabatic Effects in Noncollinear Magnetic Molecules},
author = {Bistoni, Oliviero and Mauri, Francesco and Calandra, Matteo},
journal = {Phys. Rev. Lett.},
volume = {126},
issue = {22},
pages = {225703},
numpages = {5},
year = {2021},
month = {6},
publisher = {American Physical Society},
doi = {10.1103/PhysRevLett.126.225703},
url = {https://link.aps.org/doi/10.1103/PhysRevLett.126.225703}
}

@article{Fachin2025,
author = {Fachin, Paolo and Macheda, Francesco and Barone, Paolo and Mauri, Francesco},
title = {Infrared markers of topological phase transitions in quantum spin Hall insulators},
journal = {npj Computational Materials},
year = {2025},
volume = {11},
number = {1},
pages = {307},
month = {10},
doi = {10.1038/s41524-025-01780-6},
issn = {2057-3960},
url = {https://doi.org/10.1038/s41524-025-01780-6}
}

@article{PhysRevLett.134.206701,
title = {Nonreciprocal Phonons in $\mathcal{P}\mathcal{T}$-Symmetric Antiferromagnets},
author = {Ren, Yafei and Saparov, Daniyar and Niu, Qian},
journal = {Phys. Rev. Lett.},
volume = {134},
issue = {20},
pages = {206701},
numpages = {7},
year = {2025},
month = {May},
publisher = {American Physical Society},
doi = {10.1103/PhysRevLett.134.206701},
url = {https://link.aps.org/doi/10.1103/PhysRevLett.134.206701}
}

@article{PhysRevB.106.125403,
  title = {Exciton-phonon interaction calls for a revision of the ``exciton'' concept},
  author = {Paleari, Fulvio and Marini, Andrea},
  journal = {Phys. Rev. B},
  volume = {106},
  issue = {12},
  pages = {125403},
  numpages = {16},
  year = {2022},
  month = {Sep},
  publisher = {American Physical Society},
  doi = {10.1103/PhysRevB.106.125403},
  url = {https://link.aps.org/doi/10.1103/PhysRevB.106.125403}
}

@article{PhysRevB.102.045136,
  title = {First-principles description of the exciton-phonon interaction: A cumulant approach},
  author = {Cudazzo, Pierluigi},
  journal = {Phys. Rev. B},
  volume = {102},
  issue = {4},
  pages = {045136},
  numpages = {11},
  year = {2020},
  month = {Jul},
  publisher = {American Physical Society},
  doi = {10.1103/PhysRevB.102.045136},
  url = {https://link.aps.org/doi/10.1103/PhysRevB.102.045136}
}

@article{BaymKadanoff1961,
  author    = {Baym, Gordon and Kadanoff, Leo P.},
  title     = {Conservation Laws and Correlation Functions},
  journal   = {Phys. Rev.},
  volume    = {124},
  issue     = {2},
  pages     = {287--299},
  year      = {1961},
  month     = {Oct},
  doi       = {10.1103/PhysRev.124.287},
  url       = {https://link.aps.org/doi/10.1103/PhysRev.124.287},
  publisher = {American Physical Society}
}

@article{Hyrkas2019,
  author  = {Hyrk{\"a}s, M. J. and Karlsson, D. and van Leeuwen, R.},
  title   = {Contour calculus for many-particle functions},
  journal = {Journal of Physics A: Mathematical and Theoretical},
  volume  = {52},
  pages   = {215303},
  year    = {2019},
  doi     = {10.1088/1751-8121/ab165d}
}

@article{guandalini_nanoletter,
author = {Guandalini, Alberto and Senga, Ryosuke and Lin, Yung-Chang and Suenaga, Kazu and Ferretti, Andrea and Varsano, Daniele and Recchia, Andrea and Barone, Paolo and Mauri, Francesco and Pichler, Thomas and Kramberger, Christian},
title = {Excitonic Effects in Energy-Loss Spectra of Freestanding Graphene},
journal = {Nano Lett.},
volume = {23},
number = {24},
pages = {11835-11841},
year = {2023},
doi = {10.1021/acs.nanolett.3c03863},
   
URL = { 
        https://doi.org/10.1021/acs.nanolett.3c03863
},
}

@book{kadanoff1976quantum,
  title={Quantum Statistical Mechanics: Green's Function Methods in Equilibrium and Nonequilibrium Problems},
  author={Kadanoff, L.P. and Baym, G.},
  series={Frontiers in physics},
  url={https://books.google.it/books?id=Zsm-0QEACAAJ},
  year={1976},
  publisher={W. A. Benjamin}
}

@article{PhysRevB.69.115110,
  title = {First-principles approach to the electron-phonon interaction},
  author = {van Leeuwen, Robert},
  journal = {Phys. Rev. B},
  volume = {69},
  issue = {11},
  pages = {115110},
  numpages = {20},
  year = {2004},
  month = {Mar},
  publisher = {American Physical Society},
  doi = {10.1103/PhysRevB.69.115110},
  url = {https://link.aps.org/doi/10.1103/PhysRevB.69.115110}
}

\clearpage
\onecolumngrid
\setcounter{section}{0}
\setcounter{subsection}{0}
\setcounter{equation}{0}
\setcounter{figure}{0}
\setcounter{table}{0}
\renewcommand{\thesection}{S\arabic{section}}
\renewcommand{\thesubsection}{S\arabic{section}.\arabic{subsection}}
\renewcommand{\theequation}{S\arabic{equation}}
\renewcommand{\thefigure}{S\arabic{figure}}
\renewcommand{\thetable}{S\arabic{table}}
\providecommand{\theHsection}{}
\providecommand{\theHsubsection}{}
\providecommand{\theHequation}{}
\providecommand{\theHfigure}{}
\providecommand{\theHtable}{}
\renewcommand{\theHsection}{SI.\arabic{section}}
\renewcommand{\theHsubsection}{SI.\arabic{section}.\arabic{subsection}}
\renewcommand{\theHequation}{SI.\arabic{equation}}
\renewcommand{\theHfigure}{SI.\arabic{figure}}
\renewcommand{\theHtable}{SI.\arabic{table}}
\begin{center}
  {\large\bfseries Supplementary Information to:\par}
  \vspace{0.35em}
  {\large\bfseries The principle of detailed balance between electrons and phonons
  in presence of excitonic effects\par}
  \vspace{1.2em}
  Alberto Guandalini,$^{1}$ Giovanni Caldarelli,$^{2}$ Francesco Mauri,$^{1}$
  and Francesco Macheda$^{3}$

  \vspace{0.6em}
  {\small
  $^{1}$Dipartimento di Fisica, Universit\`a di Roma La Sapienza, Roma, Italy\\
  $^{2}$Theory and Simulation of Materials (THEOS), and National Centre for
  Computational Design and Discovery of Novel Materials (MARVEL),
  \'Ecole Polytechnique F\'ed\'erale de Lausanne, Lausanne, Switzerland\\
  $^{3}$Dipartimento di Scienze e Metodi dell'Ingegneria,
  Universit\`a di Modena e Reggio Emilia, Reggio Emilia, Italy}
\end{center}
\vspace{1em}

\section{Action and effective action}
To express the action and effective action we need to employ the second-quantization formalism. We introduce the creation and annihilation operators $\hat\psi^\dagger(\mathbf r)$ and $\hat\psi(\mathbf r)$. The action of the creation operator on the non-interacting vacuum corresponding to the absence of particles is to create a bare electron at the spatial coordinate $\mathbf r$. As in the main text, spin degrees of freedom are not written explicitly. We reconsider the Hamiltonian of the main text in second quantization as
\begin{align}
\hat{H} &= \hat{H}_{\textrm{el}}+\hat{H}_{\textrm{n}}+\hat{H}_{\textrm{el-n}} 
\end{align}
\begin{align}
\hat{H}_{\textrm{el}}
&= \int d\mathbf r\,\hat{\psi}^{\dagger}(\mathbf r)\left[-\frac{\hbar^2\nabla^{2}}{2m_{\textrm{e}}}\right]\hat{\psi}(\mathbf r) +\frac{1}{2}\int d\mathbf r\,d\mathbf r'\,
\hat{\psi}^{\dagger}(\mathbf r)\hat{\psi}^{\dagger}(\mathbf r')
v(\mathbf r,\mathbf r')
\hat{\psi}(\mathbf r')\hat{\psi}(\mathbf r) 
\end{align}
\begin{align}
\hat{H}_{\textrm{n}}
&= \sum_{I=1}^{N_n}\frac{\hat{\mathbf{P}}_I^{2}}{2M_I}
+\frac{1}{2}\sum_{I\neq J}^{N_\textrm{n}}
Z_IZ_Jv(\hat{\mathbf{R}}_I,\hat{\mathbf{R}}_J)
\end{align}

\begin{align}
\hat{H}_{\textrm{el-n}}
&= -\int d\mathbf r\,\hat{n}(\mathbf r)\sum_{J=1}^{N_\textrm{n}}
Z_Jv(\mathbf r,\hat{\mathbf{R}}_J)
\end{align}
where $v(\mathbf r,\mathbf r')=e^2/|\mathbf r-\mathbf r'|$ and $\hat n(\mathbf r)=\hat\psi^{\dagger}(\mathbf r)\hat\psi(\mathbf r)$, which assumes a given equilibrium value $n^0(\mathbf r)$ for nuclei clamped at given positions.

\subsection{Action}
We now rewrite the Hamiltonian formulation in terms of a zero-temperature
real-time path integral. All time integrations are performed on the ordinary
real-time axis, $t\in(-\infty,\infty)$. The infinitesimal convergence
factors selecting the interacting ground state are left implicit. The
relevant propagators are the usual time-ordered Green's functions.

We work in the grand-canonical ensemble and define
\begin{equation}
    \hat{\mathcal H}^{\textrm{low-en}}
    =
    \hat H^{\textrm{low-en}}-\mu \hat N ,
\end{equation}
where $\mu$ is the electronic chemical potential. In the path-integral
representation, $\mathcal H^{\textrm{low-en}}(\bar\psi,\psi,U,P)$ denotes the corresponding
classical grand-canonical Hamiltonian obtained by replacing the operators
$\hat\psi,\hat\psi^\dagger,\hat U,\hat P$ with the path-integral fields, which depend on both spatial and temporal coordinates.

We start from the second-quantization representation of the Hamiltonian decomposition:
\begin{align}
\hat{H}^{\textrm{low-en}}
&= \hat{H}_{0,\textrm{el}}+\hat{H}_{0,\textrm{ph}}+\hat{H}_{\textrm{el-el}}+\hat{H}_{1}+\hat{H}_{2},
\label{eq:si-lowenH}
\end{align}
\begin{align}
\hat{H}_{0,\textrm{el}}
&= \int d\mathbf r\,\hat{\psi}^{\dagger}(\mathbf r)
\left[-\frac{\hbar^2\nabla^2}{2m_{\textrm{e}}}+V^{\textrm{BO}}(\mathbf r)\right]
\hat{\psi}(\mathbf r) ,
\end{align}

\begin{align}
\hat{H}_{0,\textrm{ph}}
&= \sum_{I=1}^{N_{\textrm{n}}}\frac{\hat{\mathbf{P}}_I^2}{2M_I}
+\frac{1}{2}\sum_{I\alpha,J\beta}
\hat{U}_{I,\alpha}C_{I,\alpha;J,\beta}\hat{U}_{J,\beta} ,
\end{align}

\begin{align}
C_{I,\alpha;J,\beta}
&\equiv \int d\mathbf r\left. \frac{\partial n^{\textrm{BO}}(\mathbf r,\mathbf{R})}{\partial R_{J,\beta}}\right|_{\mathbf{R}^{0,\textrm{BO}}}g_{I,\alpha}(\mathbf r)+\left.
\frac{\partial^2 E_{\textrm{n-n}}(\mathbf{R})}
{\partial R_{I,\alpha}\partial R_{J,\beta}}
\right|_{\mathbf{R}^{0,\textrm{BO}}}
+\int d\mathbf r\,n^{0,\textrm{BO}}(\mathbf r)g^{\mathrm{DW}}_{I,\alpha;J,\beta}(\mathbf{r}) ,
\end{align}
\begin{align}
\hat{H}_{1}
&= \sum_{I\alpha}
\left[- \int d\mathbf r\,g_{I,\alpha}(\mathbf r)n^{0,\textrm{BO}}(\mathbf r)+\int d\mathbf r\,g_{I,\alpha}(\mathbf r)\hat{n}(\mathbf r) \right] \hat{U}_{I,\alpha}=\sum_{I\alpha}
\int d\mathbf r\,g_{I,\alpha}(\mathbf r)\Delta \hat{n}(\mathbf r) \hat{U}_{I,\alpha},
\end{align}
\begin{align}
\hat H_{2}&\simeq-\frac{1}{2}\sum_{I\alpha,J\beta}\hat{U}_{I,\alpha}\int d\mathbf r\left. \frac{\partial n^{\textrm{BO}}(\mathbf r,\mathbf{R})}{\partial R_{J,\beta}}\right|_{\mathbf{R}^{0,\textrm{BO}}}g_{I,\alpha}(\mathbf r)\hat{U}_{J,\beta} = \frac{1}{2} \sum_{I\alpha,J\beta}\hat{U}_{I,\alpha} \mathcal{B}_{I,\alpha;J,\beta} \hat{U}_{J,\beta}.
\end{align}
Here the approximation sign denotes the neglect of the density-dependent Debye--Waller fluctuation, consistently with Sec.~II of the main text.
We introduce the phase-space path integral
\begin{equation}
    Z =
    \int \mathcal D[\bar\psi,\psi]\mathcal D[U,P]\,
    \exp\left\{
    \frac{i}{\hbar}
    S[\bar\psi,\psi,U,P]
    \right\},
\end{equation}
where $\mathcal D$ stands for the path-integral measure and
\begin{align}
    S[\bar\psi,\psi,U,P]
    &=
    \int_{-\infty}^{\infty}dt
    \Bigg[
    \int d\mathbf r\,
    \bar\psi(\mathbf r,t)
    i\hbar\frac{\partial}{\partial t}
    \psi(\mathbf r,t) +
    \sum_{I\alpha}
    P_{I\alpha}(t)\dot U_{I\alpha}(t)
    -
    \mathcal H(\bar\psi,\psi,U,P)
    \Bigg],
\end{align}
where the term $P\dot{U}$ is the standard Legendre term.
The electronic quadratic part can be written as
\begin{equation}
    S_{0,\textrm{el}}
    =
    \int d1\,d2\,
    \bar\psi(1)G_0^{-1}(1,2)\psi(2),
\end{equation}
where
\begin{equation}
    G_0^{-1}(1,2)
    =
    \left[
    i\hbar\frac{\partial}{\partial t_1}
    -
    h_0(\mathbf r_1)
    \right]\delta(1,2),
\end{equation}
with
\begin{equation}
    h_0(\mathbf r)
    =
    -\frac{\hbar^2\nabla^2}{2m_{\textrm{e}}}
    +
    V^{\textrm{BO}}(\mathbf{r})
    -
    \mu
\end{equation}
and
\begin{equation}
    \delta(1,2)
    =
    \delta(t_1-t_2)
    \delta(\mathbf r_1-\mathbf r_2).
\end{equation}

The phonon part is initially written in phase-space form as
\begin{align}
    S_{0}^{\textrm{ph}}[U,P]
    &=
    \int_{-\infty}^{\infty}dt
    \Bigg[
    \sum_{I\alpha}
    P_{I\alpha}\dot U_{I\alpha}
    -
    \sum_{I\alpha}
    \frac{P_{I\alpha}^2}{2M_I}  -
    \frac12
    \sum_{I\alpha,J\beta}
    U_{I\alpha}
    C_{I,\alpha;J,\beta}
    U_{J\beta}
    \Bigg].
\end{align}
Their stationary value for the momenta is
\begin{equation}
    P_{I\alpha}=M_I\dot U_{I\alpha}.
\end{equation}
Substitution gives the configuration-space phonon action
\begin{align}
    S_{0}^{\textrm{ph}}[U]
    &=
    \frac12
    \int_{-\infty}^{\infty}dt
    \Bigg[
    \sum_{I\alpha}
    M_I\dot U_{I\alpha}^2    -
    \sum_{I\alpha,J\beta}
    U_{I\alpha}
    C_{I,\alpha;J,\beta}
    U_{J\beta}
    \Bigg].
\end{align}
After integrating by parts in time, this becomes
\begin{equation}
    S_{0}^{\textrm{ph}}[U]
    =
    \frac12
    \int da\,db\,
    U(a)D_0^{-1}(a,b)U(b),
\end{equation}
where
\begin{equation}
    D_0^{-1}(a,b)
    =
    -
    \left[
    M_I\delta_{IJ}\delta_{\alpha\beta}
    \frac{\partial^2}{\partial t_a^2}
    +
    C_{I,\alpha;J,\beta}
    \right]
    \delta(t_a-t_b).
\end{equation}

The interaction terms in the action are obtained from
\begin{equation}
    S_{\textrm{int}}
    =
    -
    \int_{-\infty}^{\infty}dt\,
    H_{\textrm{int}}(t).
\end{equation}
The electron-electron interaction gives
\begin{equation}
    S_{\textrm{el-el}}
    =
    -
    \frac12
    \int d1\,d2\,
    \bar\psi(1)\bar\psi(2)
    v(1,2)
    \psi(2)\psi(1).
\end{equation}

The linear electron-phonon coupling gives
\begin{equation}
    S_{\textrm{el-ph}}
    =
    -
    \int d1\,da\,
    \Delta n(1)g(1,a)U(a),
\end{equation}
where
\begin{equation}
    n(1)=\bar\psi(1)\psi(1),
    \qquad
    \Delta n(1)=n(1)-n^{0,\textrm{BO}}(\mathbf r_1),
\end{equation}
Finally, the quadratic term retained in the low-energy Hamiltonian
\begin{equation}
    H_2
    =
    \frac12
    \sum_{I\alpha,J\beta}
    U_{I\alpha}
    \mathcal B_{I,\alpha;J,\beta}
    U_{J\beta}
\end{equation}
contributes
\begin{equation}
    S_{\mathcal B}
    =
    -
    \frac12
    \int da\,db\,
    U(a)\mathcal B(a,b)U(b),
\end{equation}
with
\begin{equation}
    \mathcal B(a,b)
    =
    \mathcal B_{I,\alpha;J,\beta}\delta(t_a-t_b).
\end{equation}

Collecting all terms, the microscopic action reads
\begin{equation}
    S[\bar\psi,\psi,U]=S_{0}^{\textrm{el}}+S_{0}^{\textrm{ph}}+S_{\textrm{el-el}}+S_{\textrm{el-ph}}+S_{\mathcal B}.
\end{equation}

Given the microscopic action, the expectation value of a functional
observable $\mathcal O[\bar\psi,\psi,U]$ is defined as its
zero-temperature time-ordered expectation value in the interacting
vacuum,
\begin{equation}
    \left\langle
    \mathcal O
    \right\rangle
    \equiv
    \frac{
    \left\langle
    \Omega
    \left|
    T\,\hat{\mathcal O}
    \right|
    \Omega
    \right\rangle
    }{
    \left\langle\Omega|\Omega\right\rangle
    },
\end{equation}
where $|\Omega\rangle$ denotes the interacting ground state of the
full Hamiltonian. In the real-time path-integral representation, this
expectation value is written as
\begin{equation}
    \left\langle
    \mathcal O[\bar\psi,\psi,U]
    \right\rangle
    =
    \frac{1}{Z}
    \int
    \mathcal D[\bar\psi,\psi]\mathcal D[U]\,
    \mathcal O[\bar\psi,\psi,U]\,
    \exp\left\{
    \frac{i}{\hbar}
    S[\bar\psi,\psi,U]
    \right\},
\end{equation}
where
\begin{equation}
    Z
    =
    \int
    \mathcal D[\bar\psi,\psi]\mathcal D[U]\,
    \exp\left\{
    \frac{i}{\hbar}
    S[\bar\psi,\psi,U]
    \right\}.
\end{equation}
The infinitesimal convergence factors implicit in the real-time
functional integral select the interacting vacuum $|\Omega\rangle$
and fix the time-ordered boundary conditions. This construction
assumes that the Hamiltonian is bounded from below and possesses a
well-defined ground state.

In the real-time formulation used here, this path-integral average
generates ordinary time-ordered correlation functions. The zero-temperature time-ordered propagators are introduced as
\begin{equation}
    G(1,2)
    =
    -\frac{i}{\hbar}
    \left\langle
    T\psi(1)\bar\psi(2)
    \right\rangle ,
\end{equation}
and
\begin{equation}
    D(a,b)
    =
    -\frac{i}{\hbar}
    \left\langle
    T U(a)U(b)
    \right\rangle .
\end{equation}

\subsection{Effective action}

The functional used later in this work is an effective action
whose independent variables are the full propagators $G$ and $D$.

The exact effective action is obtained by introducing bilocal sources
coupled to the electronic and phononic two-point functions. We denote
these sources by $J(1,2)$ and $K(a,b)$, respectively. We define the
source-dependent generating functional as
\begin{equation}
    Z[J,K]
    =
    \int
    \mathcal D[\bar\psi,\psi]\mathcal D[U]\,
    \exp\left\{
    \frac{i}{\hbar}
    \left[
    S[\bar\psi,\psi,U]
    +
    S_{\textrm{src}}[\bar\psi,\psi,U;J,K]
    \right]
    \right\},
\end{equation}
where the source term is
\begin{align}
    S_{\textrm{src}}[\bar\psi,\psi,U;J,K]
    &=
    \frac{i}{\hbar}\int d2 d1
    J(2, 1)
    \bar\psi(2)\psi(1)  -\int da db
    \frac{i}{2\hbar}
    K(b,a)
    U(a)U(b).
\end{align}
We do not add the linear terms in the sources to the Hamiltonian in this section. These would describe the evolution of  $\langle \psi \rangle$ and $\langle U \rangle$ with the interactions. We assume that all the fields are expanded around their equilibrium value, i.e. $\langle \psi \rangle =0$ and $\langle U \rangle = 0$, and that such minima do not change due to interactions. We discuss the eventual introduction of the linear terms coupling to atomic displacements in section \ref{app:linear_sources}.

For any functional observable
$\mathcal O[\bar\psi,\psi,U]$, we define its source-dependent path-integral
average as
\begin{align}
    \left\langle
    \mathcal O[\bar\psi,\psi,U]
    \right\rangle_{J,K}
    =&\frac{1}{Z[J,K]}
    \int \mathcal D[\bar\psi,\psi]\mathcal D[U]\,\mathcal O[\bar\psi,\psi,U]\,\exp\left\{
    \frac{i}{\hbar}
    \left[
    S[\bar\psi,\psi,U]
    +
    S_{\textrm{src}}[\bar\psi,\psi,U;J,K]
    \right]
    \right\}.
\end{align}
At vanishing sources we have
$ \langle \mathcal O\rangle=\left\langle \mathcal O\right\rangle_{J=0,K=0}$.

Introducing
\begin{equation}
    W[J,K]
    =
    -i\hbar\ln Z[J,K],
\end{equation}
one obtains, at vanishing sources,
\begin{align}
    \left.
    \frac{\delta W[J,K]}{\delta J(2,1)}
    \right|_{J=K=0}
    &=-
    \frac{i}{\hbar}
    \left\langle
    T\psi(1)\bar \psi(2)
    \right\rangle=G(1,2), \label{eq:dWdJG}
    \\
    \left.
    \frac{\delta W[J,K]}{\delta K(b,a)}
    \right|_{J=K=0}
    &=
    -
    \frac{i}{2\hbar}
    \left\langle
    T U(a)U(b)
    \right\rangle = \frac12 D(a,b). \label{eq:dWdKD}
\end{align}

The exact effective action is then defined as the
double Legendre transform
\begin{equation}
    \Gamma[G,D]
    =
    W[J,K]
    -
    J(\bar 2,\bar 1)G(\bar 1,\bar 2)
    -
    \frac12
    K(\bar b,\bar a)D(\bar a,\bar b).
\end{equation}
The source fields are now understood as functionals of $G$ and $D$.
Taking a variation gives
\begin{align}
    \delta \Gamma[G,D]
    &=
    -J(\bar 2,\bar 1)\delta G(\bar 1,\bar 2)
    -
    \frac12
    K(\bar b,\bar a)\delta D(\bar a,\bar b),
\end{align}
and therefore
\begin{align}
    \frac{\delta \Gamma[G,D]}{\delta G(2,1)}
    =
    -J(1,2), \quad \frac{\delta \Gamma[G,D]}{\delta D(a,b)}
    =
    -\frac12 K(b,a).
\end{align}
In the physical limit of vanishing external sources ($J=0,K=0$) the physical propagators are therefore determined by the stationarity
conditions
\begin{align}
    \frac{\delta \Gamma[G,D]}{\delta G(2,1)}
    =0, \quad \frac{\delta \Gamma[G,D]}{\delta D(a,b)}
    =0.
\end{align}

\subsection{Origin of the trace-log terms in the effective action}
\label{app:trace_log_derivation}

We now derive the one-particle trace-log terms appearing in the effective
action. We use throughout the bilocal sources $J$ and $K$ introduced in
the previous section, namely
\begin{align}
S_{\textrm{src}}[\bar\psi,\psi,U;J,K]
&=
\frac{i}{\hbar}
\int d2\,d1\,
J(2,1)\bar\psi(2)\psi(1)
-
\frac{i}{2\hbar}
\int db\,da\,
K(b,a)U(a)U(b).
\label{eq:Ssrc_trace_log}
\end{align}
No additional source convention is introduced below.

For clarity, we first evaluate the Legendre transform for the Gaussian
electronic and phononic theories. The remaining interacting contribution
is subsequently collected in the two-particle-irreducible functional
$\Phi[G,D]$.

\subsubsection{Non-interacting electronic contribution}

For the noninteracting electronic theory, the source-dependent action
defined in the previous section becomes
\begin{align}
S_{0,\textrm{el}}+S_{\textrm{src},\textrm{el}}
&=
\int d1\,d2\,
\bar\psi(1)
\left[
G_0^{-1}(1,2)
+
\frac{i}{\hbar}J(1,2)
\right]
\psi(2).
\end{align}
We define the source-dependent inverse electronic kernel
\begin{equation}
\mathcal A_J
\equiv
G_0^{-1}
+
\frac{i}{\hbar}J.
\label{eq:AJ_def}
\end{equation}
The corresponding Gaussian generating functional is
\begin{equation}
Z_{0,\textrm{el}}[J]
=
\int
\mathcal D[\bar\psi,\psi]\,
\exp
\left\{
\frac{i}{\hbar}
\bar\psi\mathcal A_J\psi
\right\}.
\end{equation}
Gaussian integration over the Grassmann fields gives
\begin{equation}
Z_{0,\textrm{el}}[J]
=
\mathcal N_{\textrm{el}}
\det\mathcal A_J,
\end{equation}
where $\mathcal N_{\textrm{el}}$ is a normalization factor independent of $J$ and depends on the chosen normalization of $Z$, which does not impact any observable. Consequently,
\begin{align}
W_{0,\textrm{el}}[J]
&=
-i\hbar\ln Z_{0,\textrm{el}}[J]=
-i\hbar\operatorname{Tr}\ln\mathcal A_J
+
\textrm{const.}
\label{eq:W0el_existing_source}
\end{align}

Using
\begin{equation}
\delta\operatorname{Tr}\ln A
=
\operatorname{Tr}\left(A^{-1}\delta A\right)
\end{equation}
and
\begin{equation}
\delta\mathcal A_J
=
\frac{i}{\hbar}\delta J,
\end{equation}
we obtain
\begin{align}
\delta W_{0,\textrm{el}}
=
-i\hbar
\operatorname{Tr}
\left[
\mathcal A_J^{-1}
\frac{i}{\hbar}\delta J
\right]=
\operatorname{Tr}
\left[
\mathcal A_J^{-1}\delta J
\right].
\end{align}
Using Eq. \eqref{eq:dWdJG} it follows that
\begin{equation}
G
=
\mathcal A_J^{-1}
=
\left(
G_0^{-1}
+
\frac{i}{\hbar}J
\right)^{-1}.
\label{eq:G_AJ_relation}
\end{equation}
Equivalently,
\begin{equation}
G^{-1}
=
G_0^{-1}
+
\frac{i}{\hbar}J,
\end{equation}
and therefore
\begin{equation}
J
=
i\hbar
\left(
G_0^{-1}-G^{-1}
\right).
\label{eq:J_existing_convention}
\end{equation}

The electronic part of the effective action is obtained from the Legendre
transform defined in the previous section,
\begin{equation}
\Gamma_{0,\textrm{el}}[G]
=
W_{0,\textrm{el}}[J]
-
\operatorname{Tr}(JG).
\end{equation}
Using Eqs.~\eqref{eq:W0el_existing_source} and
\eqref{eq:J_existing_convention}, we find
\begin{align}
\Gamma_{0,\textrm{el}}[G]
&=
-i\hbar\operatorname{Tr}\ln G^{-1}
-
i\hbar\operatorname{Tr}
\left[
\left(
G_0^{-1}-G^{-1}
\right)G
\right]
+
\textrm{const.}
\nonumber\\
&=
-i\hbar
\operatorname{Tr}
\left[
\ln G^{-1}
+
G_0^{-1}G
-
I
\right]
+
\textrm{const.}
\end{align}
After adding the $G$-independent quantity
$i\hbar\operatorname{Tr}\ln G_0^{-1}$ to the constant, this can be written
as
\begin{equation}
\Gamma_{0,\textrm{el}}[G]
=
-i\hbar
\operatorname{Tr}
\left\{
G_0^{-1}G
+
\ln\left[G_0G^{-1}\right]
-
I
\right\}
+
\textrm{const.}
\label{eq:Gamma0el_dimensionful}
\end{equation}

\subsubsection{Non-interacting phononic contribution}

We choose the harmonic Born--Oppenheimer theory as the free phononic
reference. The corresponding action is
\begin{equation}
S_{0,\textrm{ph}}[U]
=
\frac{1}{2}
\int da\,db\,
U(a)D_0^{-1}(a,b)U(b),
\label{eq:S0ph_BO}
\end{equation}
where
\begin{equation}
D_0^{-1}(a,b)
=
-
\left[
M_I\delta_{IJ}\delta_{\alpha\beta}
\frac{\partial^2}{\partial t_a^2}
+
C_{I,\alpha;J,\beta}
\right]
\delta(t_a-t_b).
\label{eq:D0_BO}
\end{equation}
Here \(C\) is the Born--Oppenheimer force-constant matrix evaluated at
the Born--Oppenheimer equilibrium geometry. The Born--Oppenheimer
configuration defines a stable harmonic reference theory, apart from
the zero modes associated with rigid translations, that allows the existence of the non-interacting phononic ground state $|0_{\textrm{ph}}\rangle$. The bilocal source \(K\) may also
be chosen to lift the translational zero modes during the derivation.
The source-dependent kernel is assumed to remain invertible in the
domain of sources considered. The vanishing-source limit is taken after the
effective-action relations have been derived.

The remaining quadratic term
\begin{equation}
S_{\mathcal B}[U]
=
-\frac{1}{2}
\int da\,db\,
U(a)\mathcal B(a,b)U(b)
\label{eq:SB_quadratic_interaction}
\end{equation}
is not included in the free phononic action. It originates from the
addition and subtraction used to organize the expansion of the
Hamiltonian around the Born--Oppenheimer reference and is treated as a
quadratic interaction. Accordingly, only the Born--Oppenheimer kernel
\(D_0^{-1}\) enters the Gaussian integration performed below.

For the free phononic theory, the source-dependent action is
\begin{align}
S_{0,\textrm{ph}}+S_{\textrm{src},\textrm{ph}}
&=
\frac{1}{2}
\int da\,db\,
U(a)
\left[
D_0^{-1}(a,b)
-
\frac{i}{\hbar}K(a,b)
\right]
U(b).
\label{eq:S0ph_source_BO}
\end{align}
We define the source-dependent inverse kernel
\begin{equation}
\mathcal A_K
\equiv
D_0^{-1}
-
\frac{i}{\hbar}K.
\label{eq:AK_BO}
\end{equation}
The corresponding Gaussian generating functional is
\begin{equation}
Z_{0,\textrm{ph}}[K]
=
\int
\mathcal D[U]\,
\exp
\left\{
\frac{i}{2\hbar}
U\mathcal A_K U
\right\}.
\end{equation}
Gaussian integration over the real bosonic field gives
\begin{equation}
Z_{0,\textrm{ph}}[K]
=
\mathcal N_{\textrm{ph}}
\left(
\det\mathcal A_K
\right)^{-1/2},
\end{equation}
where \(\mathcal N_{\textrm{ph}}\) is independent of the source.
Consequently,
\begin{equation}
W_{0,\textrm{ph}}[K]
=
-i\hbar\ln Z_{0,\textrm{ph}}[K]
=
\frac{i\hbar}{2}
\operatorname{Tr}\ln\mathcal A_K
+
\textrm{const.}
\label{eq:W0ph_BO}
\end{equation}

Since
\begin{equation}
\delta\mathcal A_K
=
-\frac{i}{\hbar}\delta K,
\end{equation}
the variation of the free generating functional is
\begin{align}
\delta W_{0,\textrm{ph}}
&=
\frac{i\hbar}{2}
\operatorname{Tr}
\left[
\mathcal A_K^{-1}
\left(
-\frac{i}{\hbar}\delta K
\right)
\right]
\nonumber\\
&=
\frac{1}{2}
\operatorname{Tr}
\left[
\mathcal A_K^{-1}\delta K
\right].
\end{align}
Using
\begin{equation}
\frac{\delta W[J,K]}{\delta K(b,a)}
=
\frac{1}{2}D(a,b),
\end{equation}
we obtain
\begin{equation}
D
=
\mathcal A_K^{-1}
=
\left(
D_0^{-1}
-
\frac{i}{\hbar}K
\right)^{-1}.
\label{eq:D_AK_BO}
\end{equation}
Equivalently,
\begin{equation}
D^{-1}
=
D_0^{-1}
-
\frac{i}{\hbar}K,
\end{equation}
and therefore
\begin{equation}
K
=
i\hbar
\left(
D^{-1}-D_0^{-1}
\right).
\label{eq:K_BO}
\end{equation}

The free phononic contribution to the effective action follows from the
Legendre transform
\begin{equation}
\Gamma_{0,\textrm{ph}}[D]
=
W_{0,\textrm{ph}}[K]
-
\frac{1}{2}\operatorname{Tr}(KD).
\end{equation}
Substituting Eqs.~\eqref{eq:W0ph_BO} and \eqref{eq:K_BO}, we find
\begin{align}
\Gamma_{0,\textrm{ph}}[D]
&=
\frac{i\hbar}{2}
\operatorname{Tr}\ln D^{-1}
-
\frac{i\hbar}{2}
\operatorname{Tr}
\left[
\left(
D^{-1}-D_0^{-1}
\right)D
\right]
+
\textrm{const.}
\nonumber\\
&=
\frac{i\hbar}{2}
\operatorname{Tr}
\left[
\ln D^{-1}
+
D_0^{-1}D
-
I
\right]
+
\textrm{const.}
\end{align}
After absorbing the \(D\)-independent contribution
\(-\frac{i\hbar}{2}\operatorname{Tr}\ln D_0^{-1}\) into the constant,
the free contribution becomes
\begin{equation}
\Gamma_{0,\textrm{ph}}[D]
=
\frac{i\hbar}{2}
\operatorname{Tr}
\left\{
D_0^{-1}D
-
\ln\left[D_0^{-1}D\right]
-
I
\right\}
+
\textrm{const.}
\label{eq:Gamma0ph_BO}
\end{equation}
We now include the remaining quadratic term
\(S_{\mathcal B}\) in the interaction part of the effective action.
Since it is quadratic in the displacement field, its contribution to the effective action can be written as
\begin{equation}
\Phi_{\mathcal B}[D]
=
-\frac{1}{2}
\operatorname{Tr}
\left(
\mathcal B D
\right).
\label{eq:PhiB}
\end{equation}
\subsubsection{Final form}
It is convenient to distinguish this contribution from the functional
\(\Phi[G,D,v,g]\), which contains the nontrivial skeleton diagrams
generated by the electron--electron and electron--phonon interaction
vertices. We therefore introduce the complete interaction functional
\begin{align}
\Phi_{\textrm{tot}}[G,D]
&=
\Phi[G,D,v,g]
+
\Phi_{\mathcal B}[D]
\nonumber\\
&=
\Phi[G,D,v,g]
-
\frac{1}{2}
\operatorname{Tr}
\left(
\mathcal B D
\right).
\label{eq:Phi_total_B}
\end{align}
The effective action can then be written as
\begin{align}
\Gamma[G,D]
&= i\hbar \Phi_{\textrm{tot}}[G,D]+\Gamma_{0,\textrm{el}}[G]+\Gamma_{0,\textrm{ph}}[D] \nonumber\\
&=i\hbar\Phi[G,D,v,g]-i\hbar
\operatorname{Tr}
\left\{
G_0^{-1}G
+
\ln\left[G_0G^{-1}\right]
-
I
\right\}
+
\frac{i\hbar}{2}
\operatorname{Tr}
\left\{
D_0^{-1}D
-
\ln\left[D_0^{-1}D\right]
-
I
-
\mathcal B D
\right\}
+
\textrm{const.}
\label{eq:Gamma_explicit_B}
\end{align}
which is the form given in the main text.
Here and in the derivation above, $\Phi[G,D,v,g]$ denotes the exact interaction functional. After the single-phonon truncation introduced in the main text, the corresponding approximate functional is denoted by $\widetilde\Phi[G,D,v,g]$, with electron--phonon contribution $\widetilde\Phi^{\textrm{el-ph}}[G,D,v,g]$.

\subsection{Linear coupling to sources in the effective action}
\label{app:linear_sources}

The effective action introduced above assumes that the expectation value
of the atomic displacement vanishes. This restriction can be removed by
introducing a linear source coupled to the displacement field. We denote
this source by $\zeta(a)$ and generalize the source-dependent generating
functional as
\begin{equation}
Z[\zeta,J,K]
=
\int
\mathcal D[\bar\psi,\psi]\mathcal D[U]\,
\exp\left\{
\frac{i}{\hbar}
\left[
S[\bar\psi,\psi,U]
+
S_{\textrm{src}}[\bar\psi,\psi,U;\zeta,J,K]
\right]
\right\},
\end{equation}
where
\begin{align}
S_{\textrm{src}}[\bar\psi,\psi,U;\zeta,J,K]
&=
\int da\,\zeta(a)U(a)
+
\frac{i}{\hbar}
\int d2\,d1\,
J(2,1)\bar\psi(2)\psi(1)
\nonumber\\
&\quad
-
\frac{i}{2\hbar}
\int db\,da\,
K(b,a)U(a)U(b).
\label{eq:Ssrc_linear_U}
\end{align}
The source $\zeta(a)$ is a classical force coupled directly to the atomic
displacement.

Introducing
\begin{equation}
W[\zeta,J,K]
=
-i\hbar\ln Z[\zeta,J,K],
\end{equation}
the expectation value of the displacement is
\begin{equation}
\mathcal U(a)
\equiv
\frac{\delta W[\zeta,J,K]}{\delta\zeta(a)}
=
\left\langle U(a)\right\rangle_{\zeta,J,K}.
\label{eq:Ubar_def}
\end{equation}
It is then convenient to separate the displacement into its expectation
value and a fluctuating part,
\begin{equation}
U(a)=\mathcal U(a)+u(a),
\qquad
\left\langle u(a)\right\rangle_{\zeta,J,K}=0.
\end{equation}
In the presence of a nonvanishing $\mathcal U$, the propagator entering
the two-particle-irreducible effective action must be defined as the
connected phonon propagator,
\begin{align}
D(a,b)
&=
-\frac{i}{\hbar}
\left[
\left\langle
T U(a)U(b)
\right\rangle_{\zeta,J,K}
-
\mathcal U(a)\mathcal U(b)
\right]=
-\frac{i}{\hbar}
\left\langle
T u(a)u(b)
\right\rangle_{\zeta,J,K}.
\label{eq:D_connected_linear}
\end{align}
Consequently, the full two-point expectation value is
\begin{equation}
-\frac{i}{\hbar}
\left\langle
T U(a)U(b)
\right\rangle_{\zeta,J,K}
=
D(a,b)
-
\frac{i}{\hbar}
\mathcal U(a)\mathcal U(b).
\label{eq:D_full_disconnected}
\end{equation}
With the normalization of the bilocal source used above, one therefore
has
\begin{equation}
2\frac{\delta W[\zeta,J,K]}{\delta K(b,a)}
=
D(a,b)
-
\frac{i}{\hbar}
\mathcal U(a)\mathcal U(b).
\label{eq:dWdK_nonzero_U}
\end{equation}

The effective action is obtained by performing the Legendre transform
with respect to all three sources:
\begin{align}
\Gamma[\mathcal U,G,D]
&=
W[\zeta,J,K]
-
\zeta(\bar a)\mathcal U(\bar a)
-
J(\bar 2,\bar 1)G(\bar 1,\bar 2)
-
\frac{1}{2}
K(\bar b,\bar a)
\left[
D(\bar a,\bar b)
-
\frac{i}{\hbar}
\mathcal U(\bar a)\mathcal U(\bar b)
\right].
\label{eq:Gamma_linear_legendre}
\end{align}
For a symmetric bosonic source, $K(a,b)=K(b,a)$, its functional
derivatives are
\begin{align}
\frac{\delta\Gamma}{\delta\mathcal U(a)}
&=
-\zeta(a)
+
\frac{i}{\hbar}
K(a,\bar b)\mathcal U(\bar b),
\\
\frac{\delta\Gamma}{\delta G(2,1)}
&=
-J(1,2),
\\
\frac{\delta\Gamma}{\delta D(a,b)}
&=
-\frac{1}{2}K(b,a).
\end{align}
The physical quantities are therefore determined, at vanishing sources,
by the three stationarity conditions
\begin{equation}
\frac{\delta\Gamma}{\delta\mathcal U(a)}
=
\frac{\delta\Gamma}{\delta G(2,1)}
=
\frac{\delta\Gamma}{\delta D(a,b)}
=
0.
\label{eq:stationarity_with_U}
\end{equation}

We now derive explicitly the form of the effective action when the expectation value of the atomic displacement is not constrained to vanish. For compactness, we define
\begin{equation}
\mathcal Q(a,b)
\equiv
D_0^{-1}(a,b)-\mathcal B(a,b).
\label{eq:Q_shifted}
\end{equation}
The part of the microscopic action containing the atomic displacement is
then
\begin{equation}
S_U
=
\frac{1}{2}U\mathcal Q U
-
\Delta n\,g\,U.
\end{equation}
After substituting $U=\mathcal U+u$, one obtains
\begin{align}
S_U
&=
\frac{1}{2}
\mathcal U\mathcal Q\mathcal U
-
\Delta n\,g\,\mathcal U
+
\frac{1}{2}u\mathcal Q u
+
u\mathcal Q\mathcal U
-
\Delta n\,g\,u.
\label{eq:SU_shifted_simple}
\end{align}

Using $\Delta n=n-n^{0,\textrm{BO}}$, the term depending on the mean
displacement can be separated as
\begin{equation}
-\Delta n\,g\,\mathcal U
=
-n\,g\,\mathcal U
+
n^{0,\textrm{BO}}g\,\mathcal U.
\end{equation}
The first term is quadratic in the electronic fields and can therefore
be included in a displacement-dependent electronic inverse propagator,
\begin{equation}
G_{0,\mathcal U}^{-1}(1,2)
=
G_0^{-1}(1,2)
-
\delta(1,2)
g(1,\bar a)\mathcal U(\bar a).
\label{eq:G0_shifted_simple}
\end{equation}
The remaining terms containing only the mean displacement define the
classical contribution
\begin{align}
S_{\textrm{cl}}[\mathcal U]
&=
\frac{1}{2}
\mathcal U(\bar a)
\mathcal Q(\bar a,\bar b)
\mathcal U(\bar b)
+
n^{0,\textrm{BO}}(\bar 1)
g(\bar 1,\bar a)
\mathcal U(\bar a).
\label{eq:Scl_shifted_simple}
\end{align}

The shifted microscopic action can consequently be organized as
\begin{align}
S[\bar\psi,\psi,\mathcal U+u]
&=
S_{\textrm{cl}}[\mathcal U]
+
\int d1\,d2\,
\bar\psi(1)
G_{0,\mathcal U}^{-1}(1,2)
\psi(2)
\nonumber\\
&\quad
+
\frac{1}{2}
\int da\,db\,
u(a)\mathcal Q(a,b)u(b)
+
S_{\textrm{el-el}}
\nonumber\\
&\quad
-
\int d1\,da\,
\Delta n(1)g(1,a)u(a)
+
\int da\,db\,
u(a)\mathcal Q(a,b)\mathcal U(b).
\label{eq:full_shifted_action}
\end{align}
The last term is linear in $u$ and shifts the source required to impose
$\langle u\rangle=0$. Its contribution is therefore encoded in the
stationarity equation for $\mathcal U$, rather than in an additional
term of the effective action. Apart from this source shift, the action
has the same structure as before, with quadratic kernels
$G_{0,\mathcal U}^{-1}$ and $\mathcal Q$, and with the fluctuating
electron--phonon interaction $-\Delta n\,g\,u$.

We can therefore repeat the same Gaussian integrations and Legendre
transforms as before. The resulting effective action is
\begin{align}
\Gamma[\mathcal U,G,D]
&=
S_{\textrm{cl}}[\mathcal U]
+
i\hbar\Phi[G,D,v,g]
\nonumber\\
&\quad
-i\hbar
\operatorname{Tr}
\left\{
G_{0,\mathcal U}^{-1}G
+
\ln\left[G_0G^{-1}\right]
-
I
\right\}
\nonumber\\
&\quad
+
\frac{i\hbar}{2}
\operatorname{Tr}
\left\{
D_0^{-1}D
-
\ln\left[D_0^{-1}D\right]
-
I
-
\mathcal B D
\right\}
+
\textrm{const.}
\label{eq:Gamma_shifted_simple}
\end{align}
The fixed propagator $G_0$ is retained inside the logarithm so that the normalization term is independent of $\mathcal U$. The explicit dependence on the mean displacement is contained in $S_{\textrm{cl}}[\mathcal U]$ and $G_{0,\mathcal U}^{-1}$. Notice that the form of $\Phi[G,D,v,g]$, which generate the self-energy diagrams, remains unaffected by the displacement.

The interaction between the fluctuating phonon field and the electronic density has the same form as in the original action. Consequently, the two-particle-irreducible functional retains the same functional form $\Phi[G,D,v,g]$ and has no  additional explicit dependence on $\mathcal U$ at fixed $G$ and $D$.

We now differentiate Eq.~\eqref{eq:Gamma_shifted_simple} with respect to $\mathcal U(a)$ at fixed $G$ and $D$. From Eq.~\eqref{eq:G0_shifted_simple},
\begin{equation}
\frac{\delta
G_{0,\mathcal U}^{-1}(1,2)}
{\delta\mathcal U(a)}
=
-
\delta(1,2)g(1,a),
\end{equation}
and therefore
\begin{align}
\left.
\frac{\delta\Gamma}
{\delta\mathcal U(a)}
\right|_{G,D}
&=
\mathcal Q(a,\bar b)\mathcal U(\bar b)
+
g(\bar 1,a)n^{0,\textrm{BO}}(\bar 1)+
i\hbar
g(\bar 1,a)G(\bar 1,\bar 1^+).
\label{eq:dGamma_dU_simple}
\end{align}
Defining the electronic density associated with the full propagator as
\begin{equation}
n_G(1)
=
-i\hbar G(1,1^+),
\end{equation}
where we added the subscript $G$ to retain the explicit dependence on the exact $G$ (which depends on $\mathcal{U}$), and
\begin{equation}
\Delta n_G(1)
=
n_G(1)-n^{0,\textrm{BO}}(1),
\end{equation}
we obtain
\begin{equation}
\left.
\frac{\delta\Gamma}
{\delta\mathcal U(a)}
\right|_{G,D}
=
\mathcal Q(a,\bar b)\mathcal U(\bar b)
-
g(\bar 1,a)\Delta n_G(\bar 1).
\label{eq:dGamma_dU_compact}
\end{equation}

At vanishing external sources, stationarity with respect to the mean
displacement gives
\begin{equation}
\left[
D_0^{-1}(a,\bar b)
-
\mathcal B(a,\bar b)
\right]
\mathcal U(\bar b)
=
g(\bar 1,a)\Delta n_G(\bar 1).
\label{eq:mean_displacement_equation_simple}
\end{equation}
This equation determines the expectation value
\begin{equation}
\mathcal U(a)=\langle U(a)\rangle.
\end{equation}
In our notation,
\begin{equation}
D_0^{-1}-\mathcal B
=
-\left(M\partial_t^2+C+\mathcal B\right)
=
-\left(M\partial_t^2+\langle K\rangle_{\textrm{BO}}\right),
\end{equation}
where $\langle K\rangle_{\textrm{BO}}=C+\mathcal B$ is the BO-projected quadratic coefficient retained after neglecting the density-dependent Debye--Waller fluctuation. Therefore, in the static limit,
\begin{equation}
D_0^{-1}(\omega=0)-\mathcal B=-\langle K\rangle_{\textrm{BO}},
\end{equation}
and Eq.~\eqref{eq:mean_displacement_equation_simple} becomes
\begin{equation}
\langle K\rangle_{\textrm{BO}}\mathcal U=-g\Delta n_G,
\label{eq:KUgDeltan}
\end{equation}
so that $\mathcal{U}=0$ if $n_G(\mathbf r,t)=n^{0,\textrm{BO}}(\mathbf r)$.  By computing the exact $n_{G}$, one finds the equilibrium positions $\mathcal{U}$ renormalized by the interactions of the many-body Hamiltonian truncated at the second-order around the BO density.

The phonon self-energy is defined from the same functional as
\begin{equation}
\Pi_{\mathcal U}(a,b)
=
-2
\left.
\frac{\delta\Phi[G,D,v,g]}
{\delta D(a,b^+)}
\right|_{G,\mathcal U,\textrm{S}}.
\label{eq:Pi_shifted_simple}
\end{equation}
Its functional expression is unchanged. Its physical value nevertheless
depends on $\mathcal U$ through the electronic propagator. The Dyson equation for the connected phonon
propagator is
\begin{equation}
D^{-1}(a,b)
-
D_0^{-1}(a,b)
+
\mathcal B(a,b)
+
\Pi_{\mathcal U}(a,b)
=
0.
\label{eq:Dyson_shifted_simple}
\end{equation}
The electronic propagator $G$ appearing in
Eq.~\eqref{eq:Gamma_shifted_simple} is an independent variational
quantity. Its physical value at a given mean displacement $\mathcal U$
is determined by the stationarity condition. The electronic self-energy is still defined as
\begin{equation}
\Sigma(1,2)
=
\frac{\delta\Phi[G,D,v,g]}
{\delta G(2,1)},
\label{eq:Sigma_shifted}
\end{equation}
and the stationarity condition gives the electronic Dyson equation
\begin{equation}
G^{-1}(1,2)
-
G_{0,\mathcal U}^{-1}(1,2)
+
\Sigma(1,2)
=
0.
\label{eq:Dyson_G_shifted}
\end{equation}

Finally, the expectation values of the displacement and atomic-position
operators are
\begin{equation}
\left\langle
\hat U_{I\alpha}
\right\rangle
=
\mathcal U_{I\alpha},
\qquad
\left\langle
\hat R_{I\alpha}
\right\rangle
=
R_{I\alpha}^{0,\textrm{BO}}
+
\mathcal U_{I\alpha}.
\label{eq:position_expectation_shifted}
\end{equation}
If the reference geometry is stationary at the same level of theory
used to evaluate the interacting density, then
$\Delta n_G=0$ at $\mathcal U=0$, and
Eq.~\eqref{eq:mean_displacement_equation_simple} admits
$\mathcal U=0$ as its solution. In this limit, the effective action
reduces to the functional $\Gamma[G,D]$ used in the main text.

\section{Diagrammatic content and symmetry factors of
$\widetilde{\Phi}^{\textrm{el-ph}}$}
We show that every diagram generated by
\begin{align}
\widetilde{\Phi}^{\textrm{el-ph}}[G,D,v,g]
&=
\left.
\frac{d}{d\eta}
\Phi_{\textrm{xc}}^{\textrm{el}}
[G,v+\eta U_D]
\right|_{\eta=0},
\qquad
U_D=gDg,
\label{eq:Phi_eph_subset}
\end{align}
appears with the correct symmetry factor.

Write the electronic exchange-correlation functional as \cite{NegeleOrland1988}
\begin{align}
\Phi_{\textrm{xc}}^{\textrm{el}}[G,v]
&=
\sum_{\Gamma}
\frac{1}{\operatorname{Ord}(\mathcal S_\Gamma)}
\mathcal A_\Gamma[G,v],
\label{eq:Phi_el_diagram_expansion}
\end{align}
where $\Gamma$ is an unlabeled electronic skeleton diagram, $\mathcal A_\Gamma$ is its amplitude, and $\mathcal S_\Gamma$ is its symmetry group, i.e. the set of operations (permutation of time labels and interactions line extremities exchanges) that, acting on $\Gamma$, do not generate a distinct diagram. `Ord' represents the order of the group, i.e. the number of total elements. We remind that the simple expression of Eq. \eqref{eq:Phi_el_diagram_expansion} is due to the cancellation of a factor $2^n n!$ between possible Wick's contractions and the Taylor expansion coefficients of order $n$.

Let $E_v(\Gamma)$ denote the set of Coulomb lines of $\Gamma$.
Expanding to first order in $\eta$ gives
\begin{align}
\left.
\frac{d}{d\eta}
\mathcal A_\Gamma[G,v+\eta U_D]
\right|_{\eta=0}
&=
\sum_{e\in E_v(\Gamma)}
\mathcal A_{\Gamma_e}[G,v,U_D],
\label{eq:line_replacement_sum}
\end{align}
where $\Gamma_e$ is obtained by replacing the Coulomb line $e$ with
the distinct interaction line $U_D$.

The group $\mathcal S_\Gamma$ acts on the Coulomb lines of $\Gamma$. For a chosen line $e$, define its orbit
\begin{align}
\mathcal O_e
&=
\left\{
s(e):s\in\mathcal S_\Gamma
\right\},
\qquad
m_e\equiv\operatorname{Ord}(\mathcal O_e).
\label{eq:line_orbit}
\end{align}
Thus, $m_e$ is the number of Coulomb lines related to $e$ by
symmetries of the original diagram. Replacing any line belonging to
$\mathcal O_e$ produces the same unlabeled mixed diagram
$\Gamma_e$. Consequently, Eq.~\eqref{eq:line_replacement_sum}
generates $\Gamma_e$ exactly $m_e$ times.

Define also the stabilizer of $e$,
\begin{align}
\mathcal S_{\Gamma,e}
&=
\left\{
s\in\mathcal S_\Gamma:s(e)=e
\right\}.
\label{eq:line_stabilizer}
\end{align}
The orbit--stabilizer theorem applied to the action of
$\mathcal S_\Gamma$ on the Coulomb lines gives
\begin{align}
\operatorname{Ord}(\mathcal S_\Gamma)
&=
\operatorname{Ord}(\mathcal O_e)\,
\operatorname{Ord}(\mathcal S_{\Gamma,e})=
m_e\,
\operatorname{Ord}(\mathcal S_{\Gamma,e}).
\label{eq:orbit_stabilizer}
\end{align}

After the replacement $e\rightarrow U_D$, the line $e$ becomes distinguishable from all remaining Coulomb lines. Therefore, a symmetry of the mixed diagram $\Gamma_e$ must leave the substituted line fixed. Conversely, every symmetry of $\Gamma$ that leaves $e$ fixed remains a symmetry of $\Gamma_e$. Hence,
\begin{align}
\mathcal S_{\Gamma_e}
&=
\mathcal S_{\Gamma,e}.
\label{eq:mixed_symmetry_group}
\end{align}
It follows that
\begin{align}
\operatorname{Ord}(\mathcal S_\Gamma)
&=
m_e\,
\operatorname{Ord}(\mathcal S_{\Gamma_e}).
\label{eq:symmetry_order_relation}
\end{align}

The original diagram $\Gamma$ has coefficient
$1/\operatorname{Ord}(\mathcal S_\Gamma)$. Since the derivative generates $\Gamma_e$ exactly $m_e$ times, the total coefficient of the
mixed diagram is
\begin{align}
\frac{m_e}{\operatorname{Ord}(\mathcal S_\Gamma)}
&=
\frac{m_e}
{m_e\operatorname{Ord}(\mathcal S_{\Gamma_e})}
=
\frac{1}
{\operatorname{Ord}(\mathcal S_{\Gamma_e})}.
\label{eq:correct_mixed_factor}
\end{align}
This is precisely the symmetry factor of the mixed diagram. If the Coulomb lines of $\Gamma$ belong to different symmetry orbits, the argument applies separately to each orbit.

The replacement does not modify the fermionic propagator topology or the number of closed fermionic loops, and therefore it does not modify the fermionic sign.

\section{Fourier conventions}
\label{app:Fourierconv}
We define a Born--von Karman super-cell containing $N=N_1\times N_2 \times N_3$ unit cells of volume $V$, where periodic boundary conditions are imposed on the Bloch's wavefunctions. All functions are periodic in the Born--von Karman supercell. We use summations on reciprocal space grid points assuming that the reciprocal space grids are compatible with the Born--von Karman supercell. We define the Fourier transform of a function with one spatial index as
\begin{align}
f_{\mathbf{G}}(\mathbf{q},\omega)=\int_{NV} \frac{d\mathbf{r_1}}{NV} \frac{d\tau}{2\pi}  e^{-i\mathbf{(q+G)} \cdot \mathbf{r_1}} f(\mathbf{r_1},t_1)e^{i \omega t_1},\\
f(\mathbf{r_1},t_1)=\sum_{\mathbf{qG}} \int d\omega e^{i\mathbf{(q+G)}\cdot \mathbf{r_1}} f_{\mathbf{G}}(\mathbf{q},\omega) e^{-i\omega t_1}.
\end{align}
It is also useful to define
\begin{align}
f_{\mathbf{q}}(\mathbf{r_1},t_1)=e^{i\mathbf{q}\cdot\mathbf{r_1}}\sum_{\mathbf{G}} \int d\omega e^{i\mathbf{G}\cdot \mathbf{r_1}} f_{\mathbf{G}}(\mathbf{q},\omega) e^{-i\omega t_1}=e^{i\mathbf{q}\cdot\mathbf{r_1}}f^{\rm p}_{\mathbf{q}}(\mathbf{r_1},t_1)
\end{align}
where $f^{\rm p}$ is periodic with the crystal periodicity. 

We define the transform of a two index function in the plane wave basis set as
\begin{align}
f_{\mathbf{G_1G_2}}(\mathbf{q},\omega)=\int \frac{d\mathbf{r_1}}{NV} \frac{d\mathbf{r_2}}{NV} \frac{d\tau}{2\pi} e^{-i\mathbf{(q+G_1)}\cdot \mathbf{r_1}} e^{i\mathbf{(q+G_2)}\cdot \mathbf{r_2}} f(\mathbf{r_1},\mathbf{r_2},t_1-t_2)e^{i\omega (t_1-t_2)},\\
f(\mathbf{r_1},\mathbf{r_2},t_1-t_2)=\sum_{\mathbf{qG_1G_2}} \int d\omega e^{i\mathbf{(q+G_1)}\cdot \mathbf{r_1}} e^{-i\mathbf{(q+G_2)}\cdot \mathbf{r_2}}f_{\mathbf{G_1G_2}}(\mathbf{q},\omega) e^{-i\omega (t_1-t_2)},
\end{align}

where we have supposed that at equilibrium the function depends only on the time difference. We define the spatial Fourier transform of a four point function as
\begin{align}
f(\mathbf{r_1},\mathbf{r_2},\mathbf{r_3},\mathbf{r_4})=\sum_{\substack{\mathbf{q_1q_2q_3}\\
\mathbf{G_1G_2G_3G_4}}}e^{i(\mathbf{q_1+G_1})\cdot \mathbf{r_1}}e^{i\mathbf{(q_2+G_2)}\cdot\mathbf{r_2}}f_{\mathbf{G_1G_2G_3G_4}}(\mathbf{q_1},\mathbf{q_2},\mathbf{q_3})e^{-i\mathbf{(q_3+G_3)}\cdot \mathbf{r_3}}e^{-i\mathbf{(q_1+q_2-q_3+G_4)}\cdot \mathbf{r_4}}\delta(\mathbf{r_3}-\mathbf{r_4})
\end{align}
\begin{align}
f_{\mathbf{G_1,G_2,G_3,G_4}}(\mathbf{q_1},\mathbf{q_2},\mathbf{q_3})=\int d\mathbf{r_1}d\mathbf{r_2}d\mathbf{r_3}d\mathbf{r_4}e^{-i(\mathbf{q_1+G_1})\cdot \mathbf{r_1}}e^{-i\mathbf{(q_2+G_2)}\cdot\mathbf{r_2}}f(\mathbf{r_1},\mathbf{r_2},\mathbf{r_3},\mathbf{r_4})e^{i\mathbf{(q_3+G_3)}\cdot \mathbf{r_3}}e^{i\mathbf{(q_1+q_2-q_3+G_4)}\cdot \mathbf{r_4}}
\end{align}
where we used the discrete translation symmetry $f(\mathbf{r_1+T},\mathbf{r_2+T},\mathbf{r_3+T},\mathbf{r_4+T})=f(\mathbf{r_1},\mathbf{r_2},\mathbf{r_3},\mathbf{r_4})$. In the case, as for our scopes, where we have two identical Cartesian coordinates such that $f(\mathbf{r_1},\mathbf{r_2},\mathbf{r_3},\mathbf{r_4})=f(\mathbf{r_1},\mathbf{r_2},\mathbf{r_3})\delta(\mathbf{r_4}-\mathbf{r_3})$, the transform becomes
\begin{align}
f_{\mathbf{G_1,G_2,G_3,G_4}}(\mathbf{q_1},\mathbf{q_2},\mathbf{q_3})=\int \frac{d\mathbf{r_1}}{NV}\frac{d\mathbf{r_2}}{NV}\frac{d\mathbf{r_3}}{NV}\frac{d\mathbf{r_4}}{NV}e^{-i(\mathbf{q_1+G_1})\cdot \mathbf{r_1}}e^{-i\mathbf{(q_2+G_2)}\cdot\mathbf{r_2}}f(\mathbf{r_1},\mathbf{r_2},\mathbf{r_3})\delta(\mathbf{r_4-r_3})e^{i\mathbf{(q_3+G_3)}\cdot \mathbf{r_3}}e^{i\mathbf{(q_1+q_2-q_3+G_4)}\cdot \mathbf{r_4}} \nonumber \\
\int \frac{d\mathbf{r_1}}{NV}\frac{d\mathbf{r_2}}{NV}\frac{d\mathbf{r_3}}{NV}e^{-i(\mathbf{q_1+G_1})\cdot \mathbf{r_1}}e^{-i\mathbf{(q_2+G_2)}\cdot\mathbf{r_2}}f(\mathbf{r_1},\mathbf{r_2},\mathbf{r_3})\delta(\mathbf{r_4-r_3})e^{i\mathbf{(q_1+q_2+G_3)}\cdot \mathbf{r_3}}=f_{\mathbf{G_1,G_2,G_3}}(\mathbf{q_1},\mathbf{q_2})
\end{align}
where we used that $\mathbf{G_3+G_4}$ is still a reciprocal lattice vector. It follows that we define the Fourier transform for functions where two coordinates are equal (diagonal part with respect to the last two indices) as
\begin{align}
f(\mathbf{r_1},\mathbf{r_2},\mathbf{r_3})=\sum_{\substack{\mathbf{q_1q_2}\\\mathbf{G_1G_2G_3}}}e^{i\mathbf{(q_1+G_1)}\cdot\mathbf{r_1}}e^{i\mathbf{(q_2+G_2)}\cdot\mathbf{r_2}}f_{\mathbf{G_1G_2G_3}}(\mathbf{q_1,q_2})e^{-i(\mathbf{q_1+q_2+G_3})\cdot \mathbf{r_3}}\\
f_{\mathbf{G_1G_2G_3}}(\mathbf{q_1,q_2})=\int \frac{d\mathbf{r_1}}{NV}\frac{d\mathbf{r_2}}{NV}\frac{d\mathbf{r_3}}{NV} e^{-i\mathbf{(q_1+G_1)}\cdot \mathbf{r_1}}e^{-i\mathbf{(q_2+G_2)}\cdot \mathbf{r_2}}f(\mathbf{r_1,r_2,r_3})e^{i\mathbf{(q_1+q_2+G_3)}\cdot \mathbf{r_3}}
\end{align}
For the time variables we start performing similar transforms. We start from
\begin{align}
f(t_1,t_2,t_3,t_4)=\int d\omega_1 d\omega_2 d\omega_3 e^{-i\omega_1 t_1} e^{-i\omega_2 t_2} f(\omega_1,\omega_2,\omega_3) e^{i\omega_3 t_3}e^{i(\omega_1+\omega_2-\omega_3) t_4} \\
f(\omega_1,\omega_2,\omega_3)=\int \frac{dt_1}{2\pi} \frac{dt_2}{2\pi} \frac{dt_3}{2\pi} \frac{dt_4}{2\pi} e^{i\omega_1 t_1} e^{i\omega_2 t_2} f(t_1,t_2,t_3,t_4) e^{-i\omega_3 t_3}e^{-i(\omega_1+\omega_2-\omega_3) t_4}
\end{align}
where we used $f(t_1,t_2,t_3,t_4)=f(t_1+t,t_2+t,t_3+t,t_4+t)$.
For three point functions such that $f(t_1,t_2,t_3,t_4)=f(t_1,t_2,t_3)\delta(t_3-t_4)$, we have
\begin{align}
f(\omega_1,\omega_2,\omega_3)=\int \frac{dt_1}{2\pi} \frac{dt_2}{2\pi} \frac{dt_3}{2\pi} \frac{dt_4}{2\pi} e^{i\omega_1 t_1} e^{i\omega_2 t_2} f(t_1,t_2,t_3)\delta(t_3-t_4) e^{-i\omega_3 t_3}e^{-i(\omega_1+\omega_2-\omega_3) t_4}=\nonumber\\
\int \frac{dt_1}{2\pi} \frac{dt_2}{2\pi} \frac{dt_3}{2\pi}  e^{i\omega_1 t_1} e^{i\omega_2 t_2} f(t_1,t_2,t_3) e^{-i(\omega_1+\omega_2) t_3}=f(\omega_1,\omega_2)\\
f(t_1,t_2,t_3)=\int d\omega_1 d\omega_2 e^{-i\omega_1 t_1} e^{-i\omega_2 t_2} f(\omega_1,\omega_2)e^{i(\omega_1+\omega_2) t_3}
\end{align}
It is now worth to perform a change of variables to the aims of our work. 
Keep in mind that we work in a different way from Eq. (14.17) of Ref. \cite{Reining_16} for $L$, since $L$ is horizontal while here we deal with vertical functions.
We perform the following change of variables
\begin{align}
\omega_1 = \omega'_1-\omega'_2, \quad \omega_2=-\omega'_1
\end{align}
so that we can finally write
\begin{align}
f(t_1,t_2,t_3)=\int d\omega'_1 d\omega'_2  f(\omega'_1,\omega'_2) e^{-i\omega'_1(t_1-t_2)}e^{i\omega'_2(t_1-t_3)} \\
f(\omega_1,\omega_2)=\int dt_1 dt_2 dt_3  f(t_1,t_2,t_3) e^{i\omega'_1(t_1-t_2)}e^{-i\omega'_2(t_1-t_3)}
\end{align}
In this way, we rewrote the dependence in terms of relative times.
To make connection with the main text, we define 
\begin{align}
I=(s\mathbf{T})
\end{align}
Then, atomic positions and equilibrium ones read
\begin{align}
R_{I\alpha}=R_{s\alpha \mathbf{T}}=R^{0,\textrm{BO}}_{s\alpha \mathbf{T}}+u_{s\alpha}, \quad R^{0,\textrm{BO}}_{s\alpha \mathbf{T}}=T_{\alpha}+\tau_{s\alpha}
\end{align}
where $u$ is the displacement from the equilibrium positions and $T$ are direct lattice vectors. We define the Fourier transform for phonon propagator as \cite{Giustino2017}
\begin{align}
D_{s\alpha,s'\beta}(\mathbf{T-T'},\omega)=\frac{1}{N}\sum_{\mathbf{q}\mu} D_{\mathbf{q}\mu}(\omega)e^{i\mathbf{q}\cdot (\mathbf{R}^{0,\textrm{BO}}_{s\mathbf{T}}-\mathbf{R}^{0,\textrm{BO}}_{s'\mathbf{T'}})}\frac{e_{\mu \mathbf{q}}^{s \alpha}}{2\omega_{\mu\mathbf{q}}\sqrt{M_s}}\frac{[e_{\mu \mathbf{q}}^{s' \beta }]^*}{2\omega_{\mu\mathbf{q}}\sqrt{M_{s'}}}\\
D_{\mathbf{q}\mu}(\omega)=\sum_{\substack{s\alpha,s'\beta \\ \mathbf{T-T'}}}e^{i\mathbf{q}\cdot (-\mathbf{R}^{0,\textrm{BO}}_{s\mathbf{T}}+\mathbf{R}^{0,\textrm{BO}}_{s'\mathbf{T'}})}D_{s\alpha,s'\beta}(\mathbf{T-T'},\omega) [e_{\mu \mathbf{q}}^{s \alpha }]^*\sqrt{2\omega_{\mu\mathbf{q}}M_s}e_{\mu \mathbf{q}}^{s' \beta }\sqrt{2\omega_{\mu\mathbf{q}}M_{s'}}
\end{align}
where we have included the basis vector explicitly in the exponential.

\section{Bloch basis set}
\label{app:BLochbasis}
We consider the matrix elements between Bloch's functions, that are solution of the self-consistent Hamiltonian, with no linewidth coming from electron-electron dressing. We disregard temporal dependence in this section to state the definition more concisely.
We define the matrix elements for the electronic Green's function as
\begin{align}
G(\mathbf{r_1},\mathbf{r_2})=\sum_{n\mathbf{k}}\psi_{n\mathbf{k}}(\mathbf{r_1})\psi^*_{n\mathbf{k}}(\mathbf{r_2})G_{n\mathbf{k}}\\
G_{n\mathbf{k}}=\int \frac{d\mathbf{r_1}}{NV}\frac{d\mathbf{r_2}}{NV} \psi^*_{n\mathbf{k}}(\mathbf{r_1})\psi_{n\mathbf{k}}(\mathbf{r_2})G(\mathbf{r_1,r_2})
\end{align}
For notation, we will indicate with capital indices the couple of band and momentum index. For example the above relation can be rewritten as
\begin{align}
G(\mathbf{r_1},\mathbf{r_2})=\sum_{I}\psi_{I}(\mathbf{r_1})\psi^*_{I}(\mathbf{r_2})G_{I}\\
G_{I}=\int \frac{d\mathbf{r_1}}{NV}\frac{d\mathbf{r_2}}{NV} \psi^*_{I}(\mathbf{r_1})\psi_{I}(\mathbf{r_2})G(\mathbf{r_1,r_2})
\end{align}
Following Eq. (12.24) of Ref. \cite{Reining_16}, i.e.
\begin{align}
F(\mathbf{r_1,r_2,r_3,r_4})=\sum_{KILJ}\psi_{K}(\mathbf{r_1})\psi_{I}(\mathbf{r_2}) F_{KILJ}\psi^*_{L}(\mathbf{r_3}) \psi^*_{J}(\mathbf{r_4})\\
F_{KILJ}=\int \frac{d\mathbf{r_1}}{NV}\frac{d\mathbf{r_2}}{NV}\frac{d\mathbf{r_3}}{NV}\frac{d\mathbf{r_4}}{NV} \psi^*_{K}(\mathbf{r_1})\psi^*_{I}(\mathbf{r_2}) F(\mathbf{r_1,r_2,r_3,r_4})  \psi_{L}(\mathbf{r_3}) \psi_{J}(\mathbf{r_4})
\end{align}
we have
\begin{align}
W(\mathbf{r_1},\mathbf{r_2})=W(\mathbf{r_1},\mathbf{r_2},\mathbf{r_2},\mathbf{r_1})\\
W(\mathbf{r_1},\mathbf{r_2})=\sum_{KILJ}\psi_{K}(\mathbf{r_1}) \psi^*_{J}(\mathbf{r_1})W_{KILJ}\psi^*_{L}(\mathbf{r_2})\psi_{I}(\mathbf{r_2})\\
W_{KILJ}=\int \frac{d\mathbf{r_1}}{NV}\frac{d\mathbf{r_2}}{NV}\psi^*_{K}(\mathbf{r_1})\psi_{J}(\mathbf{r_1})W(\mathbf{r_1},\mathbf{r_2})\psi_{L}(\mathbf{r_2})\psi^*_{I}(\mathbf{r_2})
\end{align}
The following notations are used equivalently
\begin{align}
F^{JI}_{KL}\equiv F_{KILJ}
\end{align}
For the vertex $\Gamma'$ we have, in agreement with Appendix F.2 of Ref. \cite{Reining_16} (see precisation below Eq. F.17):
\begin{align}
\Gamma'(\mathbf{r_3,r_2,r_4,r_4})=\sum_{KILJ}\psi_{K}(\mathbf{r_3})\psi^*_{I}(\mathbf{r_2}) \Gamma'_{KILJ}\psi_{L}(\mathbf{r_4}) \psi^*_{J}(\mathbf{r_4})\\
\Gamma'_{KILJ}=\int \frac{d\mathbf{r_3}}{NV}\frac{d\mathbf{r_4}}{NV}\frac{d\mathbf{r_2}}{NV} \psi^*_{K}(\mathbf{r_3})\psi_{I}(\mathbf{r_2}) \Gamma'(\mathbf{r_3,r_2,r_4,r_4})  \psi^*_{L}(\mathbf{r_4}) \psi_{J}(\mathbf{r_4})
\end{align}
The different definition of the complex conjugate is due to practical diagrammatic graphical contraction reasons. 
\section{Computational scheme}
\subsection{Generalized DFT}
For the practical graphene calculation, we evaluate the interacting electronic structure and response within the generalized Kohn--Sham framework (GDFT) introduced in Refs.~\onlinecite{PhysRevB.111.075118, PhysRevB.111.075137,y1dn-m6pc}. The formal development of that GDFT construction closely parallels the theory presented in the main and is not repeated here. Suffice it to say that the reference electronic Hamiltonian in this case is non-local and already provides, at self-consistence, the fully dressed $G$ similarly to the SX approximation case:
\begin{equation}
h^{\textrm{GDFT}}_0(\mathbf r,\mathbf r')
=
-\frac{\nabla^2}{2}
+
V^{\textrm{GDFT}}(\mathbf r, \mathbf r'),
\label{eq:DFTh0def}
\end{equation}
where $V^{\textrm{GDFT}}(\mathbf r, \mathbf r')$ contains both local and non-local interactions. We now summarize the approximations adopted in its practical implementation.

\begin{enumerate}

\item The graphene tight-binding Hamiltonian is initially fitted to the DFT band structure. Non-local, electron-hole interaction effects are then included through a static screened-exchange self-energy constructed using a cutoff screened interaction $\mathcal{W}_{\textrm{el}}$. The electronic energies and the screened interaction are determined self-consistently.

In the general GDFT construction, the nonlocal screened-exchange self-energy is accompanied by an additional local double-counting potential, which removes exchange-correlation effects already described by the underlying DFT functional. In the graphene calculation, the short-range components of $W_{\textrm{el}}$ are suppressed in order to minimize the overlap between the nonlocal screened-exchange correction and the local or semilocal DFT exchange-correlation contribution. With this cutoff prescription, the residual local GDFT double-counting potential is assumed to be small and is neglected in the band-structure calculation.  $\mathcal{W}_{\textrm{el}}$ is used consistently both in the screened-exchange self-energy and in the electron-hole interaction kernel. Besides reducing the residual GDFT double counting, the cutoff suppresses the short-distance region in which the use of a static screened interaction is less justified.

\item The local GDFT double-counting potential would generate, through its functional derivative with respect to the density, a corresponding double-counting contribution to the electronic response kernel. Consistently with the approximation made for the electronic Hamiltonian, this GDFT response double-counting contribution is also neglected.

\item The starting phonon propagator $D_0$ is constructed from the complete static DFT force-constant matrix. $\Pi^{\textrm{NA,GDFT}}(z)$ is then evaluated differentially starting from $\Pi^{\textrm{GDFT}}(z)$ computed in the partial-screen--partial-screen formulation, in which the same DFT-screened electron-phonon coupling is placed on the two external sides of the interacting electronic response, i.e.
\begin{align}
\Pi^{\textrm{NA,GDFT}}(z)
&=
g^{\textrm{DFT}}
\left[
\chi^{\mathcal{W}}(z)
-
\chi^{0}_{\textrm{DFT}}
\right]
g^{\textrm{DFT}}.
\end{align}
Excitonic effects are contained in $\chi^{\mathcal{W}}$ through the electron-hole ladder generated by $\mathcal{W}_{\textrm{el}}$. The term in phonon self-energy that involves $n^{0}$ is implicitly computed by imposing the acoustic sum rule (Eq. 2 of Ref. \onlinecite{y1dn-m6pc}).

\item With a DFT starting phonon propagator,
$\Pi^{\textrm{NA,GDFT}}$ contains both dynamical effects and a differential
static correction to the DFT force constants. Setting $\Pi^{\textrm{NA}}=0$ (in the sense of the BO expansion of the main text) is equivalent to set $\Pi^{\textrm{NA,GDFT}}(z)=\Pi^{\textrm{NA,GDFT}}(0)$.
\end{enumerate}

\subsection{Computational details}
\textit{Electronic properties---} Extensive details on the modeling are demanded to Ref. \onlinecite{PhysRevB.111.075118}. The band structure is computed in the self-consistent SX approximation. The electronic temperature is 300K. The $\mathbf{k}$-mesh is telescopic, with a maximum depth of 11 and a minimum one of 6. Around 7 self consistent cycles were needed to converge the $W_{\textrm{el}}$.

\textit{Phonon properties---} For extensive details regarding the calculation of phonon properties we refer to Ref. \onlinecite{y1dn-m6pc}. The $\mathbf{k}$-mesh telescopic grid is here set to a maximum depth of 10 and a minimum one of 6. The slightly coarser grid for the phonon case still assures a good convergence, and practically allows for the calculation of the BSE on a very large number of $\mathbf{q}$-points. 

In order to converge the electronic linewidths, we have to converge the calculation on a  $\mathbf{q}$-grid. Using uniform grids is too expensive for this task. To achieve converged results, we set up a $N_{\mathbf{q}} \times N_{\mathbf{q}}$ grid, and then select the relevant subset of $q$ points  (which we call $\{q^{\tau}\}$) that satisfy the condition
\begin{align}
\mathbf{q}\in\{q\}^\tau \, \textrm{if} \, \left[\varepsilon^{\textrm{DFT}}_{\pi\mathbf{K+\{q\}}}>-E_{\textrm{lim}} \, \textrm{and} \, \varepsilon^{\textrm{DFT}}_{\pi^*\mathbf{K+\{q\}}}<E_{\textrm{lim}}\right]\, \textrm{or}  \,\left[\varepsilon^{\textrm{DFT}}_{\pi\mathbf{K'+\{q\}}}>-E_{\textrm{lim}} \, \textrm{and} \, \varepsilon^{\textrm{DFT}}_{\pi^*\mathbf{K'+\{q\}}}<E_{\textrm{lim}}\right],
\label{eq:qcondition}
\end{align}
where $\varepsilon^{\textrm{DFT}}_n$ are the $\pi,\pi^*$ energies obtained from the tight binding with parameters fitted on the DFT energy bands, with the same notation of the main text. The above restriction is very convenient since the linewidth at $n\mathbf{k}$ is substantially determined by the density of states lying around $\varepsilon_{n\mathbf{k}} \pm \omega_{\rm ph}$, where $\omega_{\rm ph}$ is the typical phonon frequency. In other words, we require  $\varepsilon_{n\mathbf{k+q}} \sim \varepsilon^{\textrm{DFT}}_{n\mathbf{k}} \pm \omega_{\rm ph}$. We evaluate $\varepsilon$ at the DFT level for practical computational reasons, and then adjust the threshold to be sure the real $\varepsilon^{\textrm{SX}}$ energies fall in the correct range.

We set $E_{\textrm{lim}}=0.4$eV. In our convergence study, we determine the sets of points that respect Eq. \eqref{eq:qcondition} for grids $N_\mathbf{q}=300,400,600,900$, and call them $\{q^{\tau}\}_{300},\{q^{\tau}\}_{400},\{q^{\tau}\}_{600},\{q^{\tau}\}_{900}$. The number of points for each of these grids is $\{q^{\tau}\}_{300}\rightarrow 585$, $\{q^{\tau}\}_{400}\rightarrow 1063$, $\{q^{\tau}\}_{600}\rightarrow 2403$, $\{q^{\tau}\}_{900}\rightarrow 5427$. The cost of the BSE solution for each q-point on 160 processors is around 2 minutes, so that larger grids require around 30K CPUhrs.

On each of these $\mathbf{q}$ points, the first operation is to compute the ab-initio phonon via QUANTUM ESPRESSO (QE) \cite{doi:10.1073/pnas.1100242108}. Since computing the phonons at 300K with high precision in graphene requires very heavy computational resources, we use a temperature of 1500K. The high temperature allows us to use coarser $\mathbf{k}$-point grids for convergence. Then we consider the dynamical matrix obtained as 
\begin{align}
\tilde{D}^{\textrm{T=1500K}}_{s\alpha,r\beta}(\mathbf{q})=D^{\textrm{QE,T=1500K}}_{s\alpha,r\beta}(\mathbf{q})-\Pi^{\textrm{DFT,T=1500K}}_{s\alpha,r\beta}(\mathbf{q})
\end{align}
where $D^{\textrm{QE,T=1500K}}_{s\alpha,r\beta}(\mathbf{q})$ is the dynamical matrix obtained by QE,  $\Pi^{\textrm{DFT,T=1500K}}_{s\alpha,r\beta}$ is the phonon self-energy computed: (i) using our TB model fitted on DFT data, not including vertex corrections, and using the high temperature in the occupation factors of the polarizability bubble --- and not modifying the DFT band energies since they are depend very weakly on the electronic temperature ---  (ii) using screened vertexes $g^{\textrm{s}}$ a static form fitted on DFT, exploiting the variational screen-screen formulation as explained in Ref. \onlinecite{y1dn-m6pc}.

Then, we reconstruct the dynamical matrix at $T=300K$ without (wo) or with (w) excitonic effects in the following way
\begin{align}
D^{\textrm{wo,T=300K}}_{s\alpha,r\beta}(\mathbf{q})=\tilde{D}^{\textrm{T=1500K}}_{s\alpha,r\beta}(\mathbf{q})+\Pi^{\textrm{wo,SX,T=300K}}_{s\alpha,r\beta}(\mathbf{q})\label{eq:DRPA300}\\
D^{\textrm{w,T=300K}}_{s\alpha,r\beta}(\mathbf{q})=\tilde{D}^{\textrm{T=1500K}}_{s\alpha,r\beta}(\mathbf{q})+\Pi^{\textrm{w,SX,T=300K}}_{s\alpha,r\beta}(\mathbf{q})\label{eq:DBSE300}
\end{align}
where $\Pi^{\textrm{wo,SX,T=300K}}_{s\alpha,r\beta}$ is obtained excluding vertex corrections from the response, but employing SX bands, while $\Pi^{\textrm{w,SX,T=300K}}$ includes both vertex corrections  and SX bands. We find
\begin{align}
D^{\textrm{wo,T=300K}}_{s\alpha,r\beta}(\mathbf{q}) \sim  D^{\textrm{QE,T=300K}}_{s\alpha,r\beta}(\mathbf{q}),
\end{align}
so that $D^{\textrm{wo,T=300K}}_{s\alpha,r\beta}(\mathbf{q}) $ is then used, for practical computational reasons, to compute the phonon frequencies that in the main text we called $\omega^{\textrm{DFT}}$. To show that this method works we compare on selected points the phonons obtained by diagonalizing Eqs. \ref{eq:DRPA300} and \ref{eq:DBSE300}, with the ones obtained by diagonalizing
\begin{align}
\bar{D}^{\textrm{T=300K}}_{s\alpha,r\beta}(\mathbf{q})=\tilde{D}^{\textrm{T=300K}}_{s\alpha,r\beta}(\mathbf{q})+\Pi^{\textrm{TB-SX-RPA,T=300K}}_{s\alpha,r\beta}(\mathbf{q})\label{eq:barDRPA300}\\
\bar{D}^{\textrm{T=300K}}_{s\alpha,r\beta}(\mathbf{q})=\tilde{D}^{\textrm{T=300K}}_{s\alpha,r\beta}(\mathbf{q})+\Pi^{\textrm{TB-SX-BSE,T=300K}}_{s\alpha,r\beta}(\mathbf{q})\label{eq:barDBSE300}
\end{align}
We indicate the phonon frequencies as $\omega^{\textrm{wo}}$ (Eq. \eqref{eq:DRPA300}), $\omega^{\textrm{w}}$ (Eq. \eqref{eq:DBSE300}), $\bar{\omega}^{\textrm{wo}}$ (Eq. \eqref{eq:barDRPA300}),$\bar{\omega}^{\textrm{wo}}$ (Eq. \eqref{eq:barDBSE300}).

\textit{Linewidth convergence---} The linewidth convergence is shown in Fig. \ref{fig:lineconv}. We find that for $N_{\mathbf{q}}=900$ the linewidth are completely converged. 

\subsection{Further results}
Fig. \ref{fig:linewidthSI} --- it is the same figure of the main text, where we do not add the contribution of acoustic phonons.

Fig. \ref{fig:epcSI} --- we compare the statical vertex $\mathcal{G}^{\textrm{s,SX}}$ to the dynamical one $\mathcal{G}^{\textrm{s,SX}}(\omega^{\textrm{SX}}_{\textrm{ph}})$. They are overall very similar, except a slight reduction of the dynamical vertex for electronic wavevectors really near to K.

Fig. \eqref{fig:phSI} --- we compare phonon frequencies and linewidth computed with the statix vertex and with the dynamical one $\mathcal{G}^{\textrm{s,SX}}(\omega^{\textrm{SX}}_{\textrm{ph}})$. The differences are quantitatively small and qualitatively negligible.

\begin{figure}[h]
    \centering
    \includegraphics[width=\linewidth]{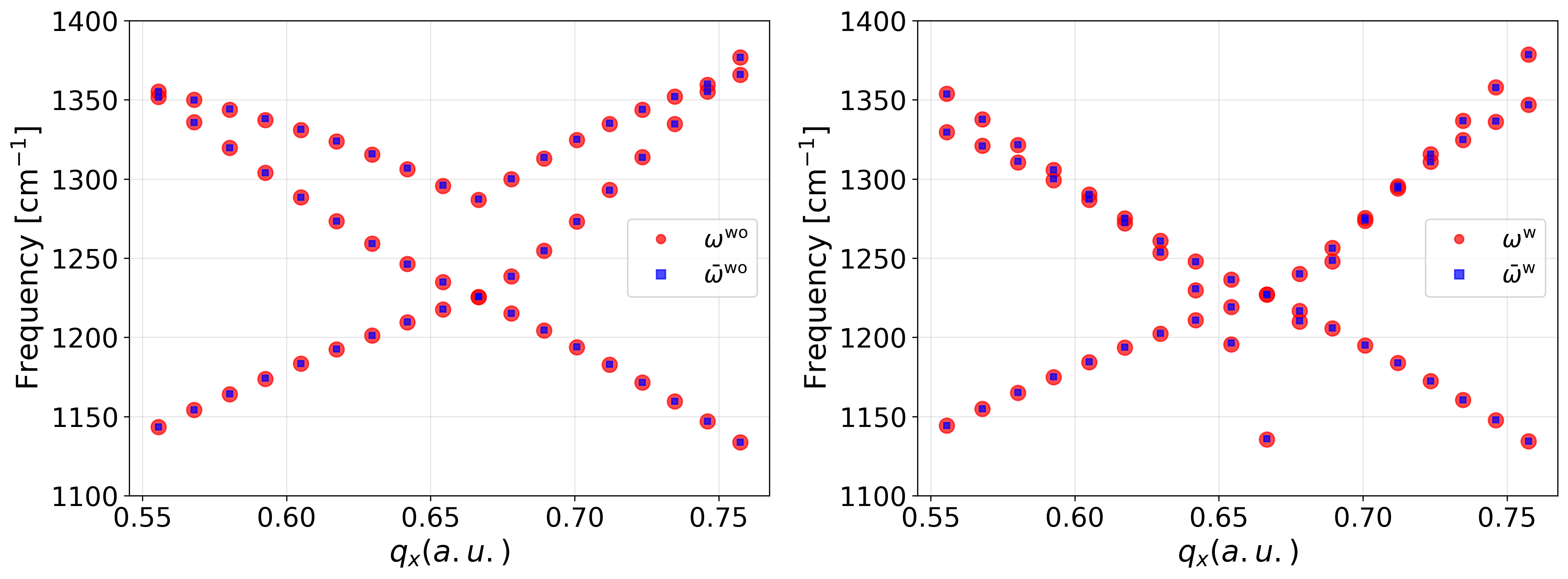}
    \caption{Comparison of the phonon frequencies as $\omega^{\textrm{wo}}$ (Eq. \eqref{eq:DRPA300}), $\omega^{\textrm{w}}$ (Eq. \eqref{eq:DBSE300}), $\bar{\omega}^{\textrm{wo}}$ (Eq. \eqref{eq:barDRPA300}),$\bar{\omega}^{\textrm{wo}}$ (Eq. \eqref{eq:barDBSE300})}
    \label{fig:wvswbar}
\end{figure}

\begin{figure}[h]
    \centering
    \includegraphics[width=1\linewidth]{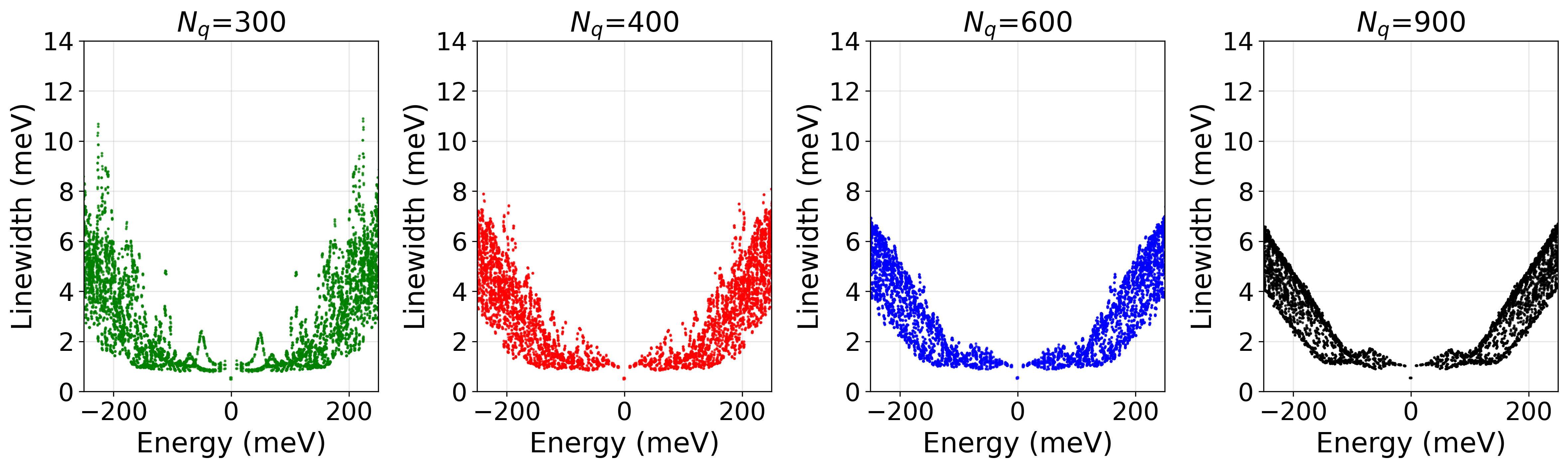}
    \caption{Linewidth convergence as a function of $N_{\mathbf{q}}$.}
    \label{fig:lineconv}
\end{figure}
\begin{figure}
    \centering
    \includegraphics[width=1\linewidth]{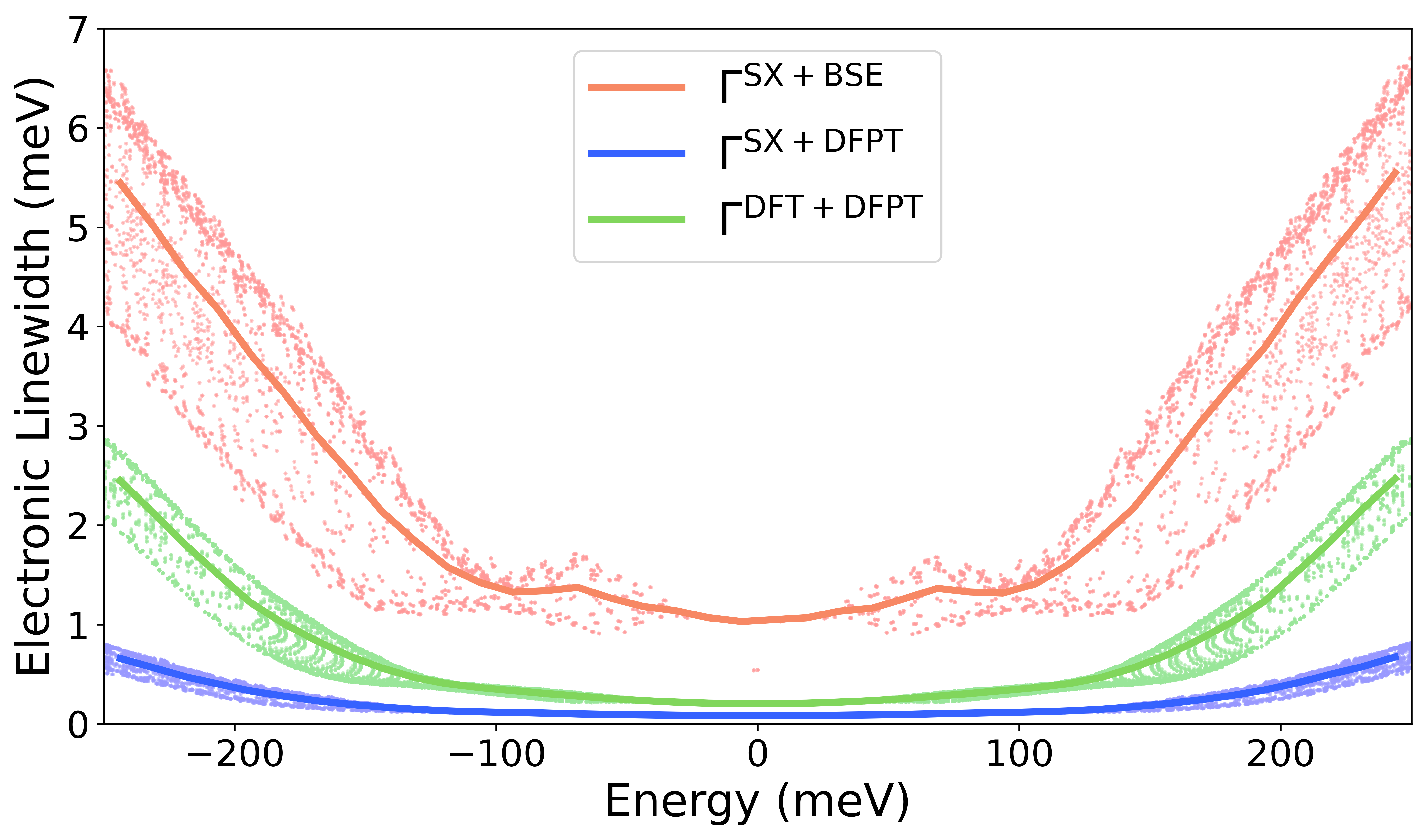}
    \caption{Same figure as the main text, but without the acoustic contribution.}
    \label{fig:linewidthSI}
\end{figure}
\begin{figure*}
    \centering
    \includegraphics[width=\linewidth]{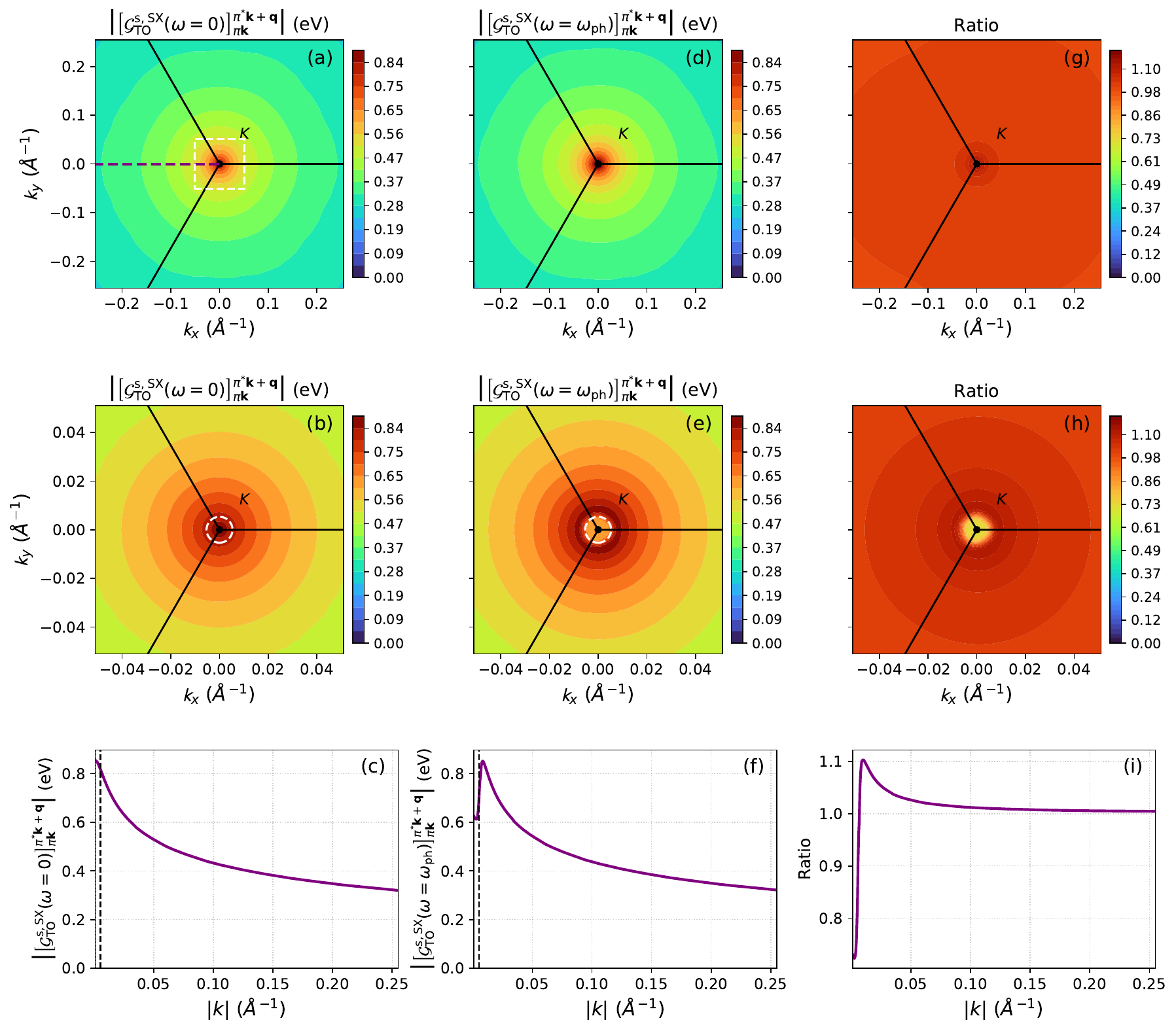}
    \caption{Same figure as the main text, but comparing dynamical and static electron-phonon vertices. They are overall very similar, except a slight reduction of the dynamical vertex at very small wavevectors, due to dynamical effects.}
    \label{fig:epcSI}
\end{figure*}
\begin{figure*}
    \centering
    \includegraphics[width=\linewidth]{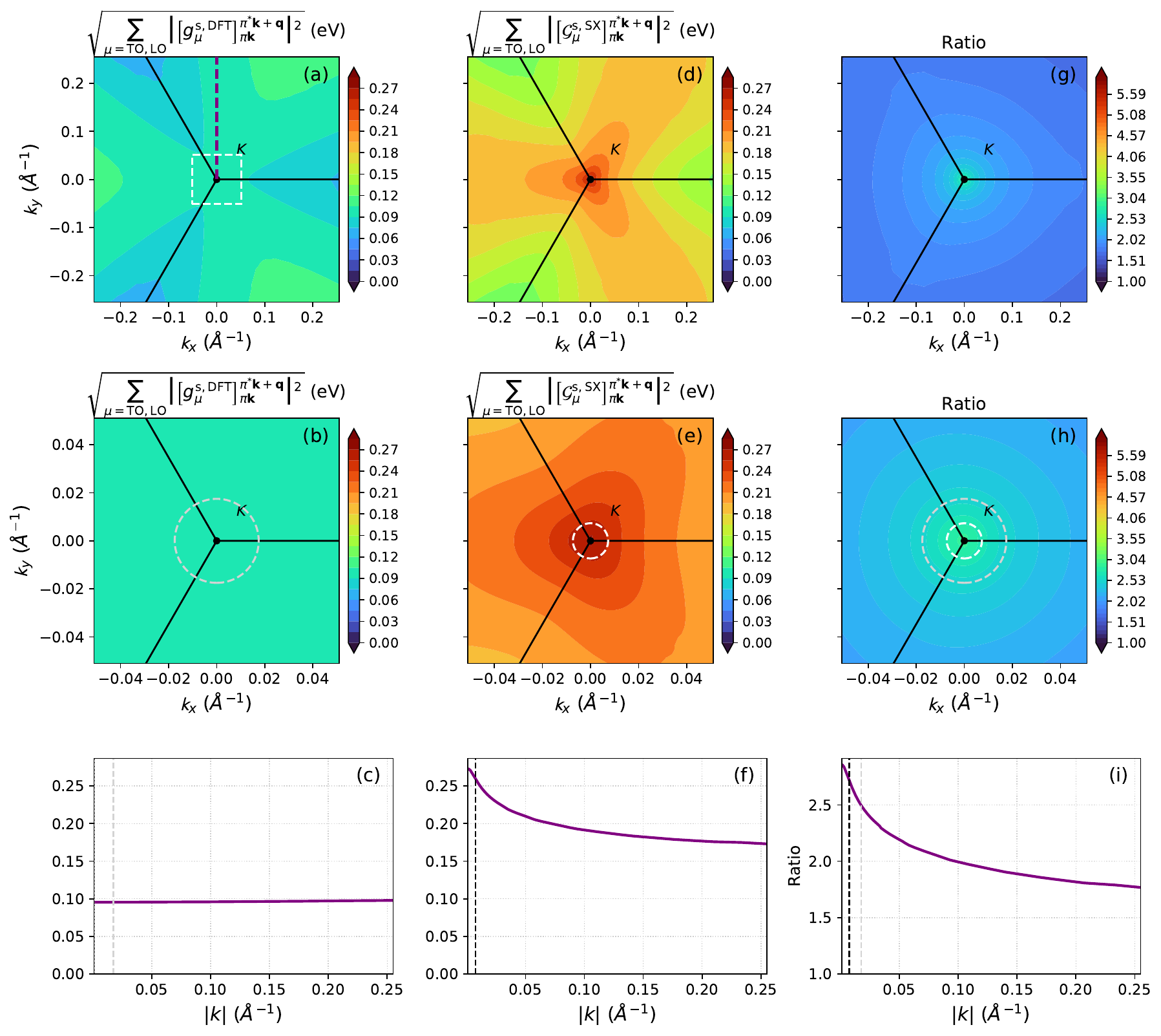}
    \caption{Same figure for the static vertex  as the main text, but for a small $q$ near $\Gamma$. The vertex is still enhanced, even if in a reduced fashion with respect to the $q=K$ case. To perform a fair comparison, we sum over both TO and LO modes, as squared moduli, and then take the square root. (c) and (f) display on the y axis the quantities plotted in (b) and (e), respectively.}
    \label{fig:epcSIGamma}
\end{figure*}
\begin{figure*}
    \centering
    \includegraphics[width=\linewidth]{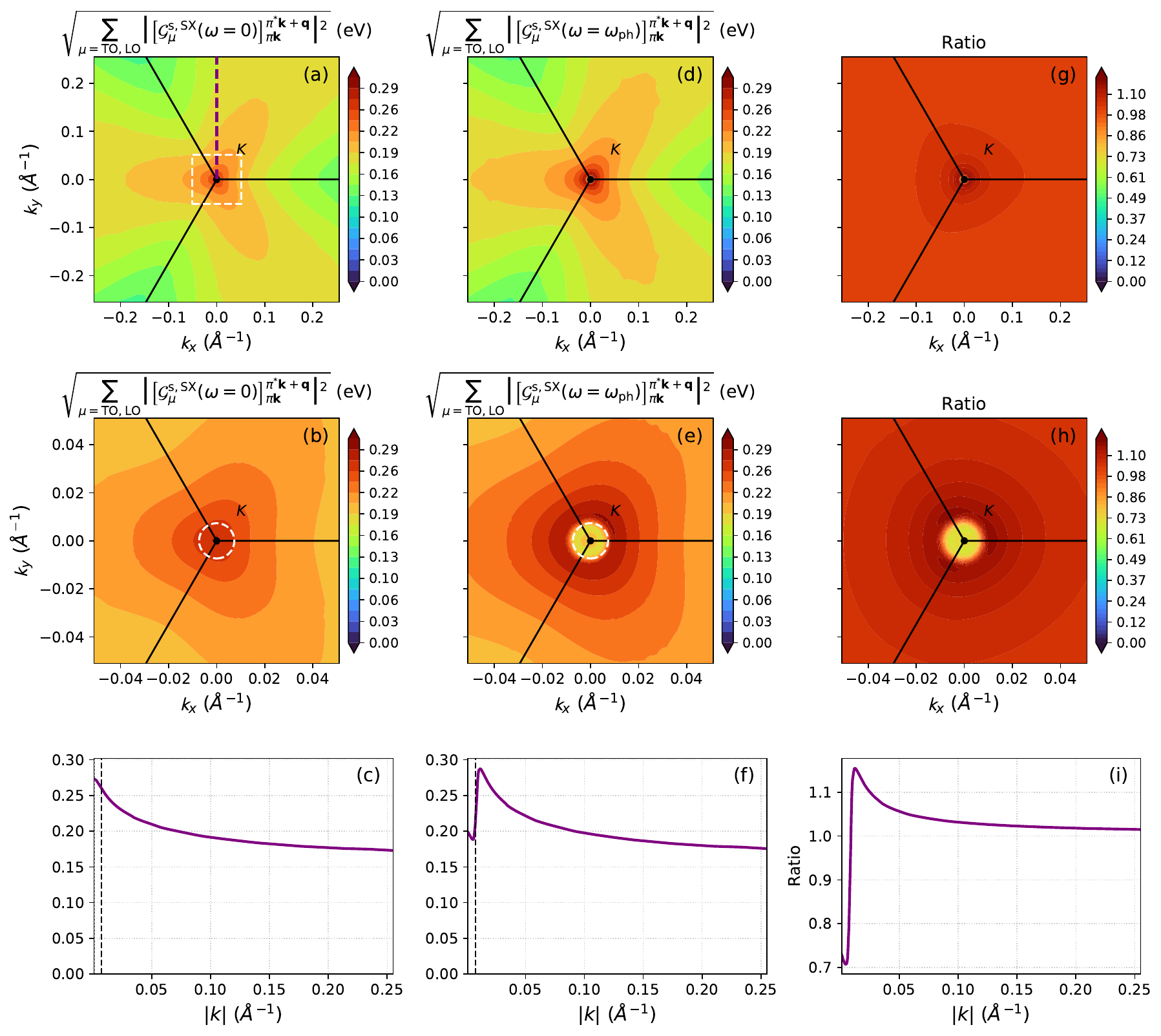}
    \caption{Same as previous figure, but the comparison is between static and dynamical vertex.  (c) and (f) display on the y axis the quantities plotted in (b) and (e), respectively.}
    \label{fig:epcSIGamma}
\end{figure*}
\begin{figure*}
    \centering
    \includegraphics[width=\linewidth]{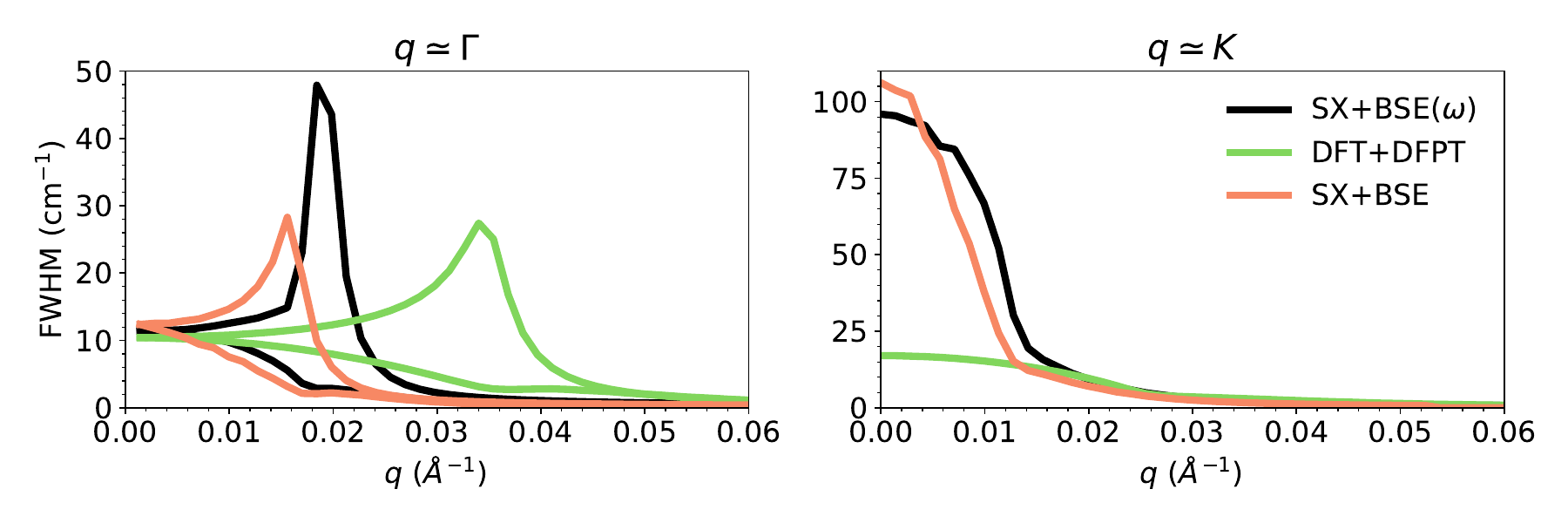}
    \caption{Linewidths for zone-center and zone-boundary computed in different approximations. SX+BSE and DFT+DFPT exactly as in the main text, while SX+BSE($\omega$) means that we perform computations with the dynamical vertex $\mathcal{G}^{\textrm{s,SX}}(\omega^{\textrm{SX}}_{\textrm{ph}})$. $q$ in both plots indicates the distance from, respectively, $\Gamma$ and K. The difference between SX+BSE and SX+BSE($\omega$) is negligible for $\mathbf{q}=\mathbf{K}$, while it slightly shifts resonances near $\Gamma$ when crossing from the interband transition region to the region where electronic transitions are not allowed.}
    \label{fig:phSI}
\end{figure*}

\end{document}